\documentclass[prd, 
floatfix,nofootinbib,longbibliography,notitlepage,superscriptaddress]{revtex4-2}

\pdfoutput=1

\usepackage{amsmath, amssymb, amsfonts, amsthm, latexsym, mathrsfs, bbm, slashed}
\usepackage{booktabs}
\usepackage{array}
\usepackage{makecell}
\usepackage{multirow}
\usepackage{mathtools}

\usepackage[usenames,dvipsnames]{xcolor}
\usepackage[title,titletoc]{appendix}
\usepackage{graphicx}
\usepackage{physics}

\usepackage{xifthen}
\usepackage{tensor}
\usepackage{dsfont}
\usepackage{subcaption}
\usepackage[inline]{enumitem}

\usepackage{setspace}
\usepackage[marginal, multiple]{footmisc}

\usepackage[T1]{fontenc}
\usepackage[utf8]{inputenc}
\usepackage{lmodern}
\usepackage[normalem]{ulem}

\usepackage[unicode, colorlinks, allcolors=blue!70!black, linktocpage, pdfusetitle]{hyperref}
\definecolor{CiteColor}{rgb}{0.55,0,0}
\hypersetup{citecolor=CiteColor}
\definecolor{RefColor}{rgb}{0,0.5,0}
\hypersetup{linkcolor=RefColor}
\usepackage[all]{hypcap}
\usepackage{color}

\usepackage{orcidlink}

\usepackage{verbatim}

\numberwithin{equation}{section}

\usepackage{cancel}

\usepackage{microtype}

\usepackage{standalone}
\usepackage{tikz}

\usepackage{ragged2e}

\usetikzlibrary{arrows.meta, decorations.markings}

\ifdefined\myext
\usetikzlibrary{external}
\fi

\newcommand{\nn}{\nonumber}

\newcommand{\Eq}[1]{Eq.~\eqref{#1}}
\newcommand{\Eqs}[1]{Eqs.~\eqref{#1}}
\newcommand{\ty}[1]{\text{#1}}
\newcommand{\pob}[1]{\left\lbrace  #1  \right\rbrace}

\newcommand{\lseff}{\mathbf{l}\cdot \mathbf{s}_{\text{eff}}}
\newcommand{\hso}{h_{\text{SO}}}
\newcommand{\hsss}{h_{\text{SS}}}
\newcommand{\seff}{\mathbf{s}_{\text{eff}}}
\newcommand{\hone}{ h_{1\text{PN}}}
\newcommand{\htwo}{ h_{2\text{PN}}}

\DeclareMathOperator{\sn}{sn}

\DeclareMathOperator{\am}{am}

\newcommand{\np}{(\mathbf{n}\!\cdot\!\mathbf{p})^2}

\allowdisplaybreaks

\usepackage{graphicx} 
\usepackage{caption}
\usepackage{subcaption} 
\usepackage{physics}

\newcommand{\Meudon}{\affiliation{Laboratoire d'étude de l'Univers et des phénomènes eXtrêmes (LUX), Observatoire de Paris, Université PSL, Sorbonne Université, CNRS, 92190 Meudon, France}}

\newcommand{\utec}{\affiliation{
Universidad de Ingenieria y Tecnologia – UTEC, Jr. Medrano Silva 165 - Barranco 15063, Lima, Peru
}}

\begin{document}
	
\title{Analytical solution of spinning, eccentric binary black hole dynamics at the second post-Newtonian order}	

\author{Tom Colin\,\texorpdfstring{\orcidlink{0009-0004-8392-2808}}{}}	
	\email{tom.colin@obspm.fr }
	\Meudon
	
	\author{Sashwat Tanay\,\texorpdfstring{\orcidlink{0000-0002-2964-7102}}{}}
	\email{stanay@utec.edu.pe}
	\Meudon
	\utec
	
	\author{Laura Bernard\,\texorpdfstring{\orcidlink{0000-0003-2856-1662}}{}}
	\email{laura.bernard@obspm.fr }
	\Meudon

	\begin{abstract}
		\noindent

Recent gravitational wave (GW) detections showing signatures of 
eccentricity and spin precession underscore the need to model 
binary black holes (BBHs) possessing these features simultaneously. 
Most efforts over the past fifteen years to model spinning BBHs and 
their corresponding GWs have relied on heuristically twisting waveforms 
from nonprecessing systems. This approach is based on empirical 
observations rather than first principles.
This article aims to model the dynamics of spinning and eccentric BBHs 
from a first-principles approach
using the post-Newtonian (PN) approximation within general relativity. 
Building on the already-existing 1.5PN solution, 
we construct an analytical solution for the time evolution
of the relative separation vector, the individual black hole 
spin vectors, and the orbital angular momentum vector
at  2PN order for BBHs with arbitrary spins and
eccentricity. Such a solution is not fully 2PN accurate in that
the tiny orbital timescale fluctuations in the solutions
for the spins are only leading 1.5PN order accurate, instead 
of 2PN. However, it is shown that our new 2PN solution is still
an order of magnitude improvement over the earlier
1.5PN solution, underlining the subdominant nature of 
the neglected next-to-leading-order 
oscillations in the spin solutions.

	\end{abstract}

\maketitle

%--------------------------------
%--------------------------------	
\section{Introduction}
%--------------------------------
%--------------------------------

Gravitational wave (GW) science has entered the era of precision astronomy. 
The advent of the next generation of gravitational wave detectors, such 
as Einstein Telescope and the space-based LISA interferometer
will introduce significant data analysis challenges if we wish to extract 
the maximal scientific return from observations. To achieve this goal, 
a key requirement is the ability to model all possible binary systems 
of interest with very high precision.

Following many GW detections that carry
signatures of spinning or eccentric black holes
(BHs) \cite{Hannam:2022pit, LIGOScientific:2020stg,
Romero-Shaw:2020thy},
 the interest in the GW community is 
converging toward such systems. 
Binary black holes (BBHs) with spins
or eccentricity are considered interesting 
because their detection sheds light
on the astrophysical environment 
where these systems evolved and merged 
(isolated or dynamical formation channels,
chemically homogeneous formation, supernova kicks,
hierarchical merger or merger inside 
active galactic nuclei)
\cite{Gerosa:2021mno, Mandel:2018hfr, Romero-Shaw:2025vbc}.
Just as interesting is 
the interplay between the
 spins and the eccentricities. 
While there is a danger that 
the two properties can mimic 
each other, leading to degeneracies
\cite{10.1093/mnras/stad031, Tibrewal:2026jci},
 there have also been efforts 
 to make eccentricity detections 
 sharper with the help of spin measurements
 \cite{Stegmann_2025, Baibhav:2025mzw}.
Given this interesting and treacherous 
terrain, it is desirable to construct
from first principles,
analytical solutions of the BBH orbital dynamics 
and the associated GWs. 
This is so because analytical solutions,
although generally elusive,
expose mathematical structures and relations
between quantities
to a  greater degree than numerical 
or empirically constructed phenomenological ones.
This gives hope to ameliorate the 
degeneracy issue between spins and eccentricity. 
Another aspect is the quick evaluation and 
implementation of these solutions owing to their analytical nature.

Gravitational waveform modeling relies on different techniques to capture 
the distinct phases of compact binary coalescences. While numerical relativity 
is essential for the nonlinear merger regime~\cite{Lehner2001,Sperhake2015} and
black hole perturbation theory is used for the ringdown stage~\cite{Kokkotas1999,Berti2009}, 
the post-Newtonian (PN) formalism remains the standard to determine the early 
inspiral dynamics~\cite{Blanchet2014}. The PN framework's analytic nature allows 
for rapid waveform generation across arbitrary parameter spaces, making it a key 
ingredient to generate full inspiral-merger-ringdown (IMR) models, including the 
effective-one-body (EOB)~\cite{Buonanno1999} and Phenom~\cite{Khan2016} classes 
of IMR waveforms.

To generate gravitational waveforms,
one must first solve
 the equations of motion (EOMs), 
ideally using analytical methods,
before using them  
to derive the expressions 
for the energy, linear and angular momenta fluxes
and the GW modes.
In the absence of analytical orbital solutions 
for spinning BBHs, a certain ``twist up'' method has 
become the standard in the GW community
\cite{Schmidt:2010it, Schmidt:2012rh, Boyle:2011gg}.
This method arose from
the empirical observation
that GWs
from spinning BBHs can be very well 
approximated by applying certain time-dependent 
rotation on GWs originating
from nonprecessing BBHs with the help of
Wigner matrices\footnote{This rotation
approximately tracks the orbital angular 
momentum.}.
Various GW models have been 
built adopting this approach---the state-of-the-art
includes both eccentricity and spin 
simultaneously
\cite{Klein2017b, Arredondo:2024nsl, Morras:2025nlp}.

On the other hand, the construction of
GWs from first principles for spinning BBHs
has been stalled for a long time due to
the unavailability of orbital and spin 
solutions for precessing and eccentric BBHs.
The standard framework for 
solving the dynamics of nonspinning systems
has been the quasi-Keplerian 
parametrization.
Originally established by Damour and Deruelle~\cite{Damour1985} 
for the 1PN dynamics, it has been successfully generalized to 4PN order for nonspinning systems~\cite{Damour2014,Cho2022} and to 3PN for aligned-spin configurations~\cite{Tessmer2010,Tessmer2013}.
However, while these standard parametrizations are typically derived via an analytical integration of the EOMs, the inclusion of precessing 2PN spin-spin 
effects renders analytical integration difficult. 
Reference \cite{Cho2019} by Cho and Lee has been one major
recent breakthrough in this direction.

The goal of the present work is to
 address this challenge by providing an analytical solution
 $\mathbf{R}(t),~\mathbf{P}(t),~\mathbf{S}_1(t),~\mathbf{S}_2(t)$ 
and $\mathbf{L}(t))$ 
 for the orbital and spin motions for
precessing, eccentric BBHs at 2PN order\footnote{We do not provide the explicit solution of $\vb{P}(t)$,
but only give the recipe to construct it.}. 
In particular, we construct our ``hybrid solution''
for the slow angular momenta variables
$\vb{L}, \vb{S}_1, \vb{S}_2$ by hybridizing 
the traditional 2PN orbit-averaged (OA) and
1.5PN full (non-OA) solutions. For the fast variables $\vb{R}, \vb{P}$, we employ the 
ansatz-based extension of the quasi-Keplerian
solution.
Before going to the core of the work,
in the following subsection, we review previous 
attempts to solve the problem, highlighting their respective contributions 
and limitations. We will then provide an executive summary of the results obtained.

%--------------------------------
\subsection{Previous analytical solutions}
%--------------------------------

The development of analytical solutions for the relative separation $\mathbf{R}$ 
and the angular momenta $\mathbf{L}$, $\mathbf{S}_1$, and $\mathbf{S}_2$ 
for spinning binaries on eccentric orbits has proceeded step by step, with 
different works addressing various sectors of the  dynamics. Below, we 
review the key results that provide the foundation for the present work.

\textit{Nonspinning 2PN dynamics.} 
The 1PN
solution was  proposed 
by Wagoner and Will in 1976
\cite{1976ApJ...210..764W}.
Thereafter, a more elegant quasi-Keplerian 
parametric (QKP) solution for the 
relative separation $\vb{R}$ 
was originally established by Damour and Deruelle, again at 1PN order~\cite{Damour1985}.
This was later extended to 2PN by Schäfer and Wex~\cite{Schafer1993}\footnote{The QKP solution for nonspinning eccentric BBHs has been extended to 3PN and 4PN orders too~\cite{Memmesheimer:2004cv, Cho2022}.}. Working with the Arnowitt-Deser-Misner (ADM) Hamiltonian~\cite{Arnowitt2008}, they derived the explicit expressions for 
radial separation magnitude $R \equiv \Vert  \mathbf{R}  \Vert$,
and the orbital phase through 
analytical integration 
of the EOMs.
A defining feature of the nonspinning conservative dynamics is the conservation of the orbital angular momentum vector $\vb{L} \equiv \vb{R} \times \vb{P}$,
which restricts the motion to a fixed plane. However, unlike in Newtonian gravity, the orbits are not closed ellipses; they exhibit a relativistic periastron advance, characterized by the quantity $k$. 
Consequently, the radial and angular motions acquire different frequencies
 necessitating distinct eccentricity parameters ($e_r$, $e_t$, $e_\phi$) to describe respectively the time evolution of the radial separation and the orbital phase via the 
 generalized Kepler equation.
Both the 1PN and 2PN QKP solutions
share the same basic structure
but the 2PN solution
introduces additional periodic corrections (with respect to the 1PN solution) proportional to the higher harmonics of a true anomaly. This framework forms the foundation upon which spin effects will be incorporated later in 
this article.

\textit{Spin-orbit dynamics at 1.5PN.} 
The work by Cho and Lee was a major breakthrough toward
incorporating the 1.5PN spin-orbit couplings 
for systems with arbitrary mass ratios, spin orientations and
eccentric orbits~\cite{Cho2019}.
Using the 1.5PN 
ADM Hamiltonian along with the 
Newton-Wigner-Pryce (NWP)
spin supplementary condition (SSC),
they constructed the solution for 
$\vb{R}, \vb{L}, \vb{S}_1, \vb{S}_2$
in terms of elliptic integrals,
while ignoring the 1PN terms 
in the Hamiltonian\footnote{Reference \cite{Cho2019} ignored the 1PN terms in the Hamiltonian because it considered the 1.5PN spin-orbit terms to be the central
difficulty.}.
Their solution for $\vb{R}$  was provided using a QKP.
While their solution of $R$ was similar to the corresponding 1PN solution 
of Ref.~\cite{Damour1985}, their
solution for the orbital phase  $\phi(t)$ had
additional terms (with respect to
the 1PN solution) involving 
incomplete elliptic integrals of the third kind. These  terms capture
the precession 
of the orbital plane induced by spin-orbit coupling. Reference \cite{Samanta2022} 
later completed this 1.5PN solution
of Ref.~\cite{Cho2019}
by incorporating the 1PN nonspinning orbital corrections 
that were initially omitted by the latter.

\textit{Orbit-averaged 2PN spin dynamics.} 
The inclusion of 2PN spin-spin interactions makes the obtention of  analytical solutions difficult. To overcome this difficulty, orbit averaging has been used in the past. 
Orbit averaging filters out 
the tiny and short orbital timescale
oscillations that the full solution 
possesses, thereby simplifying the EOMs,
and making the construction of the 
OA solution possible.
The OA EOMs at 2PN 
were first worked out by Schnittman in Ref.~\cite{Schnittman2004}.
Racine identified a new constant of motion (COM)
for these EOMs, 
which ultimately led to
the construction of the 
OA solution 
in Refs.~\cite{Gerosa:2015hba, Gerosa:2015tea, Klein2017b}.
Note that the unaveraged solutions at
1.5PN of Refs.~\cite{Cho2019,Samanta2022}
 do contain these tiny oscillations.
%(to be discussed in more detail later).

Another interesting development 
was the discovery of two
2PN perturbative 
COMs
in Ref.~\cite{Tanay2020} for the 
full 2PN EOMs (without orbit-averaging),
thereby establishing 
the system's
2PN perturbative integrability.
Just as Racine's COM
\cite{Racine2008} 
helped construct the
exact solution for 2PN OA EOMs,
we will use the 2PN unaveraged 
COMs of Ref.~\cite{Tanay2020} 
to construct our solution
in later sections.

\textit{2PN spin effects in orbital motion.} 
Gergely~\cite{Gergely:1999pd} and later
Klein and Jetzer~\cite{Klein2010} worked out the solution
for the magnitude $R$
of the orbital position vector
at 2PN order
including both spin-orbit and spin-spin effects. 
Working in harmonic coordinates 
with Kidder's Lagrangian formulation~\cite{Kidder1995}, Ref.~\cite{Klein2010} adopted 
an SSC different from the NWP SSC used in the ADM-based
approaches of Refs.~\cite{Cho2019, Samanta2022}\footnote{As demonstrated by Barker and O'Connell~\cite{Barker1974}, different spin supplementary 
conditions correspond to different choices of the center-of-mass location and are related by coordinate 
transformations; they represent physically equivalent descriptions.}.

\subsection{Motivation and scope of the present work}    \label{sec:motivation}

When it comes to solving the 
dynamics of BBHs with arbitrary 
spins and eccentricity, the current state of the art is
the solution presented in
 Refs.~\cite{Cho2019, Samanta2022},
which is valid at 1.5PN order\footnote{The
 solution of Ref.~\cite{Klein2010} 
does not track the precession of the 
orbital angular momentum relative to the 
inertial frame, leading to apparent
ambiguities 
in the definition and 
evolution of the orbital phase $\phi$, in our opinion.}.
This paper is an attempt to
extend the solution to 2PN order 
with the ADM Hamiltonian and 
NWP SSC.

An interesting aspect of our work is that the 2PN spinning BBH system has been shown to be integrable only in a perturbative sense~\cite{Tanay2020}\footnote{This means that there exist the requisite number (half the number of phase space variables) of quantities which are conserved and are in mutual involution up to  2PN order. This conservation and involution is not exact.}. 
In a strict sense, it could be chaotic---see Ref.~\cite{Hartl:2004xr} and the Introduction of Ref.~\cite{konigs:2005zz} for a nice summary of the debate that lasted over many decades. This possible chaotic nature is a departure from the traditional systems for which solutions have been worked out in the past in that most of them were exactly integrable. Examples include the Newtonian system, 1-4PN system (without spins), or even the 1.5PN spinning BBH~\cite{Cho2019, Samanta2022}.
The key idea is that just as one needs the exact COMs to solve exactly, an exactly integrable system, we will need the perturbative COMs (discovered in Ref.~\cite{Tanay2020}) of the perturbatively integrable 2PN spinning BBH system.

Our calculations can broadly be divided into
two parts.
First, Secs.~\ref{sec sep timescale}--\ref{Hybrid def}
deal with orbital and spin angular momenta
($\mathbf{L}$, $\mathbf{S}_1$, and $\mathbf{S}_2$).
There, we construct a solution that goes beyond the usual
orbit-averaging by hybridizing
the non-OA 1.5PN solution of Ref.~\cite{Samanta2022}
with the 2PN OA
solution of 
Ref.~\cite{Klein2017b}. 
The resulting hybrid solution retains 
tiny orbital timescale variations 
which are normally 
washed away during the process of 
orbit averaging while constructing 
the OA solution. 
Our resulting ``hybrid solution''
is fully 1.5PN accurate but only
partially 2PN accurate. Because we 
do account for the dominant 2PN effects
(while ignoring certain subdominnant 2PN effects), our hybrid solution
can be considered ``nearly 2PN accurate.''

After having tackled the angular momenta,
in Sec.~\ref{sec:QKP0}, taking inspiration 
from the ansatz-based approach of 
Klein and Jetzer~\cite{Klein2010}, we construct the
solution of position vector
magnitude $R(t)$ and phase $\phi(t)$ in a noninertial frame centered on $\mathbf{L}$.
 While these expressions
are themselves 2PN accurate,
the final determination of the position vector $\mathbf{R}$
is, again, only ``nearly 2PN accurate.''
This is due to the fact that the 
definition of $\phi$ hinges on 
the definition of $\mathbf{L}$, whose 
hybrid solution is also nearly 2PN accurate, as mentioned above.
Our solution is accompanied by plots so as to let the reader visually inspect its performance relative to the older 1.5PN solutions.

Later in Sec.~\ref{sec:four_frequencies}, we give the four frequencies of the 2PN spinning BBH system, before 
verifying in Sec.~\ref{sec:limits} that our solution converges toward the older results in the literature for the circular and the spin-aligned limits.
Finally, for plotting and implementation purposes, 
the analytical solutions derived in this work have been implemented in a \textit{Mathematica} notebook. This notebook, along with equivalent scripts in other accessible file formats, is available in a public GitHub repository \cite{Tom_Colin_2026_BBH_Notebook}; see Sec.~\ref{sec:conclusion}
for more details.

\subsection{Notation and conventions}
\label{sec:notation_conv}

\textit{Mass parameters.}
We consider a BBH system with individual masses $m_1 > m_2$, total mass $m = m_1 + m_2$, reduced mass $\mu = m_1 m_2/m$, and symmetric mass ratio $\nu = \mu/m$. We further define the auxiliary mass-weighted parameters $\delta_a = 2\nu(1 + 3m_b/4m_a)$ and $\nu_a =  m_b/m$, with indices $(a,b) \in \{(1,2), (2,1)\}$.

\textit{Orbital variables.}
We work in the center-of-mass frame and employ reduced variables, denoted by lowercase letters. The relative separation vector is rescaled by the product of the gravitational constant $G$ and the total mass $m$, $\mathbf{r} = \mathbf{R}/(Gm)$, while the canonical momentum is rescaled by the reduced mass, $\mathbf{p} = \mathbf{P}/\mu$. The physical time $t'$ is rescaled as $t = t'/(Gm)$.
 Overdots denote derivatives with respect to this reduced time, $\dot{x} \equiv dx/dt$. The unit separation vector is $\mathbf{n} = \mathbf{r}/r$, with $r = \|\mathbf{r}\|$. 
 We also define the reduced Hamiltonian $h=H/\mu$.

\textit{Angular momenta.}
The reduced orbital angular momentum is defined as $\mathbf{l} = \mathbf{r} \times \mathbf{p}$. The individual spin vectors are also rescaled by $\mu G m$, such that $\mathbf{s}_a = \mathbf{S}_a/(\mu G m)$. The total reduced angular momentum is $\mathbf{j} = \mathbf{l} + \mathbf{s}_1 + \mathbf{s}_2$. We characterize the spin interactions using two specific linear combinations: the effective spin $\mathbf{s}_{\mathrm{eff}} = \delta_1 \mathbf{s}_1 + \delta_2 \mathbf{s}_2$ and the auxiliary vector $\mathbf{s}_0 = \nu_1 \mathbf{s}_1 + \nu_2 \mathbf{s}_2$.

\textit{Conventions:} In this work, the post-Newtonian (PN) order is defined by the scaling parameter $x_{\text{PN}} \equiv G m/(c^2 r_p)$, where $r_p$ is the reduced relative separation at periastron. 
An $n$PN correction to a certain leading-order (LO) quantity
contains an additional factor
of  $x_{\text{PN}}^n$ with respect to the 
LO quantity. Finally, we set $G=1$ throughout this work. 
Vectors are represented by bold letters,
whereas regular versions of these letters represent their magnitudes; for example $V \equiv \Vert \vb{V} \Vert$.

%--------------------------------
\section{Hamiltonian formulation}
\label{sec:hamiltonian}
%--------------------------------

%--------------------------------
\subsection{The 2PN Hamiltonian}
%--------------------------------
We consider a BBH system consisting of two spinning BHs with masses 
$m_1$ and $m_2$ in a 
bound orbit around their common center of mass. The dynamics of the system are governed by a Hamiltonian 
expressed in the ADM coordinates~\cite{Arnowitt2008} with the NWP SSC~\cite{Pryce1948,Newton1949}, which fixes the center-of-mass worldline of each spinning body. 
The (reduced) Hamiltonian reads\footnote{The
Hamiltonian implicitly assumes
the plus signature $(-,+,+,+)$
for the spacetime metric.
} \cite{Damour2001, Hartl:2004xr, Barker:1966zz, PhysRevD.12.329}
\begin{equation}\label{eq:hamiltonian}
	h = h_N + \frac{1}{c^2} \hone + \frac{1}{c^2} \hso +  \frac{1}{c^4} \htwo +  \frac{1}{c^2} \hsss,
\end{equation}
where 
\begin{subequations}\label{eq:hamiltonian_terms}
	\begin{align}
		h_N &= \frac{p^2}{2} - \frac{1}{r} \,, \label{eq:hN}\\
		\hone &=  {\frac{3\nu-1}{8} p^4 - \frac{1}{2r}\big[(3+\nu)p^2+\nu(\mathbf{n}\cdot\mathbf{p})^2\big]}+ \frac{1}{2r^2} \,,\label{eq:h1PN}\\
		\hso &=  \frac{\lseff}{r^3} \,, \label{eq:hSO}\\
		\htwo &= \frac{1-5\nu+5\nu^2}{16}p^6+ \frac{1}{8r}\big[ (5-20\nu-3\nu^2)p^4-2\nu^2(\mathbf{n}\cdot\mathbf{p})^2 p^2-3\nu^2(\mathbf{n}\cdot\mathbf{p})^4 \big]\nonumber\\
		&\qquad + \frac{1}{2r^2}\big[3\nu(\mathbf{n}\cdot\mathbf{p})^2 + (5+8\nu)p^2\big]  -\frac{1+3\nu}{4r^3}  \,, \label{eq:h2PN}\\
		\hsss &= \frac{1}{2r^3}\left[3(\mathbf{n}\cdot \mathbf{s}_0)^2-s_0^2 \right]\,.  \label{eq:hSS}
	\end{align}
\end{subequations}
The terms $h_N$, $\hone$, and $\htwo$ correspond, respectively, to the Newtonian, 1PN, 
and 2PN nonspinning contributions. The 1.5PN spin-orbit effects are captured in $\hso$, while $\hsss$ 
accounts for the 2PN spin-spin interactions. Up to 2PN, there is no emission of gravitational waves and the system is therefore conservative.

For a Kerr black hole, the spin angular momentum is given by
\cite{Blanchet2014}
\begin{equation}
\mathbf{S}_a = \boldsymbol{\chi}_a\frac{G m_a^2}{c},
\end{equation}
where $|\boldsymbol{\chi}_a| \leq 1$,
 with equality corresponding to a maximally spinning black hole. 
The implicit factor of $1/c$ hiding inside $\mathbf{S}_a$
 alters the PN power counting: as a consequence, the spin-orbit coupling 
formally enters at 1.5PN order, while the spin-spin coupling appears at 2PN order.

In our Hamiltonian framework, we have the following phase-space variables:
$\vb{R}, \vb{P}, \vb{S}_1, \vb{S}_2$.
The EOMs for any function $f$ of the phase-space variables are obtained from its Poisson bracket with the Hamiltonian
\begin{equation}\label{eq:PB}
	\frac{df}{dt'} = \{f, H\}\,.
\end{equation}
The Poisson brackets
for the phase-space variables read
\cite{Steinhoff:2011sya}
\begin{subequations}\label{eq:PB_fundamental}
	\begin{align}
		\{R_i, P_j\} &= \delta_{ij}\,, \\
		\{S_{ai}, S_{bj}\} &= \delta_{ab}\,\epsilon_{ijk}\,S_{bk}\,,
	\end{align}
\end{subequations}
where $\delta_{ij}$ is the Kronecker delta, $\epsilon_{ijk}$ is the Levi-Civita symbol, and the indices $a,b \in \{1,2\}$ 
label the two bodies while $i,j,k \in \{1,2,3\}$ label spatial components. 
Additionally, for functions $f$ and $g$ of 
phase-space variables, the Poisson brackets obey 
the antisymmetry property and the chain rule,
\begin{align}
	\{f,g\} &= -\{g,f\} \, ,\\
	\{f, g(\mathbf{V})\} &= \{f, \mathbf{V}\}\frac{\partial g}{\partial \mathbf{V}}\, .      \label{eq: PB chain rule}
\end{align}
Any other Poisson bracket
not lying within
the purview of Eqs.~\eqref{eq:PB_fundamental}-\eqref{eq: PB chain rule} simply vanishes; an example is 
$\pob{S_{1x}, R_y} = 0$.
Note that the above rules 
also allow us to compute 
the Poisson bracket of any  two
functions of the phase-space variables.

%--------------------------------
\subsection{Equations of motion for the angular momenta}
\label{eq:EOM-for-ang-moemnta}
%--------------------------------

Equations (\ref{eq:PB})--(\ref{eq: PB chain rule}), along with
Eq.~(\ref{eq:hamiltonian})
yield the EOMs for the three angular momenta~\cite{Racine2008}\footnote{Note that the EOMs in Eqs.~(2.8)-(2.10) of Ref.~\cite{Racine2008} have a typo.
The sign associated with $( \hat{\mathbf{n}}\cdot \mathbf{S}_0)\hat{\mathbf{n}}\times  \mathbf{S}_0 $ in their Eq.~(2.10) should be flipped to a minus. 
The same typo propagates to its Eq.~(3.2c).}, 
\begin{subequations}\label{eq:spin_precession}
\begin{align}
\frac{d\mathbf{s}_1}{dt} &= \frac{1}{c^2 r^3}\left\{\delta_1 \mathbf{l} + \nu_1\left[3(\mathbf{n}\cdot \mathbf{s}_0)\mathbf{n}-\mathbf{s}_0\right] \right\} \times \mathbf{s}_1\,, \label{eq:ds1dt}\\
\frac{d\mathbf{s}_2}{dt} &= \frac{1}{c^2 r^3}\left\{\delta_2 \mathbf{l} + \nu_2\left[3(\mathbf{n}\cdot \mathbf{s}_0)\mathbf{n}-\mathbf{s}_0\right] \right\} \times \mathbf{s}_2\,, \label{eq:ds2dt}\\
\frac{d\mathbf{l}}{dt} &= -\frac{1}{c^2 r^3}\left\{ \mathbf{l}\times \seff +3 \left[(\mathbf{n}\cdot\mathbf{s}_0)(\mathbf{n} \times \mathbf{s}_0)\right]\right\}\,, \label{eq:dldt}
\end{align}
\end{subequations}
where the 2PN contributions are represented by the terms in square brackets. 
It follows from Eqs.~\eqref{eq:spin_precession} that $d\mathbf{j}/dt=0$, confirming the conservation of the total angular momentum.

These EOMs reveal some important features of the 2PN spin dynamics.
First, the timescale of variation  
of spins is
\begin{align}
\tau_{\text{sp}} \sim \frac{||\mathbf{s}_1||}{||d \mathbf{s}_1/dt||}  \sim ~  \frac{c^2 r^3}{l}  \sim ~  c^2 a^{5/2}  ,     \label{eq:T_sp}
\end{align}
where $a$ is the reduced Newtonian semimajor axis. This timescale is 1PN order 
larger than the orbital timescale $\tau_{\text{orb}}\sim a^{3/2}$.
Also, the first terms on the right-hand side (rhs) of 
Eq.~\eqref{eq:spin_precession} arise from the spin-orbit coupling $h_{\ty{so}}$, while the remaining terms originate from 
spin-spin interaction $h_{\ty{ss}}$. 
A crucial distinction from the 1.5PN case is that $l \equiv \Vert \vb{l} \Vert$ is no longer conserved at 2PN order. This can be verified by computing
\begin{equation}\label{eq:dldt_magnitude}
	\frac{d(l^2)}{dt} = 2\mathbf{l}\cdot\frac{d\mathbf{l}}{dt} = -\frac{6}{c^2 r^3}(\mathbf{n}\cdot\mathbf{s}_0)(\mathbf{l}\cdot(\mathbf{n}\times\mathbf{s}_0)) \neq 0\,.
\end{equation}
In contrast, the spin magnitudes $s_1 = \|\mathbf{s}_1\|$ and $s_2 = \|\mathbf{s}_2\|$ remain constant, as can 
be shown by taking $\mathbf{s}_a \cdot d\mathbf{s}_a/dt = 0$ from the cross-product structure 
of Eqs.~\eqref{eq:ds1dt} and \eqref{eq:ds2dt}.
Another important consequence of the 2PN spin-spin interactions is that the quantity $\lseff$, which is conserved 
at 1.5PN order, now varies with time \cite{Tanay2020}. 
These features prevent a direct application of the method 
developed in Ref.~\cite{Cho2019}.
However, as we will demonstrate in 
Sec.~\ref{sec:averaged}, 
orbit averaging restores certain 
COMs, which then
allows one to get an analytical solution~\cite{Klein2017b}.
Our hybrid approach builds on top of 
the orbit-averaging approach.

%--------------------------------
\section{Spin sector solution}
\label{sec:averaged}
%--------------------------------

%--------------------------------
\subsection{General characteristics of spin dynamics}
\label{sec:From 1.5PN to 2PN solution}
%--------------------------------
\begin{figure}[h!]  
\centering
\includegraphics[width=1\textwidth]{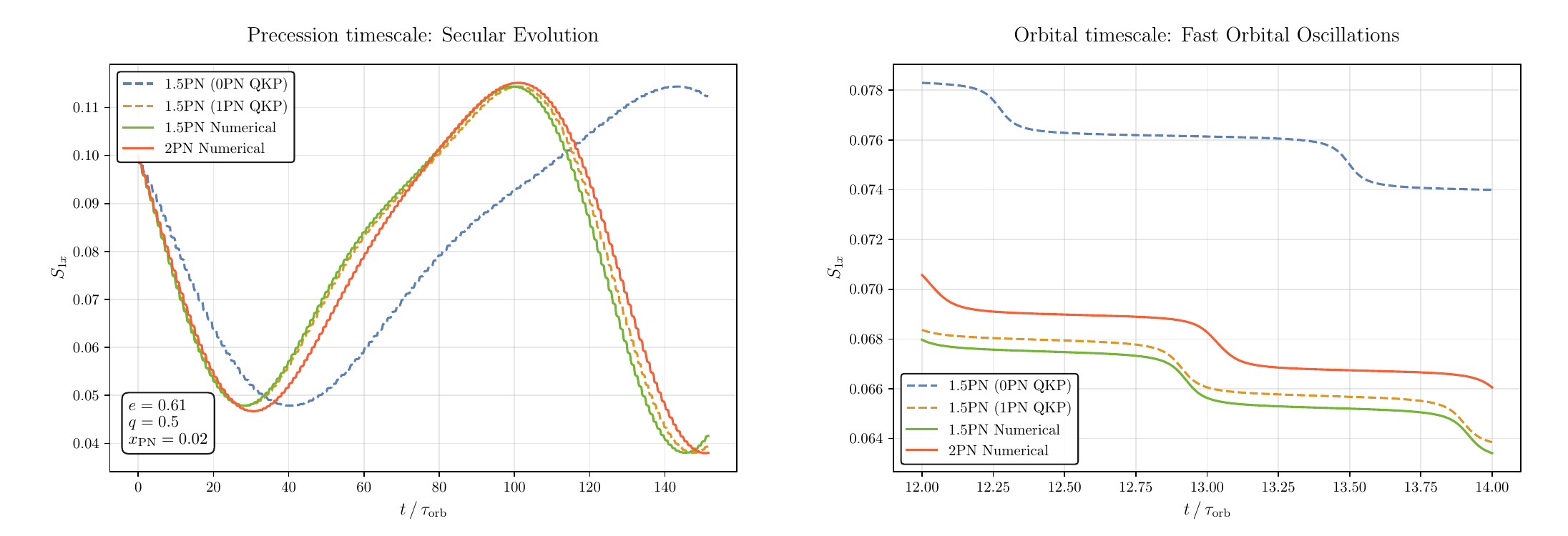}
\caption{\justifying
Evolution of 
$\mathbf{S}_{1x}$ for an eccentric BBH with
$e = 0.61$, $q \equiv m_2/m_1 = 0.5$, and $x_{\mathrm{PN}} = 2 \times 10^{-2}$. Initial angles: 
$\langle \mathbf{L}, \mathbf{S}_{1} \rangle = 32^\circ$,
$\langle \mathbf{L}, \mathbf{S}_{2}\rangle = 82^\circ$,
$\langle \mathbf{S}_{1}, \mathbf{S}_{2}\rangle = 54^\circ$.
We compare the evolution under four different evolution
schemes: 1.5PN solution with Newtonian $r$ injected
 (0PN QKP, blue dashed), the same analytical solution 
with 1PN $r$ injected (1PN QKP, orange dashed), 
1.5PN numerical solution
(green), and 2PN numerical solution (red).
Left: evolution over one slow precession period.
Right: two-orbit zoomed-in view into the time interval $[12\tau_{\mathrm{orb}}, 14\tau_{\mathrm{orb}}]$,
highlighting the fast orbital-timescale oscillations.
}
\label{plot:S1xA}
\end{figure}

Before presenting our solution,
 let us review some general characteristics of the spin 
 sector solution. This will help us strategize 
 the construction of our solution. 
 This subsection aims at presenting the general behavior 
 of the different existing and to-be-presented solutions, 
 while the necessary mathematical justifications of our 
 statements is delayed to later subsections.

We start with Fig.~\ref{plot:S1xA},
where we plot $S_{1x}$ for a certain initial condition. 
The green and red plots  correspond to the numerical solutions to the EOMs
obtained from the 1.5PN and 2PN Hamiltonians, respectively.
The plots immediately reveal the two-timescale 
feature of the solutions. The left panel 
of the figure clearly shows 
that $S_{1x}$ primarily changes at 
the slower spin precession timescale $\tau_{\ty{sp}}$, 
which is 1PN order
larger than the orbital timescale.
On the other hand, the zoomed-in version 
in the right panel shows the subdominant 
oscillations occurring on  
a much shorter orbital timescale $\tau_{\text{orb}}$.

The oscillations at these two timescales 
have already been successfully incorporated in the 1.5PN 
solution by Cho and Lee in Ref.~\cite{Cho2019},
while omitting the 1PN terms throughout.
One crucial aspect is that
while constructing the spin sector solution, or later in Sec.~\ref{Hybrid def} our
``hybrid solution,'' one must inject a lower order (Newtonian or higher-PN-order)
 $r$ solution into a differential equation 
governing the change of the cosine 
of the angle between two of the three 
angular momenta\footnote{See 
 Eq.~\eqref{eq:dxdt-hybrid}.}. As will be made mathematically clear  after Eq.~\eqref{eq:kappa1_solution_hyb}, the PN accuracy of this injected $r$ solution determines
the accuracy of both the slow
and fast frequencies\footnote{ Eqs.~\eqref{eq:freq_nut} and \eqref{eq:freq_prec}, to be encountered later, give these frequencies.}
that build up the solution of 
$\vb{S}_1$ or $\vb{S}_2$.
This is clearly borne out in the plots of 
Fig.~\ref{plot:S1xA}, where the analytical solution 
of $S_{1x}$ obtained by the injection of Newtonian (resp. 1PN) 
$r$ is plotted in blue (resp. orange)\footnote{Injection of 
1PN $r$ in the context of Ref.~\cite{Cho2019} does not make sense since Ref.~\cite{Cho2019} omits the 1PN effects throughout for simplicity. In Fig.~\ref{plot:S1xA}, the 1.5PN analytical solution is the one that is fully 1.5PN accurate (subject matter of Ref.~\cite{Samanta2022}). 
}. It is remarkable
to note the improvement brought about by injecting the 
1PN solution for $r$, over simply injecting the Newtonian $r$.
The left and right panels of Fig.~\ref{plot:S1xA} cover long and short timescales, respectively, and hence are meant to display the slow and fast frequency dynamics.
The improvement brought about by injecting the 1PN $r$ 
is clearly  noticeable 
at both the slow (left panel) and fast timescales (right panel).

Although not shown in these plots,
the OA solution for spins captures 
only the slow, long-term, large oscillations, while filtering out the tiny, small timescale variations, as the name suggests.
Just like the non-OA solution requires an $r$ to be injected,
the OA EOMs require for their construction, a background orbital 
trajectory to average over. For this, the 
Newtonian orbit has been the traditional choice~\cite{Racine2008}. 
We will mathematically establish
later at the end of Sec.~\ref{sec:sol-cos-kappa-1}
 that this choice of background trajectory to average over determines the slow frequencies of the OA dynamics. Therefore, in light of the discussion above, when it comes to
determining the slow frequencies of spin dynamics, the analog of ``which $r$ to be injected'' for non-OA solutions is ``which background orbital trajectory to average over'' for the OA solution.

We pause here to point out that
the improved accuracy in Fig.~\ref{plot:S1xA}
due to the injection of 1PN $r$
is not fully physical. To see this, note that for constructing the 1.5PN accurate solution for the angular momenta, the injection of 
Newtonian order $r$ is sufficient, as was done in Ref.~\cite{Cho2019}.
Injecting  a 1PN $r$ would therefore 
bring about a 2.5PN effect, while our 
EOMs are only 2PN accurate. 
In our work, 
we will still continue to retain 
  such unphysical accuracy by 
  injecting an $r$ solution with 
  more-than-required accuracy because
  doing so gives us useful insights for
  increasing the real physical PN  accuracy 
  if we were to start off with 
  higher-PN-order EOMs. 
This aspect of unphysical accuracy
can also manifest with the OA solutions if we employ a 1PN orbit
to average over instead of the Newtonian one. Doing so brings about change in the slow frequencies
at a relative 1PN order, while in this work, we aim to improve the BBH dynamical solutions by only half PN order (1.5PN to 2PN).

In the following sections, we construct our solution in two steps. First, we use an OA 
approach (Sec.~\ref{sec sep timescale}) to incorporate the 2PN spin-spin 
contributions that improve the slow precessional envelope. Second, to recover the fast orbital 
oscillations while preserving this secular behavior, we introduce a hybrid solution (Sec.~\ref{Hybrid def}) 
that blends the OA dynamics with the LO oscillatory features.

\subsection{Orbit averaged solution}
\label{sec sep timescale}

\subsubsection{Orbit averaged equations of motion}
\label{sec:OA-EOM}

We now proceed to orbit-average Eqs.~\eqref{eq:spin_precession} to model the slow spin dynamics that take place on the spin precession timescale.
Following the discussion in Sec.~\ref{sec:From 1.5PN to 2PN solution}, and 
with the aim of improving the 
slow precessional frequencies by 0.5PN
order, we choose to average the 
LO spin-orbit terms of 
Eqs.~\eqref{eq:spin_precession}
over the 1PN orbit (thus  inducing unphysical PN accuracy mentioned 
at the end of Sec.~\ref{sec:From 1.5PN to 2PN solution}), while the next-to-leading-order (NLO) 
spin-spin terms in the square brackets are averaged over a Newtonian orbit.

The averages are computed by treating the angular momenta as constant over an orbital period. For the 
LO terms, we employ the 1PN quasi-Keplerian parametrization of Ref.~\cite{Damour1985},
\begin{align}
	r &= a_r(1-e_r\cos u), \label{eq:kepler1PN-a} \\
	n(t-t_0) &= u - e_t \sin u, \label{eq:1PN-n}
\end{align}
where the reduced quasi-Keplerian orbital elements are 
\begin{align}
	a_r &= -\frac{1}{2h}\left(1 - \frac{(\nu-7)h}{2c^2}\right), \\
	e_r^2 &= 1 + 2hl^2 - \frac{2(6-\nu)h}{c^2} - \frac{5(3-\nu)h^2l^2}{c^2}, \\
	n &= (-2h)^{3/2}\left(1 + \frac{(15-\nu)h}{4c^2}\right), \\
	e_t^2 &= 1 + 2hl^2 + \frac{4(1-\nu)h}{c^2} + \frac{(17-7\nu)h^2l^2}{c^2}. \label{eq:kepler1PN-b}
\end{align}
Additionally, we define $d_\text{N} \equiv a \sqrt{1 - e^2}$, where $a$ and $e$ are the Newtonian limits of $a_r$ and $e_r$. 

The two orbit averages required to orbit-average Eqs.~\eqref{eq:spin_precession} are
\begin{subequations}  \label{eq:two_averages}
    \begin{alignat}{2}
        & \left\langle \frac{1}{r^3} \right\rangle_{\text{1PN}} && = \frac{1}{d^3}, \\
        & \left\langle \frac{n^i n^j}{r^3} \right\rangle_{\text{N}}      && = \frac{1}{2d_N^3}\left(\delta^{ij} - \frac{l^i l^j}{l^2}\right),
    \end{alignat}
\end{subequations}
with $d$ given by
\begin{align}
     d \equiv a_r\sqrt{1-e_\theta^2}, \quad e_\theta \equiv \frac{3e_r-e_t}{2}. \label{def:d}
\end{align}

Equations \eqref{eq:spin_precession} when 
averaged over in this manner, yield the  OA EOMs\footnote{Strictly speaking,
averaging Eqs.~\eqref{eq:spin_precession}
with the aid of Eqs.~\eqref{eq:two_averages}
leads to a factor of $1/d_\text{N}^3$ (instead of $1/d^3$) for the NLO terms.  However, replacing $d_\text{N}$ with $d$ induces a relative 1PN order effect, or 3PN order absolute effects, which we can neglect.}
\begin{subequations}
\label{eq:OA-1PN-b}
\begin{align}
\left\langle \frac{d\mathbf{s}_1}{dt} \right\rangle &= \frac{1}{c^2 d^3} \left\{ \delta_1\,\mathbf{l} + \frac{\nu_1}{2}\left[\mathbf{s}_0- 3 \frac{(\mathbf{l}\cdot \mathbf{s}_0)}{l^2}\mathbf{l}\right]\right\} \times \mathbf{s}_1, \\
\left\langle \frac{d\mathbf{s}_2}{dt} \right\rangle &= \frac{1}{c^2 d^3} \left\{ \delta_2\,\mathbf{l} + \frac{\nu_2}{2}\left[\mathbf{s}_0- 3 \frac{(\mathbf{l}\cdot \mathbf{s}_0)}{l^2}\mathbf{l}\right]\right\} \times \mathbf{s}_2, \\
\left\langle \frac{d\mathbf{l}}{dt} \right\rangle &= \frac{1}{c^2 d^3} \left\{ \mathbf{s}_{\mathrm{eff}} - \left[ \frac{3}{2}\frac{(\mathbf{l}\cdot \mathbf{s}_0)}{l^2} \mathbf{s}_0\right] \right\} \times \mathbf{l}. \label{eq:dldt_avg}
\end{align}
\end{subequations}
The traditional OA EOMs obtained by averaging
over the Newtonian orbit are the same as Eqs.~\eqref{eq:OA-1PN-b}, but with the replacement
$d \rightarrow d_N$~\cite{Racine2008, Klein2017b, Gerosa:2015tea}.
Henceforth, for notational brevity, we drop the explicit orbit-average symbol $\langle\cdot\rangle$, 
with the understanding that all quantities represent orbit-averaged variables, up until the end of Sec.~\ref{sec sep timescale}.

Equations \eqref{eq:OA-1PN-b} admit a COM
\begin{align}
\lambda \equiv \frac{(\mathbf{l}\cdot\mathbf{s}_0)}{l^2}, \label{eq:lambda}
\end{align}
first identified in Ref.~\cite{Racine2008}. This COM arises after orbit-averaging and plays a crucial role in the integrability of the dynamics. A key feature of these equations is their cross-product structure---the time derivatives of the angular momenta are given by cross products with themselves, which further implies that the magnitudes $s_1$, $s_2$, and $l$ remain constant. 
Moreover, the mathematical form of these EOMs is similar to the analogous
non-OA 1.5PN EOMs
that were solved in Ref.~\cite{Cho2019}, allowing its solution techniques to be directly extended to the present 2PN OA dynamics.

%--------------------------------
\subsubsection{Solution of $\cos \kappa_1, \cos \kappa_2$ and $\cos \gamma$}
\label{sec:sol-cos-kappa-1}
%--------------------------------

Here  we work out the solution of \Eqs{eq:OA-1PN-b}
in the style of Ref.~\cite{Cho2019} which differs in 
presentation from an equivalent solution presented in Ref.~\cite{Klein2017b}.
First, we define three angles
$\gamma, \kappa_1$ and $\kappa_2$,
such that $0 \leq \gamma, \kappa_1, \kappa_2 \leq \pi$ and
\begin{subequations} \label{defangles}
	\begin{align}
		&\cos{\gamma}\equiv\frac{\mathbf{s_1}\cdot \mathbf{s_2}}{s_1 s_2}\,,\\
		&\cos{\kappa_1}\equiv\frac{\mathbf{l}\cdot \mathbf{s_1}}{l s_1}\,,\label{eq:kappa1}\\
		&\cos{\kappa_2}\equiv\frac{\mathbf{l}\cdot \mathbf{s_2}}{l s_2}\,.
	\end{align}
\end{subequations}
Using Eqs.~\eqref{eq:OA-1PN-b},
 we can derive the time evolution of these cosines. 
A key observation is that all three derivatives are proportional to the same quantity 
$\mathbf{l}\cdot(\mathbf{s}_1\times\mathbf{s}_2)$
\begin{subequations}\label{eq:angle_rates}
	\begin{align}
		\frac{d\cos\gamma}{dt} &= \frac{(\delta_1-\delta_2)}{c^2 d^3}\frac{(1-\lambda)}{s_1 s_2}\,\mathbf{l}\cdot(\mathbf{s}_1\times\mathbf{s}_2)\,, \label{eq:dgamma_dt}\\
		\frac{d\cos\kappa_1}{dt} &= \frac{(\delta_2-\nu/2)}{c^2 d^3}\frac{(1-\lambda)}{l s_1}\,\mathbf{l}\cdot(\mathbf{s}_1\times\mathbf{s}_2)\,, \label{eq:dkappa1_dt}\\
		\frac{d\cos\kappa_2}{dt} &= -\frac{(\delta_1-\nu/2)}{c^2 d^3}\frac{(1-\lambda)}{l s_2}\,\mathbf{l}\cdot(\mathbf{s}_1\times\mathbf{s}_2)\,. \label{eq:dkappa2_dt}
	\end{align}
\end{subequations}
This proportionality allows us to construct two additional COMs by taking appropriate linear combinations. Following Cho and Lee~\cite{Cho2019}, we define
\begin{subequations}\label{eq:Sigmas}
	\begin{align}
		\Sigma_1 &= \cos\gamma + \frac{m_1-m_2}{m_1}\frac{l}{s_2}\cos\kappa_1\,, \label{eq:Sigma1}\\
		\Sigma_2 &= \cos\kappa_2 + \frac{m_2}{m_1}\frac{s_1}{s_2}\cos\kappa_1\,. \label{eq:Sigma2}
	\end{align}
\end{subequations}
We note that $\Sigma_2$ is proportional to $\lambda$. 

In the absence of GW emission, the total angular momentum vector 
$\mathbf{J} = \mathbf{L} + \mathbf{S}_1 + \mathbf{S}_2$ is  conserved. 
Under the OA dynamics, the magnitudes $L$, $S_1$, and $S_2$ 
are also conserved. Together with the two additional COMs
$\Sigma_1$ and $\Sigma_2$ introduced above, these five scalar quantities 
completely determine the magnitude $j^2$ since 
\begin{align}
j^2=l^2+s_1^2+s_2^2+ 2 s_1 s_2 \Sigma_1 + 2 l s_2 \Sigma_2\, .
\end{align}
Therefore, the conservation of the vector $\mathbf{J}$ furnishes only 
two additional independent constraints. This yields a total of seven independent COMs constraining the nine-dimensional phase space spanned 
by the components of $\mathbf{L}$, $\mathbf{S}_1$, and $\mathbf{S}_2$.

As a result, the evolution of the angular momenta is governed by  
two independent dynamical degrees of freedom. The first degree of freedom 
captures the internal relative configuration of the $(\vb{L}, \vb{S}_1, \vb{S}_2)$ triad, 
which we parametrize by  $\cos\kappa_1$ of Eq.~\eqref{eq:kappa1}. 
The second degree of freedom corresponds to the global azimuthal precession 
of this evolving triad around the invariant axis $\mathbf{J}$ (to be quantified by $\phi_L$ later). Crucially, because 
the instantaneous precession rate ($d \phi_L/dt$) is driven solely by the mutual angles between the angular momenta [as seen in the  $\cos\kappa_1$ dependence in 
Eq.~\eqref{eq:dthetaL_dt}], the evolution of these two degrees of freedom 
can be sequentially decoupled: the internal nutation is integrated first, 
followed by the global precession, thereby leading to the solutions of $\cos \kappa_1$ (and $\cos \kappa_2, \cos \gamma$), and $\phi_L$, respectively.

Using Eq.~\eqref{eq:Sigmas}
and the standard result 
from analytical geometry\footnote{This result is valid for any three 3D Cartesian vectors $\vb{l}, \vb{s}_1$ and $\vb{s}_2$.
The sign in~\Eq{eq:triple_prod_a} is decided by the handedness of the $\vb{l}, \vb{s}_1, \vb{s}_2$ triad.}
\begin{align}   \label{eq:triple_prod_a}
\mathbf{l}\cdot(\mathbf{s}_1\times\mathbf{s}_2) = \pm\, l s_1 s_2 (1 + 2 \cos \gamma \cos \kappa_1 \cos \kappa_2 - \cos^2 \gamma- \cos^2 \kappa_1- \cos^2 \kappa_2)^{1/2},
\end{align}
we can express $\mathbf{l}\cdot(\mathbf{s}_1\times\mathbf{s}_2)$ 
in terms of $\cos\kappa_1$. 
%(see Appendix~\ref{app:Triple}). 
We get
\begin{equation}\label{eq:triple_product}
	\frac{\mathbf{l}\cdot(\mathbf{s}_1\times\mathbf{s}_2)}{l s_1} = \pm \sqrt{A_3 x^3 + A_2 x^2 + A_1 x + A_0}\,,
\end{equation}
where $x \equiv \cos\kappa_1$ and 
\begin{subequations}\label{eq:cubic_coeffs}
\begin{align}
	A_3&= \frac{2 (m_1-m_2)m_2}{m_1^2}\, l \, s_1\,, \label{eq:A3}\\
	A_2&=-\frac{1}{m_1^2}\Big( (m_1 - m_2)^2\,l^2 + m_2^2\, s_1^2 + m_1^2\, s_2^2 + 2 m_1 m_2\, s_1 s_2 \Sigma_1 + 2 m_1 (m_1 - m_2)\,l\,s_2\, \Sigma_2 \Big)\, , \label{eq:A2}\\
	A_1&=\frac{2 s_2}{m_1}\Big( (m_1 - m_2)\,l \,\Sigma_1 +(m_2\, s_1 + m_1\, s_2\, \Sigma_1) \Sigma_2  \Big)\,, \label{eq:A1}\\
	A_0&= ( 1- \Sigma_1^2 - \Sigma_2^2)\,s_2^2\,. \label{eq:A0}
\end{align}
\end{subequations}

We define the three roots of the cubic polynomial in Eq.~\eqref{eq:triple_product} 
to be $x_-, x_+$ and $x_3$,
such that
 $x_- \leq x_+  \leq x_3$.
 The roots are given in Appendix~\ref{app:rootsx}.
 We will soon see [via Eq.~\eqref{eq:kappa1_solution}]
 that this specific ordering of roots implies that $x \equiv \cos \kappa_1$
is bound within
$x_- \leq x \leq x_+$. Substituting Eq.~\eqref{eq:triple_product} into 
Eq.~\eqref{eq:dkappa1_dt} yields a separable differential equation
\begin{equation}\label{eq:dx_separated}
	\frac{dx}{\sqrt{A(x-x_-)(x-x_+)(x-x_3)}} = \frac{\pm dt}{c^2 d^3}\,,
\end{equation}
where
\begin{equation}\label{eq:A_constant}
	A = \frac{9}{2}\frac{m_2(m_1-m_2)}{(m_1+m_2)^2}(1-\lambda)^2 \,l \,s_1 >0\,.
\end{equation}

The key difference from the 
non-OA 1.5PN case treated by Ref.~\cite{Cho2019} is that 
instead of Eq.~\eqref{eq:dx_separated}, they had
\begin{equation}\label{eq:dx_separated_CL}
	\frac{dx}{\sqrt{A(x-x_-)(x-x_+)(x-x_3)}} = \frac{\pm dt}{c^2 r^3}\,,
\end{equation}
which differs from Eq.~\eqref{eq:dx_separated}
by the substitution $d \rightarrow r$\footnote{The cubic polynomial in $\cos \kappa_1$ and its roots have different expressions and numerical values, but the overall structure remains the same.}. 
It is at this point that
a Newtonian or 1PN order $r$ solution needs to be 
injected before continuing with the integration; we discussed 
this $r$ injection to construct spin sector solution in Sec.~\ref{sec:From 1.5PN to 2PN solution}.

The left-hand side (lhs) of Eq.~\eqref{eq:dx_separated} can be 
integrated using elliptic integrals, which yields the
solution of $\cos \kappa_1$
\begin{equation}\label{eq:kappa1_solution}
	\cos\kappa_1(t) = x_- + (x_+-x_-)\,\text{sn}^2(\Upsilon(t), \beta)\,,
\end{equation}
where $\text{sn}(u,k)$ denotes the Jacobi elliptic sine function with argument $u$ and modulus $k$, defined through its relation to the incomplete elliptic integral of the first kind
\begin{equation}
	\text{sn}(u, k) = \sin(\text{am}(u, k)) = \sin\varphi\,, \quad \text{where} \quad u = F(\varphi, k) = \int_0^\varphi \frac{d\theta}{\sqrt{1-k^2\sin^2\theta}}\,.
\end{equation}
Here $\text{am}(u,k)$ is the Jacobi amplitude function, which gives the angle $\varphi$ corresponding to the elliptic integral value $u$. The quantities appearing in Eq.~\eqref{eq:kappa1_solution} are
\begin{subequations}\label{eq:Upsilon_params}
	\begin{align}
		\Upsilon(t) &= \frac{\sqrt{A(x_3-x_-)}}{2}\left(\alpha + \frac{t}{c^2 d^3}\right)\,, \label{eq:Upsilon}\\
		\beta &= \sqrt{\frac{x_+-x_-}{x_3-x_-}}\,, \label{eq:beta}\\
		\alpha &= \pm\frac{2}{\sqrt{A(x_3-x_-)}} F\left(\arcsin\sqrt{\frac{x(t_0)-x_-}{x_+-x_-}}, \beta\right)\,, \label{eq:alpha}
	\end{align}
\end{subequations}
with $F(\varphi, k)$ being the incomplete elliptic integral of the first kind as defined above. 
The sign in Eq.~\eqref{eq:alpha} is chosen to match the initial condition: $+$ if 
$d\cos\kappa_1/dt > 0$ at initial time $t_0$, and $-$ otherwise. More details are present in Ref.~\cite{Cho2019}. 

The main distinction from the solution presented in Ref.~\cite{Cho2019} is that our $\Upsilon$ in Eq.~\eqref{eq:Upsilon_params}
varies linearly with time. As a result, the solution in Eq.~\eqref{eq:Upsilon_params} not only neglects the fast oscillations appearing at 2PN order, but also fails to reproduce the 
1.5PN oscillations obtained in Ref.~\cite{Cho2019}. Nevertheless, all the secular effects 
are correctly captured up to 2PN order.
The time evolution of $\cos{\kappa_2}$ and $\cos{\gamma}$ follows directly from 
Eqs.~\eqref{eq:Sigmas} and~\eqref{eq:kappa1_solution}.

This is a good time to reflect on what the above solution tells us.
$\Upsilon$ of \Eq{eq:Upsilon} acts like the phase 
of the $\sn$ function.
And since  $1/d^3$ is a  prefactor
multiplying time $t$, we conclude that
one of the slow frequencies of the system
depends on $d$, which further depends on
the background orbital trajectory over which 
we average,
as already stated in 
Sec.~\ref{sec:From 1.5PN to 2PN solution},
but without justification. 
See Sec.~\ref{sec:four_frequencies}
for the explicit expressions of all four 
frequencies.

%--------------------------------
\subsubsection{Orbital angular momentum solution}
\label{sec:orb_L_vec_sol}
%--------------------------------

Having determined the evolution of the angles between the angular momenta vectors, 
we now construct the explicit time dependence of $\mathbf{L}$ in an inertial frame. 
First, we define two frames
\begin{itemize}
	\item Inertial frame (IF): an orthonormal basis $(\mathbf{x}, \mathbf{y}, \mathbf{z})$ such that 
    $ \vb{z}  \parallel \mathbf{J}$.
    $\vb{x}$ may point in any fixed     direction, while $\vb{y} = \vb{z} \times \vb{x}$.
	\item Noninertial frame (NIF): an orthonormal basis $(\mathbf{i}, \mathbf{j}, \mathbf{k})$ 
    % obtained by rotating the IF such that the $$k$-axis aligns with $\mathbf{L}$.
	such that $\vb{k} \parallel \vb{L}$, $\mathbf{i} \parallel \mathbf{J} \cross \mathbf{L}$, and $\mathbf{j} = \mathbf{k} \cross \mathbf{i} $.
\end{itemize}

\begin{figure}[t]
	\centering
	\includegraphics[width=0.5\textwidth]{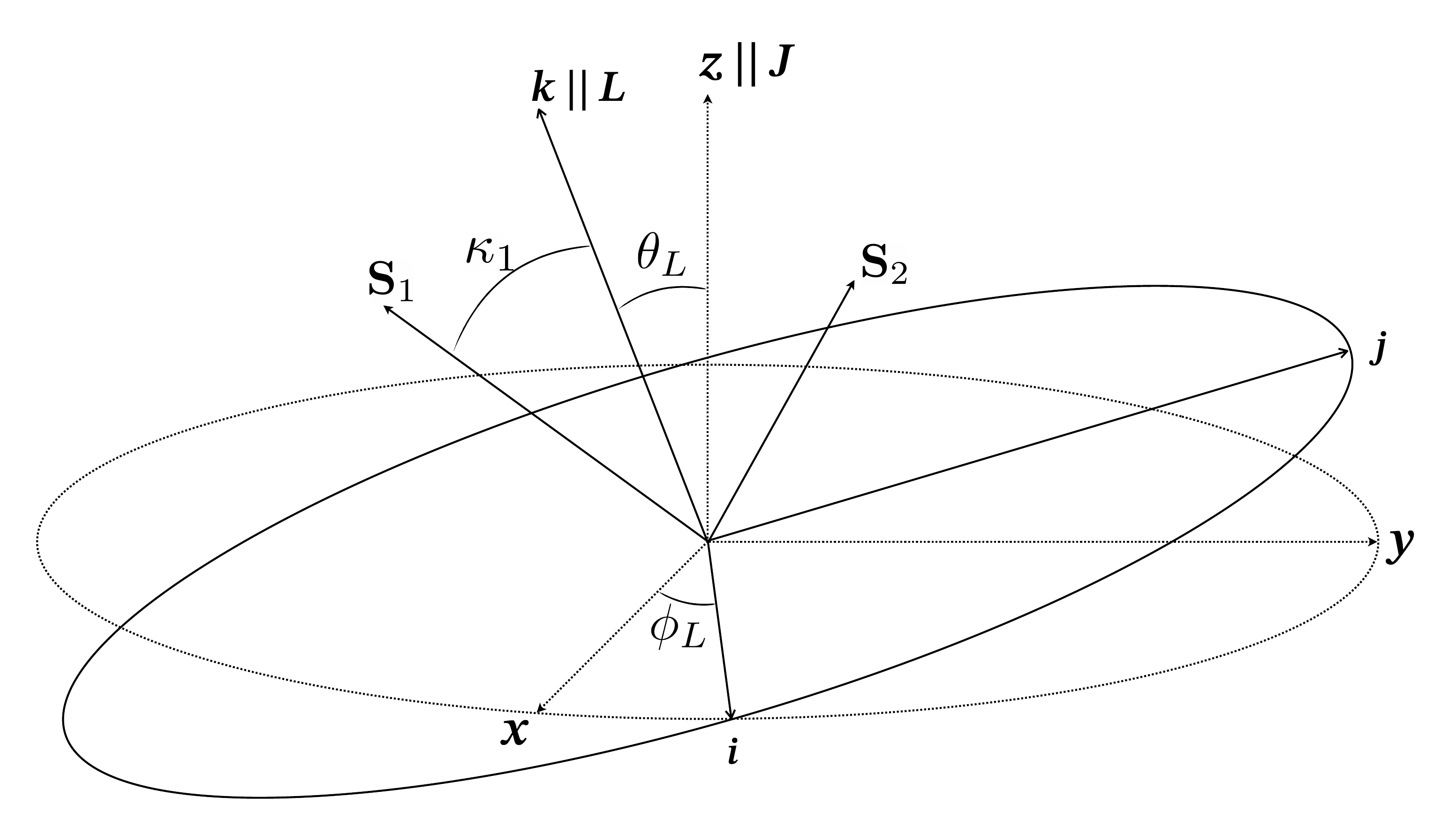}
	\caption{\justifying The dashed frame represents the inertial frame (IF) {$(\mathbf{x},\mathbf{y},\mathbf{z})$} centered around $\mathbf{J}$. 
	The solid frame is the noninertial frame (NIF) $(\mathbf{i},\mathbf{j},\mathbf{k})$ centered around $\mathbf{L}$.
	}
	\label{IF-NIF}
\end{figure}

\begin{figure}[t]
    \centering
\includegraphics[width=0.65\columnwidth]{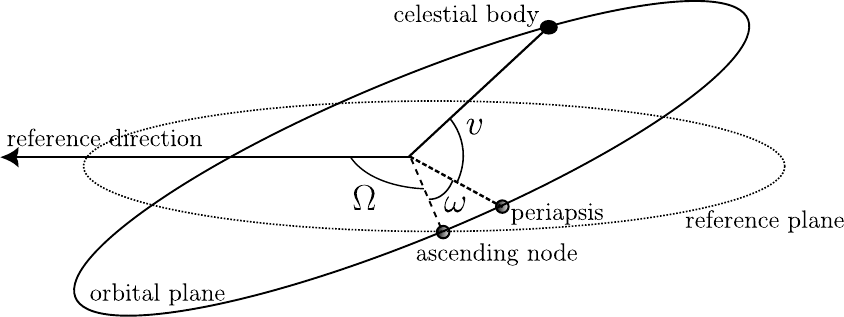}
    \caption{\justifying Geometry of the fixed Newtonian elliptical orbit  relative to a reference plane. The angles displayed are
    the longitude of the ascending node $\Omega$,
    the argument of periapsis $\omega$, and 
    the true anomaly $v$.
    Only $v$ changes with time for Newtonian ellipses.}
    \label{fig:orbital_elements}
\end{figure}

The two frames have been shown 
in Fig.~\ref{IF-NIF}, where the
polar and azimuthal angles ($\theta_L$ and 
$\phi_L$, respectively) of $\vb{L}$
are also shown.
Here $\phi_L$ can be considered the
generalization of the longitude of the ascending node (for fixed Newtonian elliptical orbits) to precessing
PN orbits~\cite{Poisson_Will_2014, SPITALE2002211, Scheeres}.
For reference, the classic orbital
elements for fixed elliptical orbits
have been shown in Fig.~\ref{fig:orbital_elements}, including the longitude of ascending node.

The corresponding Euler rotation matrix which when multiplied
with the  components of a vector in the IF
gives the corresponding components in the NIF
is
\begin{equation}\label{eq:Euler_matrix}
		\Lambda_{\text{IF}\to\text{NIF}} = \begin{pmatrix}\cos\phi_L & \sin\phi_L & 0 \\-\sin\phi_L\cos\theta_L & \cos\phi_L\cos\theta_L & \sin\theta_L \\ \sin\phi_L\sin\theta_L & -\cos\phi_L\sin\theta_L & \cos\theta_L \end{pmatrix}\,.
\end{equation}

Given that the magnitude of $\mathbf{L}$ is conserved under our OA EOMs,
its temporal evolution is entirely 
determined by the angles $\theta_L$ and $\phi_L$. 
Now in light of 
the solutions for $\cos\kappa_1(t)$ and $\cos\kappa_2(t)$ that we constructed in
Sec.~\ref{sec:sol-cos-kappa-1}, 
$\theta_L$ 
can be immediately determined from
\begin{equation}\label{eq:cos_thetaL}
	\cos\theta_L = \frac{\mathbf{l}\cdot\mathbf{j}}{lj} = \frac{l^2 + l s_1\cos\kappa_1 + l s_2\cos\kappa_2}{lj}\,.
\end{equation}

To determine $\phi_L(t)$, we first need the expressions of the angular momenta components in the NIF
\begin{align}\label{eq:vecNIF}
\mathbf{l} = l
\begin{bmatrix}
0 \\ 0 \\ 1
\end{bmatrix}_{\text{NIF}}, \quad
\mathbf{s}_1 = s_1
\begin{bmatrix}
\sin\kappa_1\cos\phi_{s1} \\ \sin\kappa_1\sin\phi_{s1} \\ \cos\kappa_1
\end{bmatrix}_{\text{NIF}}, \quad
\mathbf{s}_2 = s_2
\begin{bmatrix}
\sin\kappa_2\cos \phi_{s2} \\ \sin\kappa_2\sin \phi_{s2} \\ \cos\kappa_2
\end{bmatrix}_{\text{NIF}},
\end{align}
where $\phi_{s1}$ and $\phi_{s2}$ respectively denote the azimuthal angles of 
$\mathbf{s}_1$ and $\mathbf{s}_2$ in the NIF.

Since vector time derivatives become
different geometrical objects depending on
which frame the time derivative is being 
taken in, at this point, we establish 
that all vector time derivatives in this article
are meant to be taken in the inertial frame IF,
unless stated otherwise.
Their components may be written in either of the 
two frames, IF or NIF.
With that established, using Eq.~\eqref{eq:Euler_matrix}, 
we first write the time derivative of $\mathbf{l}$ in the IF and then 
transform the components to the NIF
\begin{equation}\label{eq:dLdtNIF}
	\frac{d\mathbf{l}}{dt}=l\begin{pmatrix}  \dot{\phi}_L \, \sin{\theta_L}\ \\ -\dot{\theta_L} \\ 0  \end{pmatrix} _{\text{NIF}}\,.
\end{equation} 
We then equate this expression to Eq.~\eqref{eq:dldt_avg}, after first evaluating the 
latter in the NIF using Eq.~\eqref{eq:vecNIF}. By rearranging the first component, 
we obtain (see Appendix~\ref{app:AzeomL})
\begin{equation}\label{eq:dthetaL_dt}
	\frac{d \phi_L}{dt }= \frac{1}{c^2\,d^3} \left( \frac{\beta_{1L}}{\alpha_{1L}+\cos{\kappa_1}}- \frac{\beta_{2L}}{\alpha_{2L}+\cos{\kappa_1}}+\beta_{3L} \right)\,,
\end{equation}
where
\begin{subequations} \label{eq:alpha_beta_L}
	\begin{align}
		\alpha_{1L} &= \frac{m_1}{m_1 - m_2} \, \frac{j + l + s_2 \Sigma_2}{s_1}, \\[4pt]
		\beta_{1L} &= \frac{3 m_1}{4 (m_1^2 - m_2^2)} \, \frac{1}{s_1} (1 - \lambda)\big[ m_1 j^2 + m_2 l^2 + (m_1 + m_2) j l - (m_1 - m_2) s_1 (s_1 + s_2 \Sigma_1)+ (m_2 l + m_1 j) s_2 \Sigma_2 \big], \\[6pt]
		\beta_{3L} &= \frac{j \nu}{2}.
	\end{align}
\end{subequations}
The coefficients $\alpha_{2L} $ and $ \beta_{2L}$ are obtained from $\alpha_{1L}$ 
and $\beta_{1L}$  by the substitution $j \rightarrow -j$.

Finally, Eq.~\eqref{eq:dthetaL_dt} can be integrated using the incomplete elliptic integral of the third kind
\begin{align}
	\Pi(a,b,c)\equiv\int_0^b\frac{1}{\sqrt{1-c^2 \sin^2{\theta}}}\frac{d\theta}{1-a\sin^2\theta}\,.
\end{align}
The solution is
\begin{align}\label{eq:phiL_solution}
	\phi_L(t)-\phi_{L}(t_0)&=\frac{2}{\sqrt{A(x_3-x_-)}}\Bigg( \frac{\beta_{1L}}{\alpha_{1L}+x_-}\Pi\left(n_1, \am(\Upsilon,\beta),\beta\right) -  \frac{\beta_{2L}}{\alpha_{2L}+x_-}\Pi\left(n_2, \am(\Upsilon,\beta),\beta \right) + \beta_{3L} \Upsilon  \Bigg)\,,
\end{align}
where  $n_i \equiv (x_- \,-\, x_+)/(\alpha_{iL} + x_-)$. Equations \eqref{eq:cos_thetaL} and~\eqref{eq:phiL_solution}, together with the conserved magnitude $L$, 
completely determine the time evolution of the averaged orbital angular momentum vector $\mathbf{L}(t)$ in IF.

%--------------------------------
\subsubsection{Spin angular momenta solution}\label{sec spins sol}
%--------------------------------

The solutions for $\mathbf{S}_1$ and $\mathbf{S}_2$ follow an analogous procedure. 
For each spin, we construct a noninertial frame with its orthonormal triad $(\vb{i'}, \vb{j'}, \vb{k'})$ such that
 $\vb{k'} \parallel \vb{S}_A$, $\mathbf{i'} \parallel \mathbf{J} \cross \mathbf{S}_A$, and $\mathbf{j'} = \mathbf{k'} \cross \mathbf{i'} $.
Analogously as for $\vb{L}$, for $\mathbf{S}_1$, the angle $\theta_{S_1}$ (inclination relative to $\mathbf{J}$) is given by
\begin{equation}\label{eq:theta_s1}
	\cos\theta_{S_1} = \frac{\mathbf{s}_1\cdot\mathbf{j}}{s_1 j} = \frac{s_1^2 + l s_1\cos\kappa_1 + s_1 s_2\cos\gamma}{s_1 j}\,,
\end{equation}
which is immediately determined from the known solutions of $\cos\kappa_1(t)$ and $\cos\gamma(t)$.
The azimuthal angle $\phi_{S_1}$ of $\vb{S}_1$
in the IF satisfies
\begin{equation}\label{eq:dthetaS1_dt}
	\frac{d\phi_{S_1}}{dt} = \frac{1}{c^2 d^3}\left(\frac{\beta_{1S_1}}{\alpha_{1S_1}+\cos\kappa_1} - \frac{\beta_{2S_1}}{\alpha_{2S_1}+\cos\kappa_1} + \beta_{3S_1}\right)\,,
\end{equation}
with coefficients
\begin{subequations}\label{eq:alpha_beta_S1}
\begin{align}
	\alpha_{1S_1} =& \frac{m_1}{m_2}\frac{1}{l}(-j+s_1+s_2\,\Sigma_1)\,,\\
	\beta_{1S_1}=&\frac{m_1}{4m_2}\frac{1}{l}\left[\left(2\delta_2-3 \nu_2\lambda\right)j^2-\left(2\delta_1-3\nu_1\lambda\right) l^2-\nu(s_1^2+s_2^2+2s_1\,s_2\,\Sigma_1+2l\,s_2\,\Sigma_2)-3\left(2\nu_2-\nu_1\right)(1-\lambda)\,j\, s_1 \right. \nn\\
	&\left.+3(\nu_2^2-\nu_1^2)(1-\lambda)(s_1^2+s_1\, s_2\, \Sigma_1) -3\left(1-\lambda \right)(\nu_2 \,j\, \Sigma_1+\nu_1\,l\,\Sigma_2)\,s_2\right]\,,\\
	\beta_{3s_1}=& \left(\delta_2-\frac{3}{2}\nu_2\,\lambda \right)j \,.
\end{align}
\end{subequations}
The coefficients $\alpha_{2S_1} $ and $ \beta_{2S_1}$ are obtained from 
$\alpha_{1S_1} $ and $\beta_{1S_1}$  by the substitution $j \rightarrow -j$. 

Integrating Eq.~\eqref{eq:dthetaS1_dt} yields
\begin{align}\label{eq:thetaS1_solution}
	\phi_{S_1}(t)-\phi_{S_1}(t_0)&=\frac{2}{\sqrt{A(x_3-x_-)}}\Bigg( \frac{\beta_{1S_1}\Pi\left(\frac{x_-\,-\,x_+}{\alpha_{1S_1}+x_-}, \am(\Upsilon,\beta),\beta \right)}{\alpha_{1S_1}+x_-} -  \frac{\beta_{2S_1}\Pi\left(\frac{x_-\,-\,x_+}{\alpha_{2S_1}+x_-}, \am(\Upsilon,\beta),\beta \right)}{\alpha_{2S_1}+x_-} + \beta_{3S_1} \Upsilon  \Bigg)\,.
\end{align}
Finally, $\mathbf{S}_2$ can be obtained either by an analogous calculation 
or by using conservation of total angular momentum
\begin{equation}
	\mathbf{S}_2 = \mathbf{J} - \mathbf{L} - \mathbf{S}_1\,.
\end{equation}

This completes the solution for the OA dynamics of the angular momenta. 
All vectors are expressed in terms of elliptic and elementary functions of time, 
with coefficients depending on the COMs of the
OA system.

%--------------------------------
\subsection{Beyond orbit averaging: The hybrid solution
for the angular momenta}    \label{Hybrid def}
%--------------------------------

\subsubsection{Introducing the hybrid model equations of motion}

Let us summarize the discussion so far. The solution presented in
Secs.~\ref{sec:sol-cos-kappa-1}--\ref{sec spins sol} is accurate through
2PN order on the precession timescale, but it does not capture the
short-timescale oscillations that are suppressed by orbit averaging. 
These fast oscillations are instead described accurately at 1.5PN order by the
solution of Ref.~\cite{Cho2019}.

The goal of this subsection is to construct a solution that is fully
1.5PN accurate for both the fast and slow dynamics, while incorporating
the 2PN spin-spin secular corrections to the dynamics. This is achieved by
hybridizing the complete 1.5PN solution of Ref.~\cite{Samanta2022} with
the 2PN orbit-averaged solution of Ref.~\cite{Klein2017b}, presented in
Secs.~\ref{sec:sol-cos-kappa-1}--\ref{sec spins sol}. We refer to the
resulting solution for the three angular momenta as the ``hybrid
model.'' As we will demonstrate, among the currently available analytic
solutions, the hybrid model provides the most faithful approximation to
the exact numerical 2PN dynamics on both the slow and fast timescales.

In what follows, we compare the evolution of the dynamical variables
under four different post-Newtonian prescriptions:
(i) the full 2PN EOMs,
Eqs.~\eqref{eq:spin_precession}, denoted by ``(2)'';
(ii) the full 1.5PN EOMs of
Refs.~\cite{Cho2019,Samanta2022}, denoted by ``(1.5)'';
(iii) the 2PN OA EOMs,
Eq.~\eqref{eq:OA-1PN-b}, denoted by ``(2, A)'';
and (iv) the hybrid model developed in this section, denoted by ``(2, H).''

The EOMs for $\mathbf{s}_1$
under the evolution systems (2), (1.5) and (2, A), 
as labeled above take the general
form of a cross product 
\begin{equation}\label{eq:precession}
	\frac{d\mathbf{s}_1^{(\cdot)}}{dt}
	= \boldsymbol{\Omega}^{(\cdot)} \times \mathbf{s}_1^{(\cdot)} ,
\end{equation}
where $\boldsymbol{\Omega}^{(\cdot)}$ 
denotes the precessional angular velocity and
$\mathbf{s}_1^{(\cdot)}$ the corresponding spin solution for the model $(\cdot)$
under consideration.

Adding the following definitions 
(where ``SO'' and ``SS''
stand for ``spin-orbit'' and ``spin-spin'')
to the 
above-established $\boldsymbol{\Omega}^{(\cdot)}$
convention
\begin{equation}
	\boldsymbol{\Omega}_{\rm SO}
	= \frac{\delta_1}{c^2 r^3}\,\mathbf{l}\,, \qquad
	\boldsymbol{\Omega}_{\rm SS}
	=\frac{\nu_1}{c^2 r^3}\left(
		 3(\mathbf{n}\cdot \mathbf{s}_0)\mathbf{n}-\mathbf{s}_0 \right),
	\label{eq:Omegas}
\end{equation}
\begin{equation}
	\bar{\boldsymbol{\Omega}}_{\rm SO}
	\equiv
	% \left\langle \boldsymbol{\Omega}_{\rm SO} \right\rangle_O	= 
	\frac{\delta_1}{c^2 d^3}\,\mathbf{l}, \qquad
	\bar{\boldsymbol{\Omega}}_{\rm SS}
	\equiv
	% \left\langle \boldsymbol{\Omega}_{\rm SS} \right\rangle_O = \frac{1}{c^2 d^3}
	\frac{\nu_1}{2c^2 d^3}\left(
		\mathbf{s}_0-3 \lambda \mathbf{l}
	\right),
\end{equation}
allows us to write the EOM
for  $\mathbf{s}_1$ and the corresponding
$\boldsymbol{\Omega}^{(\cdot)}$
 for  the leading
$1.5$PN spin dynamics 
[as per Eqs.~\eqref{eq:ds1dt}]
\begin{equation}\label{eq:angular15}
	\frac{d\mathbf{s}_1^{(1.5)}}{dt}
	= \boldsymbol{\Omega}^{(1.5)} \times \mathbf{s}_1^{(1.5)} \,, \qquad
	\boldsymbol{\Omega}^{(1.5)} = \boldsymbol{\Omega}_{\rm SO} \,.
\end{equation}
The full $2$PN EOM incorporates
both SO and SS couplings and reads [via Eqs.~\eqref{eq:ds1dt}]
\begin{equation}\label{eq:angular2}
	\frac{d\mathbf{s}_1^{(2)}}{dt}
	= \boldsymbol{\Omega}^{(2)} \times \mathbf{s}_1^{(2)} \,, \qquad
	\boldsymbol{\Omega}^{(2)} = \boldsymbol{\Omega}_{\rm SO}
	+ \boldsymbol{\Omega}_{\rm SS} \,.
\end{equation}
The OA 2PN spin evolution  EOM is given by [as per Eqs.~\eqref{eq:OA-1PN-b}]
\begin{equation}\label{eq:angular2A}
	\frac{d\mathbf{s}_1^{(2,\mathrm{A})}}{dt}
	= \boldsymbol{\Omega}^{(2,\mathrm{A})}
	\times \mathbf{s}_1^{(2,\mathrm{A})} \,, \qquad
	\boldsymbol{\Omega}^{(2,\mathrm{A})}
	= \bar{\boldsymbol{\Omega}}_{\rm SO}
	+ \bar{\boldsymbol{\Omega}}_{\rm SS} \,.
\end{equation}

Against this backdrop and convention, finally, we introduce our 2PN hybrid model defined by the following
precessional angular velocity vector
for $\vb{s}_1$
\begin{equation}\label{eq:angular2H}
	\boldsymbol{\Omega}^{(2,\mathrm{H})}
	\equiv \frac{1}{c^2 r^3}
	\left(  \delta_1\,\mathbf{l} + \frac{\nu_1}{2}\left(\mathbf{s}_0- 3\lambda\mathbf{l}\right)
	\right)
	= \boldsymbol{\Omega}_{\rm SO}
	+ \frac{d^3}{r^3}\,\bar{\boldsymbol{\Omega}}_{\rm SS} \,,
\end{equation}
rendering the EOM for $\vb{s}_1$ 
to be
\begin{align}    \label{eq:hybd_s1_EOM}
    \frac{d\mathbf{s}_1^{(2,\mathrm{H})}}{dt}
	= \boldsymbol{\Omega}^{(2,\mathrm{H})}
	\times \mathbf{s}_1^{(2,\mathrm{H})} \,.
\end{align}
By definition, 
the EOM for $\vb{s}_2$ 
under the hybrid model
is obtained
by performing the label exchange
$1 \leftrightarrow 2$ operation on Eq.~\eqref{eq:hybd_s1_EOM}. 
Once the spins are determined under this hybrid model, $\vb{L}$ can be had from
$\vb{L} = \vb{J} - \vb{S}_1 -\vb{S}_2$ since we define
$\vb{J}$ to be a constant 
vector under the hybrid model. This completes the
specification of our hybrid (2, H) model.
It helps to write out explicitly the EOMs for the three angular momenta under the hybrid model. Doing so renders the hybrid EOMs to be
\begin{subequations}
\label{eq:hybrid-EOM}
\begin{align}
 \frac{d\mathbf{s}_1}{dt}  &= \frac{1}{c^2 r^3} \left\{ \delta_1\,\mathbf{l} + \frac{\nu_1}{2}\left(\mathbf{s}_0- 3 \lambda \mathbf{l}\right)\right\} \times \mathbf{s}_1, \\
 \frac{d\mathbf{s}_2}{dt}  &= \frac{1}{c^2 r^3} \left\{ \delta_2\,\mathbf{l} + \frac{\nu_2}{2}\left(\mathbf{s}_0- 3 \lambda \mathbf{l}\right)\right\} \times \mathbf{s}_2, \\
 \frac{d\mathbf{l}}{dt}  &= \frac{1}{c^2 r^3} \left\{ \mathbf{s}_{\mathrm{eff}} -  \frac{3}{2} \lambda \mathbf{s}_0 \right\} \times \mathbf{l}.
\end{align}
\end{subequations}
Note that Eqs.~\eqref{eq:hybrid-EOM}
differ from Eqs.~\eqref{eq:OA-1PN-b}
by the substitution
$r^3 \rightarrow d^3$
in the denominator.

Equations \eqref{eq:angular2H} and \eqref{eq:hybrid-EOM} make it apparent that
the hybrid model is a 2PN perturbation of the 1.5PN model. Hence both the 
slow and fast 1.5PN dynamics are captured by the hybrid model. 
Furthermore, because orbit-averaging the hybrid EOMs over a 1PN orbit 
(by simply replacing $r^3$ with $d^3$)  recovers the 2PN OA model of 
Eq.~\eqref{eq:angular2A}, the hybrid equations inherently encode the correct 2PN secular evolution.

To explicitly solve the hybrid model while preserving this secular accuracy, 
we apply the equivalence established in Sec.~\ref{sec:From 1.5PN to 2PN solution}: for unaveraged EOMs,
injecting a higher-PN $r$ solution into evolution equation for 
$\cos \kappa_1$
 is analogous to averaging 
over a higher-PN order background orbital trajectory, as far as the slow frequency dynamics are concerned. Therefore, while solving Eqs.~\eqref{eq:hybrid-EOM},
we  strategize to inject 1PN $r$  
(rather than the Newtonian one), into the evolution 
equation for $\cos \kappa_1$. 
This yields a  solution that captures the full 1.5PN dynamics across all timescales, alongside the 2PN dynamics on the slow, secular timescale.

\subsubsection{Solution to the hybrid model equations of motion}

Note that Eqs.~\eqref{eq:hybrid-EOM} are very similar
to two other sets of EOMs:
the 2PN OA EOMs~\eqref{eq:OA-1PN-b}, and 
the 1.5PN unaveraged EOMs of the (1.5) model  
(basically Eqs.~\eqref{eq:spin_precession}, 
but with square brackets therein set to zero).
The former set features a constant $d$, while the latter 
set features a time-varying $r$. Due to this, the method of 
finding the solution of Eqs.~\eqref{eq:hybrid-EOM}
is  very similar to that for 
the (1.5) model
(subject of Refs.~\cite{Cho2019, Samanta2022}), or  
Eqs.~\eqref{eq:OA-1PN-b}. 
Since we have already 
shown how to derive the solution for
EOMs~~\eqref{eq:OA-1PN-b} using the 
techniques of Ref.~\cite{Cho2019},
we  will only provide the 
final solution of Eqs.~\eqref{eq:hybrid-EOM} rather than 
detailed derivation.

Before proceeding, we note that due to the presence of $r^3$
in Eqs.~\eqref{eq:hybrid-EOM} (instead of $d^3$), 
we obtain, in place of Eq.~\eqref{eq:dx_separated}, the following 
expression (analogous to Eq.~(3.4) of Ref.~\cite{Cho2019})
\begin{align}     \label{eq:dxdt-hybrid}
\frac{d x}{\sqrt{A\left(x-x_-\right)\left(x-x_+\right)\left(x-x_3\right)}}=\frac{ \pm d t}{c^2 r^3}
\end{align}
At this point,
we inject the 1PN $r$ 
of Eqs.~\eqref{eq:kepler1PN-a}-\eqref{eq:kepler1PN-b}, as planned above, thus  inducing unphysical PN accuracy mentioned 
in Sec.~\ref{sec:From 1.5PN to 2PN solution}.
The quantities $l$, $\lambda$, $\Sigma_1$, $\Sigma_2$, coefficients of the cubic polynomial, 
and its roots $x_k$ still remain constants with respect to
time in the hybrid model.
Their expression are still 
given by
 Eqs.~\eqref{eq:lambda}
and \eqref{eq:Sigmas}
and their numerical values
also remain the same as in the OA case of Sec.~\ref{sec sep timescale}, thus making the solution feasible. During the course of derivation, we 
 reencounter Eqs.~\eqref{eq:dx_separated}, 
\eqref{eq:dthetaL_dt}, and \eqref{eq:dthetaS1_dt},
except that $d$, in these equations, is replaced by $r$.
The final hybrid solution of Eqs.~\eqref{eq:hybrid-EOM}
is given by
\begin{subequations}\label{eq:kappa1_solution_hyb}
\begin{align}
	\cos\kappa_1^{(2,\mathrm{H})}(t)   &=   x_- + (x_+-x_-)\,\text{sn}^2(\Upsilon^{(2,\mathrm{H})}(t), \beta)\,,    \label{eq:cos_k1_hybd}   \\
\Upsilon^{(2,\mathrm{H})}(u)    &= \frac{\sqrt{A(x_3-x_-)}}{2}\Bigg[\alpha + \frac{1}{c^2 d^3}\frac{v_\theta + e_\theta \sin v_\theta}{n}  \Bigg]\,, \label{eq:Upsilonhyb}    \\
v_\theta &= u + 2 \arctan \left( \frac{\beta_\theta \sin u}{1 - \beta_\theta \cos u} \right)\, ,     \label{v_r}       \\
\beta_\theta  &= \frac{e_\theta}{1 + \sqrt{1 - e_\theta^2}}\,  ,    \label{beta-r} 
\end{align}
\end{subequations}
where the parameters $A$, $\alpha$, and $\beta$ are 
still defined via Eqs.~\eqref{eq:A_constant} and \eqref{eq:Upsilon_params}. 
$x_-$, $x_+$, and $x_3$
are the roots of the cubic polynomial ${\textstyle \sum\limits_{k=0}^3 A_k x^k}$
with the $A_k$'s  defined by Eqs.~\eqref{eq:cubic_coeffs}. Finally the 1PN QK parameters $d$ and 
$e_\theta$ were defined in Eq.~\eqref{def:d}. 
The rest of the solution is given by 
Eqs.~\eqref{eq:cos_thetaL}, \eqref{eq:phiL_solution}, \eqref{eq:theta_s1}, and 
\eqref{eq:thetaS1_solution}
but with the replacement $\Upsilon \rightarrow \Upsilon^{(2, \text{H})}$.
The evolution of $u$ above is governed by the 1PN
QKP solution of 
Eqs.~\eqref{eq:kepler1PN-a} and \eqref{eq:1PN-n}. This completes the specification of the 
solution of the hybrid model of angular momenta.

Let us now glean from Eqs.~\eqref{eq:kappa1_solution_hyb}, how 
the slow and fast scale oscillations emerge.
One can clearly see that, whereas the OA argument $\Upsilon^{(2,\mathrm{A})}$,
which is $\Upsilon$ of Eq.~\eqref{eq:Upsilon} 
grows linearly in time 
as $t/(c^2 d^3)$, the
 hybrid counterpart $\Upsilon^{(2,\mathrm{H})}$ contains additional 
oscillatory terms depending on the eccentric anomaly, which restore the fast orbital-timescale oscillations of 
the  1.5PN dynamics while preserving the correct 
secular behavior on the precession timescale.
The argument $\Upsilon^{(2, \text{H})}$
of the Jacobi $\sn$ function acts like a phase. 
The tiny, fast oscillations in this phase are brought
about  by $v_r$ and functions of $u$
 in Eq.~\eqref{eq:Upsilonhyb} with the frequency
 of 1PN $n$; this is due to Eqs.~\eqref{v_r} and~\eqref{eq:1PN-n}.
 Now when it comes to the slow frequency,
 it emerges due to the $c^2$ factor
 in the denominator in Eq.~\eqref{eq:Upsilonhyb}, and hence is 1PN order 
 smaller than the fast frequency. 
 One final point to note is that both the slow and
 fast frequencies depend on the 1PN QKP parameters
  ($e_r, e_t, a_r, n$ etc.)
 present in Eq.~\eqref{eq:Upsilonhyb}. 
 These 1PN QKP parameters are brought about
 due to the injection of 1PN-accurate $r$
 in Eq.~\eqref{eq:dxdt-hybrid}. This makes it 
 apparent that the PN accuracy of 
 the injected $r$ is essential to determine the PN accuracy of the 
 slow and fast frequencies, as already stated
 without  justification in Sec.~\ref{sec:From 1.5PN to 2PN solution}.

To further assess the accuracy of the hybrid model relative to the other models of evolution, we derive 
in Appendix~\ref{App:errorbounds}, the error bounds by comparing each approximate 
spin evolution to the full 2PN dynamics. 
We find that for $t \lesssim \tau_{\text{sp}} \sim c^2$
\begin{subequations}\label{eq:errors0}
\begin{align}
	\|\mathbf{s}_1^{(1.5)}(t)- \mathbf{s}_1^{(2)}(t)\|
	& \sim \mathcal{O}\!\left(t \, c^{-4}\right), \\
	\|\mathbf{s}_1^{(2,\mathrm{A})}(t)- \mathbf{s}_1^{(2)}(t)\|
	& \sim \mathcal{O}\!\left(c^{-3}\right), \\
	\|\mathbf{s}_1^{(2,\mathrm{H})}(t)- \mathbf{s}_1^{(2)}(t)\|
	& \sim \mathcal{O}\!\left(c^{-4}\right).
\end{align}
\end{subequations}

\subsubsection{The plots}

\begin{figure}[h!] 
	\centering
\includegraphics[width=1\textwidth]{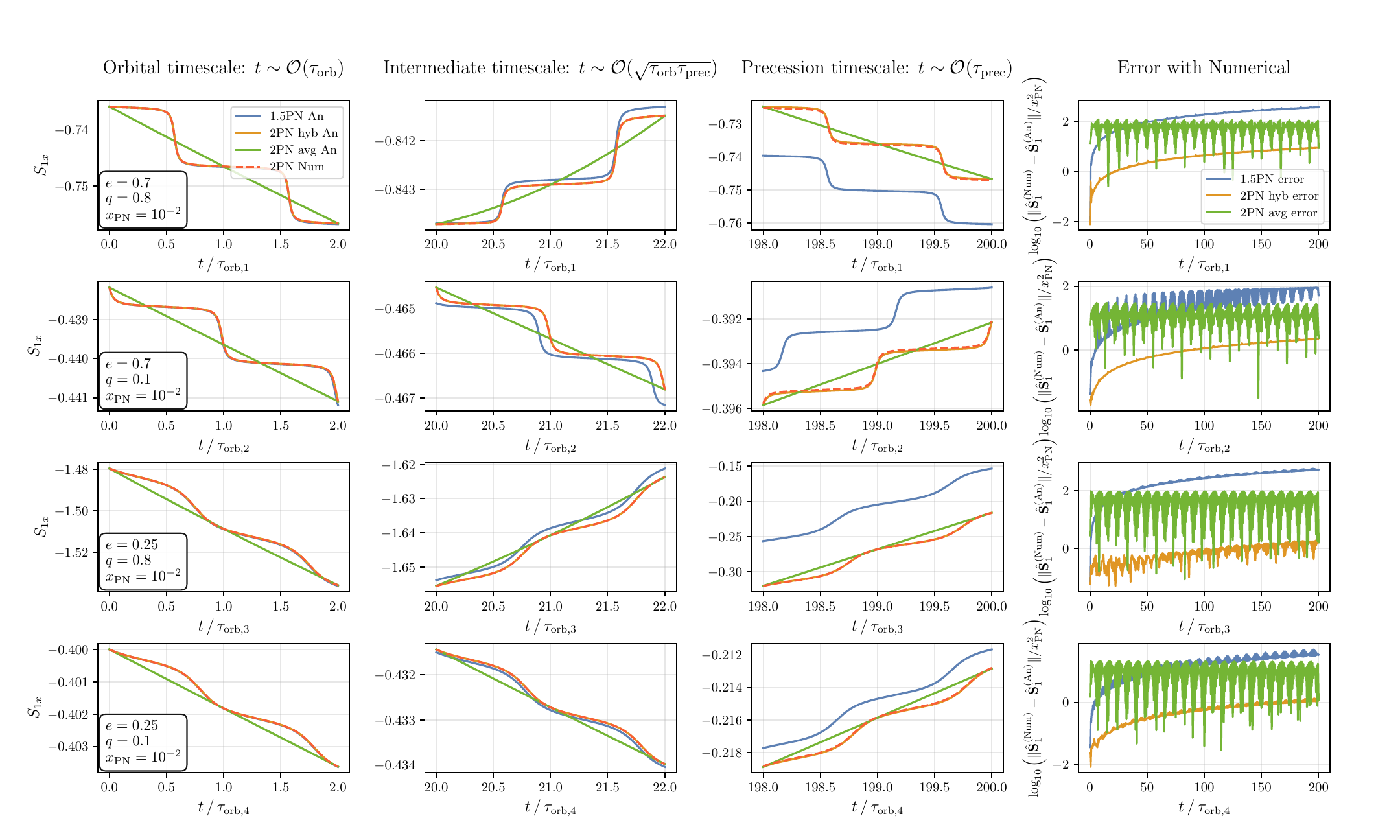} 
	\caption{\justifying
Comparison of the analytical $\mathbf{S}_1$ solutions 
from various approximate models against the full 2PN dynamics obtained 
via numerical integration (red). The $1.5$PN solution $(1.5)$ (blue), 
the hybrid model $(2,\mathrm{H})$ (orange), and the OA 2PN 
solution $(2,\mathrm{A})$ (green) are shown. 
The first three columns display the $x$-component of $\mathbf{S}_1$ 
over increasing timescales, from orbital periods to spin precession times. The fourth column shows the norm of the vectorial error of the three analytical models with respect to the 2PN numerical one; see 
Eq.~\eqref{eq:y_error}
in the main body.
The hybrid model accurately recovers the 
LO fast orbital oscillations of the 2PN numerical evolution while preserving the correct 
secular behavior. The OA model captures
only the secular evolution, and the 
$1.5$PN model captures the oscillations but misses the secular spin-spin effects. Results 
are shown for four systems with different mass ratios ($q = m_2/m_1$) and eccentricities, 
all sharing the same randomly chosen spin orientations: $\kappa_1 = 68^\circ$, $\kappa_2 = 10^\circ$, 
$\gamma = 73^\circ$. }
\label{plot:S1x} 
\end{figure}

Figure \ref{plot:S1x} provides a visual confirmation of the scalings of Eqs.~\eqref{eq:errors0}.
The first three columns display the evolution of the $x$-component of 
$\mathbf{S}_1$ over increasing timescales. The fourth column plots 
the normalized logarithmic vectorial error, defined as
\begin{align}     \label{eq:y_error}
\log_{10} \left( \frac{\lVert \hat{\mathbf{S}}_1^{\text{(2PN-Num)}} - \hat{\mathbf{S}}_1^{\text{(An)}} \rVert}{x_{\text{PN}}^2} \right)   .
\end{align}
We use the unit spin vectors here because the magnitude 
$\|\mathbf{S}_1\|$ is conserved in time for all four models.
These scalings of Eqs.~\eqref{eq:errors0}
reveal a clear, timescale-dependent hierarchy of accuracy. 
On orbital timescales ($t \sim \tau_{\mathrm{orb}} = \mathcal{O}(1)$), 
both the $1.5$PN and hybrid models accurately capture the LO 
fast orbital oscillations, achieving errors of $\mathcal{O}(c^{-4})$ (first column). 
In contrast, the OA
model smooths out these oscillations, 
resulting in a larger $\mathcal{O}(c^{-3})$ baseline error. 
At intermediate times 
($t \sim \sqrt{\tau_{\text{orb}} \tau_{\text{sp}}}$), secular spin-spin effects 
become non-negligible (second column). The OA model effectively tracks this slow drift, 
bringing its error to the same order of magnitude as that of the $1.5$PN 
model, as seen where the green and blue curves cross in the fourth column. 
On precession 
timescales ($t \sim \tau_{\mathrm{sp}} = \mathcal{O}(c^2)$), however, 
the $1.5$PN error grows linearly to $\mathcal{O}(c^{-2})$ due to the 
absence of secular spin-spin contributions (third column). Meanwhile, the OA and 
hybrid models remain bounded at $\mathcal{O}(c^{-3})$ and 
$\mathcal{O}(c^{-4})$, respectively. Therefore, the hybrid approach 
optimally combines the strengths of both methods: it preserves the 
short-timescale orbital oscillations of the $1.5$PN model while 
capturing the long-timescale secular evolution of the OA model, 
making it the most accurate approximation across the entire evolution.

To sum up, the hybrid solution is superior to the traditional OA solutions \cite{Klein2017b} in that 
(a) it possesses fast timescale, small oscillations and
(b) its slow frequency is also more accurate due to averaging over the 1PN background orbit. The hybrid solution is  superior 
to the full 1.5PN solution of Refs.~\cite{Cho2019, Samanta2022} in that it 
captures both the
slow and fast frequency 
dyanmics more accurately due to the consideration of 2PN terms at the level of EOMs.
Finally, it is important to note that the $1.5$PN and 2PN OA analytical 
solutions shown in Fig.~\ref{plot:S1x} are already enhanced (when compared against their traditional presentation in the literature) by 
injecting the 1PN radial solution $r$ and averaging over the 1PN orbit, 
respectively. Had standard Newtonian orbits been used, as is traditional in the literature, 
these two models would exhibit significantly larger phase and timing errors relative 
to the full 2PN numerical integration, as discussed in Sec.~\ref{sec:From 1.5PN to 2PN solution}. 
Yet, even with these substantial improvements, they remain 
significantly less accurate than the 2PN hybrid model.

%--------------------------------
%--------------------------------	
\section{The quasi-Keplerian solution for the orbital dynamics}\label{sec:QKP0}
%--------------------------------
%--------------------------------

We now turn to the orbital dynamics and derive a QKP
solution for eccentric orbits of the 
2PN spinning system governed by the Hamiltonian of Eq.~\eqref{eq:hamiltonian}. 
We first obtain the EOMs for
the radial and azimuthal
variables $r$ and $\phi$ 
(azimuthal angle of $\mathbf{r}$ in the NIF).
Then we construct a solution 
using an ansatz-based 
approach analogous to that of Ref.~\cite{Klein2010}. 
When deriving the two EOMs, the angular momenta are taken to satisfy the full 2PN precession equations~\eqref{eq:spin_precession}, 
rather than the hybrid EOMs~\eqref{eq:hybrid-EOM}. In this section, 
the hybrid description is used only in Eq.~\eqref{eq:dphidtApp}, and solely 
to account for the precession of the NIF with respect to the IF.

%--------------------------------
\subsection{Radial equation of motion}\label{sec:eomR}
%--------------------------------

To start, we derive the EOM 
 for the radial coordinate $r$ by computing its Poisson 
bracket with the Hamiltonian of Eq.~\eqref{eq:hamiltonian}. It yields

\begin{align}\label{eq:drdt_squared_initial}
	\left(\frac{dr}{dt}\right)^2&= p_r^2 - \frac{1}{c^2}\, \biggl[(1-3\nu)\,p_r^4 +\left(\frac{1}{r}(6+4\nu) +\frac{l^2}{r}(1-3\nu)\right)p_r^2 \biggr] +\frac{1}{c^4}\biggr[\left( \frac{1}{r}\left(8-27\nu-14\nu^2\right) +\frac{l^2}{2r^2}\left(4-21\nu+24\nu^2\right)\right)p_r^4\nn \\
	&+\frac{1}{4}\left(4-21\nu+24\nu^2 \right)p_r^6 +\left(\frac{1}{r^2}\left(19+34\nu+4\nu^2\right)+\frac{l^2}{r^3}\left(8-27\nu-10\nu^2\right)+\frac{l^4}{4r^4}\left(4-21\nu+24\nu^2 \right) \right)p_r^2 \biggl] \,,
\end{align}
where $p_r$ is the canonical momentum conjugate to the radial separation $r$, such that
\begin{align}\label{eq:p_squared}
	p_r^2  = \np  = \mathbf{p}^2 - l^2/r^2.
\end{align}
We recall that in Eqs.~\eqref{eq:drdt_squared_initial} and~\eqref{eq:p_squared}, 
the norm of $\mathbf{l}$ is no longer a COM, as shown in Eq.~\eqref{eq:dldt_magnitude}. 
To express Eq.~\eqref{eq:drdt_squared_initial} solely in terms of $r$, $h$, $l$, and the 
spin variables, we need to eliminate $p_r^2$. It is achieved by perturbatively inverting the 
Hamiltonian of Eq.~\eqref{eq:hamiltonian} to obtain $p_r^2$ as a PN expansion in powers of 
$1/c^2$. Keeping all the terms up to 2PN order, we find
\begin{equation}\label{eq:pr_squared}
	p_r^2= \tilde{A} + \frac{2\tilde{B}}{r} + \frac{\tilde{C}}{r^2}+ \frac{\tilde{D}_1}{r^3} + \frac{\tilde{D}_2}{r^4} + \frac{\tilde{D}_3}{r^5} + \mathcal{O}\left(\frac{1}{c^5}\right)\,,
\end{equation}
where the coefficients $\tilde{A},\,\tilde{B},\,\tilde{C},\,\tilde{D}_{1,2,3}$ are the following functions of 
$h$, $l$, and the spins
\begin{subequations} \label{eq:coeff_pr2}
	\begin{align}
		\tilde{A} &= 2h + \frac{1}{c^2}\, (1-3\nu) h^2 - \frac{1}{c^4}\, \nu(1-4\nu) h^3,\\[4pt]
		\tilde{B} &= 1 + \frac{1}{c^2}\, (4-\nu) h + \frac{1}{c^4}\, (2-2\nu+\nu^2) h^2,\\[4pt]
		\tilde{C} &= -l^2 + \frac{1}{c^2}\, (6+\nu) + \frac{1}{c^4}\, 15 h,\\[4pt]
		\tilde{D}_1 &= -\frac{1}{c^2}\, \big[ \nu l^2 + 2\lseff - s_0^2 + 3(\mathbf{n}\!\cdot\!\ \mathbf{s}_0)^2\big]  
		+ \frac{1}{c^4}\,\biggl[\frac{17}{2}+\frac{5\nu}{2}+2\nu^2 - \nu(1+\nu) h l^2  \biggr],\label{eq:D1tilde}\\[4pt]
		\tilde{D}_2 &= -\frac{1}{c^4}\, \nu(1+3\nu) l^2,\\[4pt]
		\tilde{D}_3 &= \frac{1}{c^4}\, \frac{3\nu^2}{4} l^4.
	\end{align}
\end{subequations}

Finally, substituting Eq.~\eqref{eq:pr_squared} into Eq.~\eqref{eq:drdt_squared_initial} 
and collecting terms, we obtain the radial EOM in a polynomial form
\begin{equation}\label{eq:drdt_final}
	\left(\frac{dr}{dt}\right)^2 = A + \frac{2B}{r}+\frac{C}{r^2}+\frac{D_{1o}+D_{1s}}{r^3}+\frac{D_2}{r^4}+\frac{D_3}{r^5}+ \mathcal{O}\left(\frac{1}{c^5}\right)\,,
\end{equation}
where
\begin{subequations}\label{eq:radial_coeffs}
	\begin{align}
		A &= 2h + \frac{1}{c^2}\, 3(-1+3\nu)\, h^2 + \frac{1}{c^4}\, (4-19\nu+16\nu^2) \,h^3\,,\\[4pt]
		B &= 1 + \frac{1}{c^2}\, (-6+7\nu)\, h + \frac{1}{c^4}\, 3(3-16\nu+7\nu^2) \,h^2\,,\\[4pt]
		C &= -l^2 + \frac{1}{c^2}\,\big[5(\nu-2)+2(1-3\nu) \,h\, l^2\big] + \frac{1}{c^4}\,\big[ (37-122\nu+36\nu^2) \,h-3(1-5\nu+5\nu^2) \,h^2 \,l^2 \big]\,,\\[4pt]
		D_{1o} &= \frac{1}{c^2}\, (8-3\nu) l^2 + \frac{1}{c^4}\,\biggl[\frac{53}{2}-\frac{83\nu}{2}+10\nu^2 - (16-73\nu+19\nu^2) h l^2 \biggr]\,, \label{eq:D1o}\\[4pt]
		D_{1s} &= -\frac{1}{c^2}\, \left[2\lseff - s_0^2 + 3 (\mathbf{n}\!\cdot\! \mathbf{s}_0)^2\right]\,, \label{eq:D1s}\\[4pt]
		D_2 &= -\frac{1}{c^4}\, 3(11-11\nu+2\nu^2) \,l^2\,,\\[4pt]
		D_3 &= -\frac{1}{c^4}\, \frac{\nu(4+\nu)}{4}\, l^4\,.
	\end{align}
\end{subequations}

When compared to the nonspinning or 1.5PN spin-orbit case, 
the main technical challenge in constructing a 2PN QKP solution for spinning 
binaries is that some coefficients in Eqs.~\eqref{eq:radial_coeffs} vary on orbital timescales. 
To determine which coefficients are time-dependent, we employ the two 
2PN perturbative COMs, 
denoted with a tilde,  derived in 
Ref.~\cite{Tanay2020}. They are 
deformed versions of $\lseff$ and $l^2$\footnote{We remark that 
$\tilde{l^2}$ of 
Eq.~\eqref{eq:ltilde2def}
still remains a COM if we drop 
the first term (proportional to $\nu$) in the square brackets.
Reference \cite{Tanay2020} retained this term to ensure that the  Poisson bracket between 
$\tilde{l^2}$ and $\widetilde{\lseff}$ vanishes up to a certain PN order to ensure 2PN integrability.}
\begin{subequations}\label{eq:ltilde_squared}
\begin{align}
\widetilde{\lseff}&=\lseff +\frac{\nu}{2}(\mathbf{s}_1 \cdot \mathbf{s}_2) + \mathcal{O}(c^{-4})\,, \\
\widetilde{l^2} &= l^2 + \frac{1}{c^2} \,\Bigg[ 2 \,\nu\,(\mathbf{s}_1\!\cdot\!\mathbf{s}_2)\,h +\frac{2}{r}(\mathbf{n}\!\cdot\! \mathbf{s}_0)^2  - (\mathbf{p}\!\cdot\!\mathbf{s}_0)^2  \Bigg]\,.\label{eq:ltilde2def}
\end{align}
\end{subequations}
These COMs reveal two key features of the dynamics. First, since $\widetilde{\lseff}$ 
is a COM and differs from $\lseff$ in Eq.~\eqref{eq:D1s}
only by the term $(\mathbf{s}_1 \cdot \mathbf{s}_2) {\nu}/{2}$
which evolves on the slow precession timescale, $\lseff$ can be treated as effectively a constant over an orbital 
period. The same argument applies to $s_0^2$ in Eq.~\eqref{eq:D1s}. 
Second, Eq.~\eqref{eq:ltilde2def} 
reveals that $\tilde{l^2}$ 
has oscillatory terms (featuring
$\mathbf{n}$ and $\mathbf{p}$)
 at relative 2PN order. 
 These oscillatory terms oscillate at the 
 orbital frequency.
 As an aside, note that, just as solving an exactly integrable system requires the use of its COMs, the use of 
 perturbative COMs [Eqs.~\eqref{eq:ltilde_squared}] appears 
 to be the key to solving
  systems that are only perturbatively integrable, such as our 2PN spinning BBH; see Ref.~\cite{Tanay2020}
 for more details.

We now examine the time dependence of each coefficient in Eqs.~\eqref{eq:radial_coeffs}. 
The coefficients $A$ and $B$ depend only on the conserved Hamiltonian $h$ and are therefore constant. 
The coefficients $D_2$ and $D_3$ depend on $l^2$, but this dependence first appears at 2PN order. 
Since $l^2$ itself possesses 2PN oscillations, these oscillations contribute to $D_2$ and $D_3$ only at 
orders beyond 2PN, which we neglect. Thus $A$, $B$, $D_2$, and $D_3$ are all effectively constant 
and retain the same functional forms as in the 2PN nonspinning case of Ref.~\cite{Schafer1993}.
In contrast, the coefficient $C$ contains $l^2$ at Newtonian order. The 2PN oscillations of $l^2$ 
therefore cannot be neglected and induce a time dependence in $C$ through the orbital-frequency terms 
in Eq.~\eqref{eq:ltilde2def}. Similarly, $D_1 \equiv D_{1o} + D_{1s}$ is time-dependent because 
it depends explicitly on $(\mathbf{n}\cdot \mathbf{s}_0)^2$ in Eq.~\eqref{eq:D1s}, which oscillates 
with the orbital frequency.

To rewrite the time-dependent dot products $(\mathbf{n}\cdot \mathbf{s}_0)$ and $(\mathbf{p}\cdot \mathbf{s}_0)$ 
appearing in $C$ and $D_1$, we work in the NIF (defined in Sec.~\ref{sec:orb_L_vec_sol}) where the orbital angular momentum $\mathbf{L}$ is 
aligned with the $k$-axis. The orbital motion is confined to the $ij$-plane (see Fig.~\ref{IF-NIF}). 
At Newtonian order, the unit orbital separation and momentum vectors are given by
\begin{subequations}\label{eq:newtonian_vectors}
	\begin{align}
		\mathbf{n} &= \{\cos\phi, \sin\phi, 0\} \, , \\
		\mathbf{p} &= \frac{1}{l} \{-\sin\phi, e+\cos\phi, 0\} \, .
	\end{align}
\end{subequations}

Next, we project the spin vectors onto the orbital plane, denoting them as $\mathbf{s}_{ap} \equiv
\mathbf{s}_a - (\mathbf{s}_a \cdot \mathbf{k}) \mathbf{k}$ with magnitude $s_{ap} = s_a \sin \kappa_a$. 
Here ``$p$'' signifies projection onto
the orbital plane.
Decomposing $\mathbf{s}_0$ into its in-plane and orthogonal components ($\mathbf{s}_{0} \equiv \mathbf{s}_{0p} + \mathbf{s}_{0z}$), we find
\begin{equation}\label{eq:calS}
	\mathbf{s}_{0p} = \nu_1 \mathbf{s}_{1p} + \nu_2 \mathbf{s}_{2p} \,\quad \text{and}\quad \mathbf{s}_{0z}=\left(\nu_1 s_1 \cos \kappa_1 + \nu_2 s_2  \cos \kappa_2\right)\frac{\mathbf{l}}{l}      \, .
\end{equation}
The magnitude $s_{0p}$
and phase $\phi_{sp}$
(measured from $\mathbf{i}$ in the NIF)
 of the spin combination $\mathbf{s}_{0p}$ 
    are explicitly given by
\begin{subequations}
	\begin{align}
		s_{0p}^2 &=\nu_1^2s_{1p}^2 +\nu_2^2s_{2p}^2 + 2 \nu\,s_{1p}\,s_{2p}\,\cos(\phi_{s1}-\phi_{s2}) \, , \\
		\phi_{sp} &= \arctan\left(\frac{\nu_1\,s_{1p}\sin\phi_{s1} + \nu_2\,s_{2p}\sin\phi_{s2}}{\nu_1 \,s_{1p}\cos\phi_{s1} + \nu_2\,s_{2p}\cos\phi_{s2}}\right) \, .
	\end{align}
\end{subequations}

With these definitions, the oscillatory terms in Eq.~\eqref{eq:ltilde_squared} simplify to 
trigonometric functions of the phase difference $(\phi - \phi_{sp})$:
\begin{subequations}\label{eq:squaredterms}
\begin{align}
	(\mathbf{n} \cdot \mathbf{s}_0)^2&=s_{0p}^2 \cos^2(\phi - \phi_{sp})\,,\\[4pt]
	(\mathbf{p} \cdot \mathbf{s}_0)^2&=\frac{s_{0p}^2}{l^2}\,\big(e\sin\phi_{sp} - \sin(\phi - \phi_{sp})\big)^2\,.
\end{align}
\end{subequations}
Substituting Eqs.~\eqref{eq:squaredterms} into Eq.~\eqref{eq:ltilde_squared}, the 2PN constant $\widetilde{l^2}$ can be rewritten as
\begin{align}\label{eq:ltildesq}
	\widetilde{l^2} =&\, l^2 +\frac{1}{2c^2 }\frac{s_{0p}^2}{l^2}\left[\left( 3e\cos(\phi - 2\phi_{sp})+ 3\cos(2\phi - 2\phi_{sp})+ e\cos(3\phi - 2\phi_{sp})\right) +\left(1-e^2+4\nu \frac{h\,l^2}{s_{0p}^2} (\mathbf{s}_1\cdot \mathbf{s}_2)+e^2\cos 2\phi_{sp} \right)  \right]\,.
\end{align}
Since $\widetilde{l^2}$ is a 2PN COM,
we can evaluate it at $\phi=0$ and a
general $\phi$, and set these two evaluations 
equal to each other to express $l^2$ in terms of $l_0^2\equiv l^2(\phi=0)$:
\begin{align}\label{eq:l0sq}
	l^2&=l_0^2-\frac{1}{2c^2 }\frac{s_{0p}^2}{l_0^2}\left[\left( 3e\cos(\phi - 2\phi_{sp})+ 3\cos(2\phi - 2\phi_{sp})+ e\cos(3\phi - 2\phi_{sp})\right) -(3+4e) \cos 2\phi_{sp}  \right]\,.
\end{align}
From this equation (also present in Ref.~\cite{Gergely:1999pd}), we observe that $l^2(\phi=2\pi)=l^2_0$, indicating that the 
magnitude of the orbital angular momentum returns to its initial value after 
one full orbital period. 

The last step is to express $l^2$ in terms of its OA value, $\bar{l}\,^2$. 
The latter can be obtained by averaging either Eq.~\eqref{eq:ltildesq} or Eq.~\eqref{eq:l0sq}:
\begin{align}\label{eq:lbar_squared}
	\bar{l}\,^2&=l^2_0+\frac{1}{2c^2 }\frac{s_{0p}^2}{l_0^2}\left(3+4e \right)\cos 2\phi_{sp} = \widetilde{l^2}-\frac{1}{c^2}\, \left[2\nu \,(\mathbf{s}_1\!\cdot\!\mathbf{s}_2)\,h+\frac{s_{0p}^2}{2\widetilde{l^2}}\left(1-e^2+e^2\cos 2\phi_{sp}\right)\right]\,.
\end{align}
Using this OA value, $l^2$ can finally be decomposed as 
\begin{align}\label{eq:l_decomposition}
	l^2=\bar{l}\,^2-\frac{1}{2c^2 }\frac{s_{0p}^2}{\bar{l}\,^2}\left[3e\cos(\phi - 2\phi_{sp})+ 3\cos(2\phi - 2\phi_{sp})+ e\cos(3\phi - 2\phi_{sp})\right]\,.
\end{align}

We note that $l^2, \widetilde{l^2}, l_0^2$, and $\bar{l}\,^2$ all differ only by a 2PN quantity. 
Equation \eqref{eq:l_decomposition} shows that $l^2$ consists of a slowly varying secular 
component $\bar{l^2}$ plus rapidly oscillating harmonics at integer multiples of the orbital 
frequency. While the structure of this decomposition was previously obtained by 
Refs.~\cite{Gergely:1999pd, Klein2010},
here we have provided
a different derivation. Finally, substituting 
Eq.~\eqref{eq:l_decomposition} into the coefficients of Eq.~\eqref{eq:radial_coeffs}, we obtain
\begin{subequations}\label{eq:radial_coeffs_final}
	\begin{align}
		C &= -\bar{l}\,^2 + \frac{1}{c^2}\,\big[5(\nu-2)+2(1-3\nu) \,h \,\bar{l}\,^2+ \frac{s_{0p}^2}{2 \,\bar{l}\,^2}\left(3e\cos(\phi - 2\phi_{sp})+ 3\cos(2\phi - 2\phi_{sp})+ e\cos(3\phi - 2\phi_{sp})\right)\big]\nonumber\\
  		&\quad + \frac{1}{c^4}\,\big[ (37-122\nu+36\nu^2) \,h-3(1-5\nu+5\nu^2) \,h^2 \,\bar{l}\,^2 \big],\\[4pt]
		D_1 &= \frac{1}{c^2}\ \left[ (8-3\nu) \,\bar{l}\,^2 - 2\lseff + s_0^2 - 3 \,s_{0p}^2\, \cos^2(\phi - \phi_{sp})\right]+  \frac{1}{c^4}\, \biggl[\frac{53}{2}-\frac{83\nu}{2}+10\nu^2 - (16-73\nu+19\nu^2) \,h\, \bar{l}\,^2 \biggr]\, . 
	\end{align}
\end{subequations}
For the remaining coefficients ($A$, $B$, $D_2$, and $D_3$), $l$ can be replaced by its 
OA counterpart $\bar{l}$, as the resulting differences contribute only at orders 
beyond 2PN. This completes the derivation of the radial EOM. 
We stress that the hybrid spin sector solution was not needed for the above manipulations, 
meaning that the radial EOM is fully 2PN accurate.

%--------------------------------
\subsection{Azimuthal equation of motion}\label{sec:eomphi}
%--------------------------------

We now derive the EOM for the 
azimuthal angle $\phi$  of $\mathbf{r}$
in the NIF, measured from the
unit vector $\mathbf{i}$,
 in the orbital plane perpendicular 
 to $\mathbf{l}$, 
following the approach of Ref.~\cite{Cho2019}. The derivation requires expressing $\mathbf{l}=\mathbf{r}\times \mathbf{p}$ in the NIF, 
 via the Euler matrix in Eq.~\eqref{eq:Euler_matrix}. 
The separation vector's components in the NIF are given by
\begin{equation}\label{eq:rvecNIF}
	\mathbf{r} = r\begin{bmatrix}\cos \phi \\ \ \sin \phi \\\ 0\end{bmatrix}_{\text{NIF}}.
\end{equation}
Here $\phi$ can be considered as the generalization of the sum of 
the argument of periapsis
(originally defined for fixed Newtonian Keplerian ellipses) 
and the true anomaly to 
PN precessing orbits---refer to Fig.~\ref{fig:orbital_elements}.
For fixed Newtonian orbits,
the argument of periapsis is defined as the angle from the ascending node to the periapsis (or pericenter), whereas the true anomaly is measured from the pericenter to the particle's position, all in the orbital plane~\cite{Scheeres, SPITALE2002211}.

To express the canonical momentum $\mathbf{p}$ in the NIF, we start 
with the EOM for $\vb{r}$
\begin{align}    \label{eq:drdt_p}
\frac{d\mathbf{r}}{dt}
&= \frac{\partial h}{\partial \mathbf{p}} \nn\\
&= \mathbf{p}
-\frac{1}{2c^2}\bigg[
(1-3\nu)p^2\mathbf{p}
+\frac{2}{r}\Big((3+\nu)\mathbf{p}+\nu(\mathbf{n}\cdot\mathbf{p})\mathbf{n}\Big)
+\frac{2}{r^2}\big(\mathbf{n}\times\seff\big)
\bigg] \nn\\
&\quad +\frac{1}{8c^4}\bigg[
3(1-5\nu+5\nu^2)p^4\mathbf{p}
+\frac{4}{r}\Big((5-20\nu-3\nu^2)p^2\mathbf{p}
-\nu^2\big((\mathbf{n}\cdot\mathbf{p})p^2\mathbf{n}
+(\mathbf{n}\cdot\mathbf{p})^2\mathbf{p}
-3(\mathbf{n}\cdot\mathbf{p})^3\mathbf{n}\big)\Big) \nn\\
&\quad
+\frac{8}{r^2}\Big((5+8\nu)\mathbf{p}+3\nu(\mathbf{n}\cdot\mathbf{p})\mathbf{n}\Big)
\bigg].
\end{align}
Inverting this relation to express $\mathbf{p}$ in terms of $\dot{\mathbf{r}}$ up to 2PN order
%, as detailed in Appendix~\ref{app:drvecdt}, 
yields
\begin{align}\label{eq:p_from_drdt}
\mathbf{p} &= \dot{\mathbf{r}}
+\frac{1}{2c^2}\left[(1-3\nu)p^2\dot{\mathbf{r}}
+\frac{2}{r}\Big(\nu(\mathbf{n}\cdot\mathbf{p})\mathbf{n}+(3+\nu)\dot{\mathbf{r}}\Big) 
+ \frac{2}{r^2}(\mathbf{n}\times \seff )\right] \nn\\
&\quad +\frac{1}{8 c^4}\bigg[(-1+3\nu+3\nu^2)p^4\dot{\mathbf{r}} +\frac{8}{r^2}\Big(\nu^2(\mathbf{n}\cdot\mathbf{p})\mathbf{n}+(4-2\nu+\nu^2)\dot{\mathbf{r}}\Big)\nn\\
&\quad +\frac{4}{r}\Big((1+4\nu-3\nu^2)p^2\dot{\mathbf{r}}+\nu(1-2\nu)(\mathbf{n}\cdot\mathbf{p})p^2\mathbf{n}+\nu^2(\mathbf{n}\cdot\mathbf{p})^2\dot{\mathbf{r}}+3\nu^2(\mathbf{n}\cdot\mathbf{p})^3\mathbf{n}\Big)\bigg]+\mathcal{O}\Big(\frac{1}{c^5}\Big)\,.
\end{align} 
The derivative of 
$\mathbf{r}$ in the NIF becomes\footnote{
Starting with Eq.~\eqref{eq:rvecNIF},
we first write $\mathbf{r}$ in the IF
with the Euler matrix transformation of Eq.~\eqref{eq:Euler_matrix}.
Take time derivative of the resulting 
components and then transform them back to the 
NIF using the inverse Euler matrix transformation.}
\begin{align}\label{rvecdotNIF}
\dot{\mathbf{r}} =
	\begin{bmatrix}
		\dot{r}\cos\phi - r\sin\phi\,(\dot{\phi}_L\cos\theta_L+\dot{\phi}) \\
		\dot{r}\sin\phi + r\cos\phi\,(\dot{\phi}_L\cos\theta_L+\dot{\phi}) \\
		r(-\dot{\phi}_L\sin\theta_L\cos\phi + \dot{\theta}_L\sin\phi)
	\end{bmatrix}_{\text{NIF}}\,.
\end{align}

We now express the $z$-component of the vector equation
\begin{equation}\label{eq:lNIF}
\mathbf{l}_{\text{NIF}} = \mathbf{r}_{\text{NIF}} \times \mathbf{p}_{\text{NIF}}\,,
\end{equation}
with the understanding that the $z$-component of the left-hand side is equal to $l$, by construction.
By substituting Eqs.~\eqref{eq:vecNIF} and \eqref{eq:rvecNIF}--\eqref{rvecdotNIF} into Eq.~\eqref{eq:lNIF}
and isolating $d\phi/dt$, we obtain the full azimuthal equation of motion
\begin{equation}\label{eq:dphidtApp}
\frac{d\phi}{dt} = \frac{F}{r^2} + \frac{I_{1\text{o}}+\lseff/(c^2 l)}{r^3}
+\frac{I_2}{r^4}+\frac{I_3}{r^5}
- \cos \theta_L \frac{d\phi_L}{dt}\, ,
\end{equation}
where the coefficients $F$, $I_{1o}$, $I_2$, and $I_3$ possess the same functional form as in the nonspinning case derived in Ref.~\cite{Schafer1993}
\begin{subequations}\label{eq:azimuthal_coeffs}
\begin{align}
F &= l + \frac{1}{c^2} (3\nu-1)\, h\, l
+ \frac{1}{2c^4} (2-9\nu+6\nu^2)\, h^2 \,l, \label{eq:coeffF}\\
I_{1o} &= \frac{2}{c^2} (\nu-2)\,l
+ \frac{2}{c^4} (2-11\nu+4\nu^2)\, h\, l,\\
I_2 &= \frac{1}{2c^4} (17-22\nu+10\nu^2) \,l,\\
I_3 &= \frac{1}{2c^4} \nu (1-2\nu) \,l^3 .
\end{align}
\end{subequations}

Although Eq.~\eqref{eq:dphidtApp} follows directly from the full 2PN dynamics,
obtaining a closed-form solution requires expressing its right-hand side
entirely in terms of analytic quantities.

We first simplify all terms in Eq.~\eqref{eq:dphidtApp}, except for the
NIF-precession contribution,
$-\cos\theta_L\,d\phi_L/dt$, while remaining entirely within the full
2PN framework. Since the norm of the orbital angular momentum $l$
appears at Newtonian order only through the coefficient $F$, we retain
its complete 2PN expression given by Eq.~\eqref{eq:l_decomposition}. Its
2PN oscillations therefore contribute to 
Eq.~\eqref{eq:dphidtApp}, playing a role analogous to that of the
coefficient $C$ in Eq.~\eqref{eq:drdt_final}. For all other
occurrences of $l$ in the higher-order terms, we replace it with the
orbit-averaged constant $\bar{l}$. The associated oscillatory corrections
would contribute only beyond the 2PN order. Consequently, $I_{1o}$,
$I_2$, and $I_3$ may be treated as constant over an orbital period,
analogous to $D_{1o}$, $D_2$, and $D_3$ in
Eqs.~\eqref{eq:drdt_final}.

Likewise, $\lseff$ may be regarded as constant on the orbital timescale.
This is justified because it differs from a 2PN constant of motion only
through a term proportional to $(\mathbf{s}_1\cdot\mathbf{s}_2)$
[see Eq.~\eqref{eq:ltilde_squared}], which evolves on the much longer
spin-precession timescale.

The hybrid approximation is introduced only in the NIF-precession
contribution. Specifically, we replace $\cos\theta_L$ and
$d\phi_L/dt$ by their corresponding hybrid solutions, given by
Eqs.~\eqref{eq:cos_thetaL} and \eqref{eq:dthetaL_dt}, with the
replacement $d^{-3}\rightarrow r^{-3}$. As a consequence,
$\Sigma_2$ and $\lambda$, which are not conserved quantities under the
full 2PN dynamics, are treated as constants within this contribution,
consistent with the hybrid prescription. No hybrid approximation is
introduced in the remaining terms of Eq.~\eqref{eq:dphidtApp}.

After applying these substitutions, $d\phi/dt$ simplifies to
\begin{align}\label{eq:dphidt}
\frac{d \phi}{dt} &=
\frac{F}{r^2}
+ \frac{I_{1o}+I_{1s}}{r^3}
+ \frac{I_2}{r^4}
+ \frac{I_3}{r^5}
+ \frac{1}{c^2 r^3}
\left(
\frac{\beta_{1L}}{\alpha_{1L}+\cos\kappa_1^{(2,\mathrm{H})}}
+
\frac{\beta_{2L}}{\alpha_{2L}+\cos\kappa_1^{(2,\mathrm{H})}}
\right).
\end{align}
where the coefficient $I_{1s}$ is given by
\begin{align}
I_{1s}
=
-\frac{1}{2c^2}
\Big[
2\Delta(1-\lambda)\bar{l}
+\lambda(\nu\bar{l}-3\nu_2s_2\Sigma_2)
\Big],
\end{align}
with the mass parameter $\Delta$ defined as
$\Delta\equiv\delta_1+\delta_2-\nu/2$.

We emphasize that $I_{1s}$ depends only on quantities varying on the
spin-precession timescale. This dependence is a direct consequence of
applying the hybrid approximation to the precession of the NIF. Thus,
the present framework retains both the secular and oscillatory dynamics
associated with $l$, as well as the leading-order oscillatory and secular
evolution of the NIF. However, it does not capture the spin-spin
oscillations generated by the NIF precession.

The final term in Eq.~\eqref{eq:dphidt} arises from the precession of the
NIF and reflects the fact that the orbital phase $\phi$ is measured with
respect to the $i$-axis of the NIF, which itself precesses relative to
the inertial frame. The appearance of the factor $1/r^3$, rather than
the orbit-averaged factor $1/d^3$, follows directly from the hybrid
prescription adopted for $d\phi_L/dt$. The coefficients $\alpha_{iL}$ and
$\beta_{iL}$ entering this term were defined previously in
Eqs.~\eqref{eq:alpha_beta_L}.

In summary, Eq.~\eqref{eq:dphidt} preserves the full 2PN treatment of
the orbital dynamics, including the oscillatory behavior of $l$. The
hybrid approximation is restricted exclusively to the NIF-precession
term and the quantities entering it. In the following subsection, we
construct a QKP solution for $\mathbf{R}$ in the NIF by finding a
solution that simultaneously satisfies
Eqs.~\eqref{eq:drdt_final} and \eqref{eq:dphidt}.

%--------------------------------
\subsection{Quasi-Keplerian parametrization of $\mathbf{R}$ vector}\label{sec:QKP}
%--------------------------------

\subsubsection{The solution}

Unlike the nonspinning case treated by Schäfer and Wex in Ref.~\cite{Schafer1993} or the 
spin-orbit case by Cho and Lee in Ref.~\cite{Cho2019}, a direct integration of 
EOMs~\eqref{eq:drdt_final} and~\eqref{eq:dphidt} is not possible because some coefficients 
vary on the orbital timescale. We therefore follow the ansatz-based approach introduced by 
Klein in Ref.~\cite{Klein2010}, constructing a parametric solution involving generalized 
Keplerian elements with free parameters that are fixed by requiring the ansatz to satisfy 
the EOMs. The key constraint is that the resulting orbital elements must remain 
constant on the orbital timescale (though they may vary on the longer precessional timescale). 
A detailed harmonic analysis justifying the functional form of our ansatz is presented in 
Appendix~\ref {app:ansatz}. 
Here we simply present
the resulting QKP solution
that satisfies
 Eqs.~\eqref{eq:drdt_final} and~\eqref{eq:dphidt}.
 It reads 
\begin{subequations}\label{eq:QK_parametrization}
\begin{align}
n(t-t_0) &= u - e_t \sin u + \frac{1}{c^4} \left[f_t \sin v_\phi + g_t (v_\phi - u) \right],\label{eq:QKPt}\\
r &= a_r (1 - e_r \cos u) + \frac{1}{c^2} h_r \cos(2 v_\phi - 2 \phi_{sp}),\label{eq:QKPr}\\
v_\phi &= u + 2 \arctan \left( \frac{\beta_\phi \sin u}{1 - \beta_\phi \cos u} \right), \quad \text{ with }\  \beta_\phi = \frac{e_\phi}{1 + \sqrt{1 - e_\phi^2}},\\
\phi(t) -\phi(t_0)&= (1+k') v_\phi + \frac{1}{c^4} \left[f_\phi \sin 2 v_\phi + g_\phi \sin 3 v_\phi\right] +\frac{1}{c^2} \left[h_{\phi1} \sin(v_\phi - 2 \phi_{sp}) + h_{\phi2} \sin(2 v_\phi - 2 \phi_{sp}) \right] +\Pi_{\text{osc}}(u) \label{eq:QKPphi},
\end{align}
\end{subequations}
where the purely oscillatory term $\Pi_{\text{osc}}(u)$ is defined as
\begin{align}\label{eq:piosc}
\Pi_{\text{osc}}(u) &= \sum_{i=1}^{2} \frac{\beta_{iL}}{\alpha_{iL} + x_-} \left[ \frac{2\,\Pi\left(n_i, \operatorname{am}(\Upsilon^{(2,\mathrm{H})}(u), \beta), \beta\right)}{\sqrt{A(x_3 - x_-)}} - \frac{\Pi(n_i, \beta)}{n c^2 d^3 K(\beta)}v_\phi \right].
\end{align}
The quasi-Keplerian parameters are given by
\begin{subequations}\label{eq:QK_parameters}
\begin{align}
n &=  (-2 h)^{3/2} \Bigg[ 1 + \frac{1}{4 c^2} (15 - \nu)\, h + \frac{1}{c^4} \Bigg( \frac{3}{2}(-5+2\nu)\frac{(-2 \,h)^{3/2}}{\bar{l}} + \frac{1}{32} (555 + 30 \nu + 11 \nu^2)\,h^2 \Bigg) \Bigg]\,,\\
e_t^2 &= 1 + 2 \,h \, \bar{l}\,^2 + \frac{1}{c^2} \Bigg[ 4 (1 - \nu) \,h + (17 - 7 \nu) \,h^2 \,\bar{l}\,^2 + \frac{h}{\bar{l}\,^2} (4 \lseff - 2 s_0^2 + 3 s_{0p}^2) \Bigg] \\
&\quad + \frac{1}{c^4} \Bigg[(11 \nu-17) \frac{h}{\bar{l}\,^2} + 2 (2 + \nu + 5 \nu^2) \,h^2 + (112 - 47 \nu + 16 \nu^2) \,h^3 \,\bar{l}\,^2  + 3 (2 \nu - 5) (1 + 2 \,h \,\bar{l}\,^2) \frac{(-2 h)^{3/2}}{\bar{l}} \Bigg]\,,\\
f_t &= -\frac{1}{8} \nu (4 + \nu) \sqrt{1 + 2\, h\, \bar{l}\,^2} \frac{(-2 \,h)^{3/2}}{\bar{l}}\,,\label{eq:param_ft}\\
g_t &= \frac{3}{2} (5 - 2 \nu) \frac{(-2\, h)^{3/2}}{\bar{l}}\,,\label{eq:param_gt}\\
a_r &= -\frac{1}{2 h} \Bigg[ 1 + \frac{1}{2c^2} \,h \Big( (7 - \nu) - \frac{1}{\bar{l}\,^2}\left(4 \lseff -2 s_0^2+3s_{0p}^2\right) \Big) + \frac{1}{4c^4}  \Big( 2 (17 - 11 \nu) \frac{h}{\bar{l}\,^2} + (1 + 10 \nu + \nu^2)\, h^2 \Big) \Bigg]\,,\\
e_r^2 &= 1 + 2\, h\, \bar{l}\,^2 - \frac{1}{c^2} \Big[ 2(6 - \nu) \,h + 5 (3 - \nu)\, h^2\, \bar{l}\,^2 -2\frac{h}{\bar{l}\,^2}(1+h \bar{l}\,^2)\left(4 \lseff -2 s_0^2+3s_{0p}^2\right) \Big] \nn\\
&\quad + \frac{1}{c^4} \Big[ -2 (17 - 11 \nu) \frac{h}{\bar{l}\,^2} + (26 + \nu + \nu^2) \,h^2 + (80 - 55 \nu + 4 \nu^2) \,h^3\, \bar{l}\,^2 \Big]\,,\\
h_r &= -\frac{s_{0p}^2}{4\,\bar{l}\,^2}\,,\label{eq:param_hr}\\
e_\phi^2 &= 1 + 2 \,h \,\bar{l}\,^2  + \frac{1}{c^2} h \Bigg[ -(12 - 4 \Delta) - (15 - \nu - 8 \Delta) \,h \,\bar{l}\,^2 + \frac{1}{\bar{l}\,^2} (3 + 4 \,h \,\bar{l}\,^2) (4 \lseff - 2 s_0^2 + 3 s_{0p}^2) \nn\\
&\qquad - 2 (1 + 2 \,h \,\bar{l}\,^2) \lambda \left( (2 \Delta - \nu) + 3 \frac{s_2}{\bar{l}} \nu_2\Sigma_2 \right) \Bigg] + \frac{1}{c^4} \Bigg[ \frac{1}{8} \Big( -408 + 232 \nu + 15 \nu^2 + 64 \Delta \Big(4 - \frac{5}{2} \nu + \Delta \Big) \Big) \frac{h}{\bar{l}\,^2} \nn\\
&\qquad - \frac{1}{2} \Big( 16 - 88 \nu - 9 \nu^2 - 12 \Delta \Big(1 - 9 \nu + \frac{14}{3}\Delta \Big) \Big)  h^2  + \frac{1}{2} \Big( 160 - 30 \nu + 3 \nu^2 - 16 \Delta (10 + 3 \nu - 3 \Delta) \Big) \,h^3 \,\bar{l}\,^2 \Bigg]\,\\
k' &= \frac{1}{c^2\bar{l}^2} \Bigg[ 3 - \Delta   -\frac{3}{4\bar{l}^2} (4 \lseff - 2 s_0^2 + 3 s_{0p}^2) + \frac{1}{2} \lambda \left( (2 \Delta - \nu) + 3 \frac{s_2}{\bar{l}} \nu_2 \Sigma_2 \right) + \frac{\bar{l}^2}{n d^3} \sum_{i=1}^{2} \frac{\beta_{iL}}{\alpha_{iL} + x_-} \frac{\Pi(n_i, \beta)}{K(\beta)}  \nn\\
&\quad + \frac{1}{4c^2\bar{l}^2} \Big( 105 - 30 \nu - 60 \Delta + 15 \Delta \nu - 2 \,h \,\bar{l}^2 \big(-15 + 6 \nu + \Delta (18 - 13 \nu)\big) \Big) \Bigg]   \label{eq:param_k}\,,\\
f_\phi &= \frac{1}{8} (\nu - 3 \nu^2 - \Delta (8 - 5 \nu + 2 \Delta)) (1 + 2 \, h\,  \bar{l}\,^2)\,  \frac{1}{\bar{l}\,^4}\,,\\
g_\phi &= - \frac{3 \nu^2}{32} (1 + 2\,  h\,  \bar{l}\,^2)^{3/2}\,  \frac{1}{\bar{l}\,^4}\,,\\
h_{\phi1} &= -\frac{1}{2} \sqrt{1 + 2\,h\,  \bar{l}\,^2} \, \frac{s_{0p}^2}{\bar{l}\,^4}\,,\label{eq:param_hphi1}\\
h_{\phi2} &= -\,  \frac{s_{0p}^2}{8\,\bar{l}\,^4}\, \label{eq:param_hphi2}.
\end{align}
\end{subequations}

\begin{figure*}[p] 
    \centering
    \includegraphics[height=0.97\textheight, width=\textwidth, keepaspectratio, trim=0 0 0 -15]{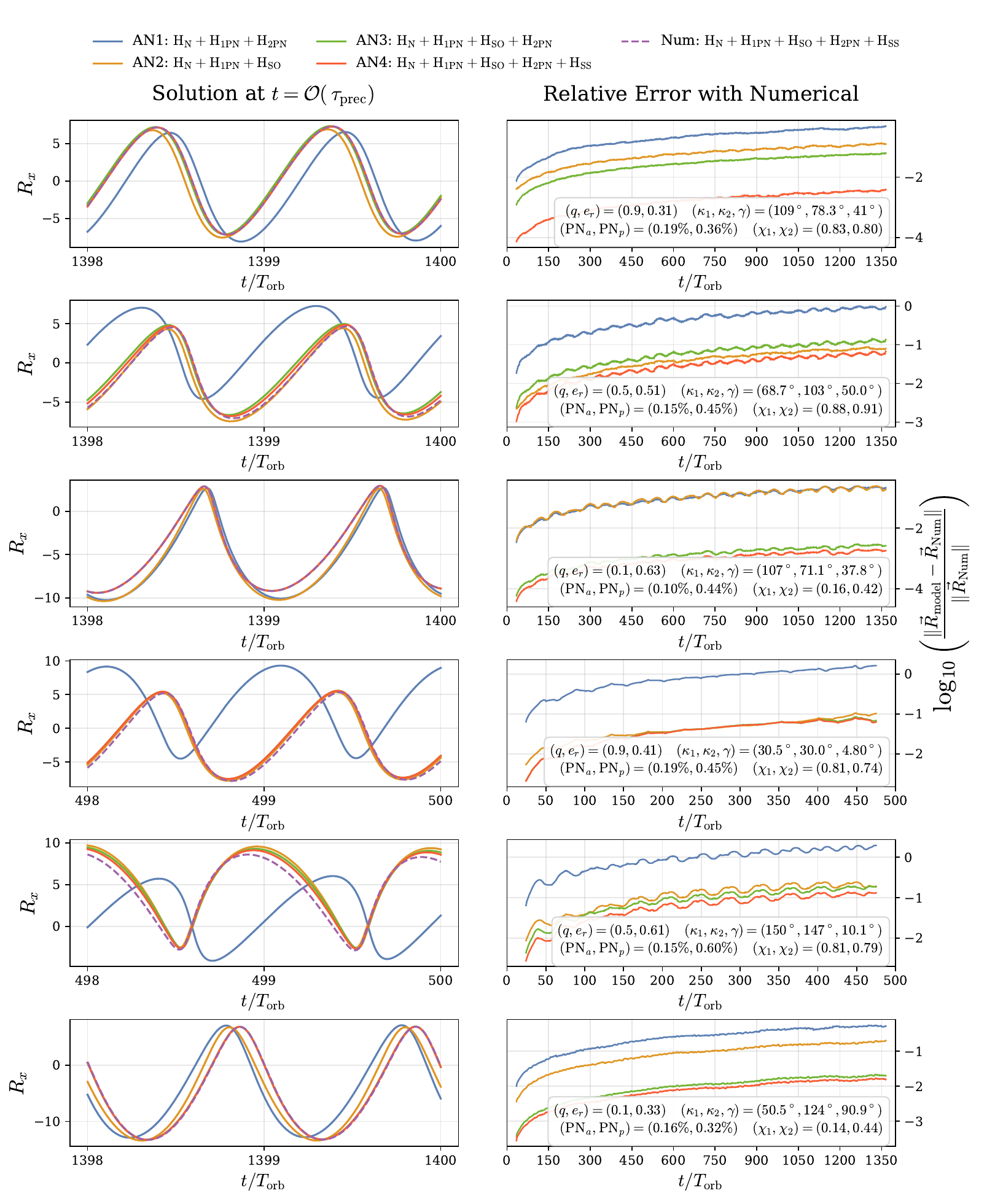}
  \caption{\justifying
  Evolution of $R_x$ (left column) and moving average of the associated log-fractional errors (right column) for six random systems. The left column shows a two-orbit window around $t \approx 500\, T_{\text{orb}}$ or
  $1400\, T_{\text{orb}}$, while the fractional errors (right) are plotted over the entire evolution
  period. System parameters ($q, e_r, \kappa_i, \gamma, \mathrm{PN}_i, \chi_i$) are given for each case. Overall, our model (AN4) yields the best agreement with the numerical result, although for some initial conditions, the improvement over the model without spin-spin terms (AN3) is modest. Agreement with the numerical solution also degrades when the spins are nearly (anti)aligned with  $\vb{L}$ (fourth and fifth rows).}
    \label{fig:r_vector_plot}
\end{figure*}

Interestingly, although
$\Delta \equiv \delta_1 + \delta_2 - \nu/2$
originates  from the 1.5PN spin-orbit and 2PN spin-spin couplings
(via Eq.~\eqref{eq:hamiltonian_terms} and $\mathbf{s}_{\mathrm{eff}} = \delta_1 \mathbf{s}_1 + \delta_2 \mathbf{s}_2$; also see 
 Eq.~\eqref{eq:angle_rates}, where both $\delta_{1/2}$ and $\nu$ are
 featured),
it enters the azimuthal EOM~\eqref{eq:dphidt} at  1PN order through 
the coefficient $I_{1s}$.
This 1PN scaling allows it to couple  with the 1PN nonspinning 
orbital terms, generating novel 2PN terms in the azimuthal quasi-Keplerian 
parameters---note multiple occurrences 
at relative 2PN order
of $\Delta$
in the definitions of $e_\phi$, $k'$
and $f_{\phi}$, which make up
the azimuthal Eq.~\eqref{eq:QKPphi}.
This  coupling represents a qualitative 
departure from the 1.5PN dynamics---in the latter, the nonspinning and the spin-orbit PN contributions remain   
additively separate in the QKP solution of $(r, \phi)$.
At 2PN order, this clean separation breaks
down in the azimuthal sector, necessitating a unified  treatment of the spin 
and orbital terms. Furthermore, because of the 2PN spin-spin dynamics, the quasi-Keplerian 
parameters that depend explicitly on $\mathbf{s}_0$ or $\lseff$ are no longer exactly constant, 
but instead exhibit slow secular variations.

We highlight a subtle but physically significant distinction between the azimuthal 
parameter $k'$ appearing in Eq.~\eqref{eq:QKPphi} and the standard periastron advance $k$
found in much of the literature (e.g. Refs.~\cite{Schafer1993,Tessmer2010}).

Because the orbital phase $\phi(t)$ is formulated 
as a phase in the NIF, 
it is measured relative to the precessing $\mathbf{i}$-axis of the NIF. 
Consequently, $k'$ physically represents the ``noninertial periastron advance'' relative to this moving axis. 
However, one can construct a meaningful ``inertial periastron advance'' parameter $k$ from $k'$, as we will do later in Sec.~\ref{sec:four_frequencies}.

Two quick comments are in order. First,
it is easy to see that $\phi(t_0)$ in Eq.~\eqref{eq:QKPphi} is the generalization of the argument of periapsis (shown in Fig.~\ref{fig:orbital_elements};  originally defined for fixed Newtonian ellipses) to precessing PN orbits.
Also, note  that
Eq.~\eqref{eq:QKPphi} cleanly
separates the secular periastron advance term 
(the first  term on the rhs)
from the oscillatory ones
(the last four terms).

In summary, Eqs.~\eqref{eq:QK_parametrization}--\eqref{eq:QK_parameters} constitute our 
2PN quasi-Keplerian solution for spinning binaries on eccentric orbits. The solution is valid for 
arbitrary mass ratios, spin magnitudes, spin orientations, and eccentricities (provided the orbit 
remains bound, i.e., $e < 1$) as long as we remain within the PN approximation.

\subsubsection{The plots}

We plot our $\vb{R}$ solution (specifically the $x$-component, in a random inertial frame) in  Fig.~\ref{fig:r_vector_plot}, along with 
three other analytical solutions and the full 2PN
numerical solution (the baseline to compare against). The other three analytical 
solutions displayed are 
(a) 2PN purely orbital (without spin effects),
(b) full 1.5PN solution, and
(c) the previous solution but with the 2PN orbital effects included; refer to the legends therein.
These solutions are plotted in the first column, whereas in the second column, we plot the moving average of their relative error against the full 2PN numerical solution for comparison on a log scale; see the $y$-axis label for 
a concrete description.
We apply the moving average because the raw error plots are highly erratic with sharp fluctuations.

A few patterns can be read from the plots of Fig.~\ref{fig:r_vector_plot}.
First, as expected, the purely 2PN solution (blue) 
has the largest error, thus 
underscoring the importance of faithfully including the spin-precession effects while modeling the spinning BBH dynamics.
Second, our 2PN solution overall 
has the lowest error and fares the best. In many of the plots, there is an order-of-magnitude improvement when we switch from the 1.5PN solution (orange) to our 2PN solution (red). This is when the plotted 1.5PN
solution (just like the 2PN) has  been artificially enhanced by ``injecting'' the 1PN $r$ solution
during the construction of the associated spin solutions, which 
makes the resulting solutions have 
unphysical PN accuracy, as already discussed before.

%--------------------------------
\subsection{Solution of $\vb{P}$ vector}   \label{sec:p_sol}
%--------------------------------

The solution of $\vb{P}$ (or rather $\vb{p}$) is straightforward to construct.
 We do not present the explicit form, but rather discuss the procedure to obtain it. 
$\vb{p}$ is given by 
Eq.~\eqref{eq:p_from_drdt}.
$d \vb{r}/dt$
is straightforward
(but possibly tedious) to construct by a simple differentiation of the $\vb{r}$
solution constructed in the previous subsections.
The terms in the square braces  of 
Eq.~\eqref{eq:p_from_drdt}
are higher PN terms which are additionally functions of $ \vb{p}$, and the spins, which can further be replaced by their 1.5PN solutions, available in Ref.~\cite{Samanta2022}. 
This completes the description of how to construct the $\vb{P}$ solution.

\section{The Four Frequencies of the System}
\label{sec:four_frequencies}
%--------------------------------

In the action-angle framework, frequencies emerge directly from the Hamiltonian $H$ 
and the action variables $\mathcal{J}_i$ via $\omega_i = \partial H/\partial \mathcal{J}_i$.
The number of frequencies or the number of action variables is half the number of phase space variables
(dimension of the symplectic manifold) in question, although some of the frequencies 
could turn out to be zero due to the symmetries of the Hamiltonian~\cite{Tanay2023}.

Our system is described by four 3D Cartesian vectors: the relative separation $\mathbf{r}$, 
the canonical momentum $\mathbf{p}$, and the two spin vectors $\mathbf{s}_1$ 
and $\mathbf{s}_2$, yielding a 12-dimensional phase space, which is not necessarily 
symplectic. Upon the imposition of the NWP SSC, the conservative dynamics 
preserves the magnitudes of the individual spins, $s_1$ and $s_2$. This leads to 
two two-dimensional spherical symplectic manifolds (one for each spin vector)~\cite{Tanay2020}, 
in addition to the six-dimensional Cartesian symplectic manifold with coordinates $(\vb{r}, \vb{p})$\footnote{See Ref.~\cite{Jose1998} for a general discussion on the symplectic geometric 
approach to classical mechanics, and Chap.~2 of Ref.~\cite{cannas} for a concise 
discussion on spherical symplectic manifolds.}. This gives us a ten-dimensional symplectic 
manifold as our phase space, containing five canonical position-momentum pairs. 
Since each of these pairs produces one frequency of the dynamics of the system, 
we have five potential frequencies. For our 2PN BBH system, due to the 
SO(3) invariance of the system, one of the frequencies 
$\partial H/\partial J_z$ vanishes. Here, $J_z$ is one of the action variables, 
just as it is for the 1.5PN system as well~\cite{Tanay2023}.
This leaves  the 2PN spinning system with four nonzero  frequencies 
governing its multitimescale evolution. Two of them correspond to the 
spin-sector dynamics, and hence are slow, whereas the other two frequencies 
correspond to the orbital sector.

The first frequency, $\omega_{\text{nut}}$, governs the internal nutation dynamics of the spin subsystem. 
Because it is a slow frequency, we can extract it directly from the solution of the OA dynamics. 
While the individual spin magnitudes are constant, spin-orbit and spin-spin interactions cause the relative 
angle between the spins ($\cos\gamma$) and their inclinations relative to the orbital angular momentum 
($\cos\kappa_1, \cos\kappa_2$) to oscillate, as detailed in Eq.~\eqref{eq:cos_k1_hybd}. The inclination of 
the orbital plane, $\theta_L$, shares this periodicity through its dependence on the $\Sigma_i$ parameters; 
see Eqs.~\eqref{eq:Sigmas} and~\eqref{eq:cos_thetaL}. This oscillation is captured by the single degree 
of freedom $\cos\kappa_1$, whose evolution is determined by the linear growth of the argument 
$\Upsilon^{(2,\mathrm{H})}(t)$ of Eq.~\eqref{eq:Upsilon_params}. Because the $\text{sn}^2(\Upsilon, \beta)$ term 
in Eq.~\eqref{eq:kappa1_solution} has a period of $2K(\beta)$ in its first argument $\Upsilon$, 
the nutation frequency ($2\pi$ divided by the time period of $\cos \kappa_1$) is computed to be
\begin{equation}\label{eq:freq_nut}
    \omega_{\text{nut}}= \frac{\pi \sqrt{A(x_3-x_-)}}{2 c^2 d^3 K(\beta)} \, ,
\end{equation}
where $K(\beta)$ is the complete elliptic integral of the first kind, $A$ 
is the  parameter defined in Eq.~\eqref{eq:A_constant}, ($x_-, x_3$) are the roots of the 
cubic polynomial detailed in Appendix~\ref{app:rootsx}, and $d$ 
(defined in Eq.~\eqref{def:d}) depends on the 1PN QKP.

The second slow frequency, $\omega_{\text{prec}}$, characterizes the secular precession 
of the orbital plane. It corresponds to the rate at which the azimuthal angle $\phi_L$ 
advances, describing the rotation of the orbital angular momentum $\mathbf{L}$ 
around the conserved vector $\mathbf{J}$. This frequency can similarly be extracted from 
the OA linear drift of $\phi_L(t)$ given in Eq.~\eqref{eq:phiL_solution}. Defined as $2\pi$ divided 
by the time required for $\phi_L$ to advance by $2\pi$, the precession frequency evaluates to
\begin{equation}\label{eq:freq_prec}
    \omega_{\text{prec}} = \frac{1}{c^2 d^3} \left[ \frac{\beta_{1L}}{\alpha_{1L} + x_-} \frac{\Pi(n_1, \beta)}{K(\beta)} - 
    \frac{\beta_{2L}}{\alpha_{2L} + x_-} \frac{\Pi(n_2, \beta)}{K(\beta)} + \beta_{3L} \right] \, .
\end{equation}

The third frequency is the radial mean motion $\omega_r = n$, which characterizes the quasi-Keplerian 
orbital motion. It governs the period of radial oscillation between successive periastron passages, 
$\tau_{\text{orb}} = 2\pi/n$, and is defined in Eq.~\eqref{eq:QK_parameters}. 

The fourth fundamental frequency tracks the intrinsic azimuthal motion of $\vb{r}$ within the orbital plane. 
Over a single radial period, the orbital phase $\phi$ in the NIF accumulates an advance of $\Delta \phi = 2\pi(1+k')$. 
This secular growth can be seen from Eq.~\eqref{eq:QKPphi}, where the first term grows linearly in $v_\phi$ while 
all remaining contributions are bounded and $2\pi$-periodic in $u$. Consequently, the fourth frequency is defined as
\begin{equation}\label{eq:peri_adv_orb}
\omega_{\phi}^{\mathrm{orb}} = n(1+k') \, .
\end{equation}
This completes our specification of the four frequencies of the system. However, to establish contact 
with the existing celestial mechanics literature (specifically the longitude of periapsis) and the spin-aligned 
post-Newtonian periastron advance, it is highly advantageous to define a new frequency constructed 
out of two of the previous four.

Just as the phases $\phi_L$ and $\phi$ yield the frequencies $\omega_{\text{prec}}$ and 
$\omega_{\phi}^{\text{orb}}$, the auxiliary phase $\phi_L + \phi$ yields the inertial frequency 
\begin{equation}
\omega_{\phi}^{\mathrm{inertial}} \equiv \omega_{\phi}^{\mathrm{orb}} + \omega_{\text{prec}} \, .
\end{equation}
Taking inspiration from Eq.~\eqref{eq:peri_adv_orb} to define the periastron advance parameter 
associated with a given frequency, we define the \textit{inertial periastron advance} parameter $k$ via
\begin{align}
\omega_{\phi}^{\mathrm{inertial}}  &  \equiv n (1 + k) ,  
\end{align}
which yields
\begin{align}   \label{eq:inertial_k_param}
k  & = k' + \frac{\omega_{\ty{prec}}}{n}.   
\end{align}

The hasty introduction of the inertial periastron advance parameter $k$ warrants some discussion. One can 
raise concern about 
$\omega_{\phi}^{\mathrm{inertial}} $ being a legitimate frequency. This is so because it is based on  the auxiliary angle
$\phi_L + \phi$,
which is not a  geometrical angle between two lines---$\phi_L$ and 
 $\phi$ are measured in two different planes, one perpendicular to $\vb{J}$, and the other perpendicular to
 $\vb{L}$.
However, such a combination of angles is not devoid of historical precedents.
In fact, 
$\phi_L + \phi$ can be considered as a generalization of the sum of longitude of the periapsis  and the true anomaly
 (for fixed Newtonian ellipses)
to our precessing PN orbit.
In the context of Fig.~\ref{fig:orbital_elements} (for fixed Newtonian ellipses),
the longitude of periapsis is
the sum of the longitude of the ascending node and the argument of periapsis\footnote{We caution the reader that
 some references define the longitude of periapsis to mean the same thing as the argument of periapsis; for example, see Ref.~\cite{Poisson_Will_2014}.}.
 Leaving
 this historical precedent aside, below we argue why 
$\omega_{\phi}^{\mathrm{inertial}} $ is indeed a meaningful frequency.

First, the introduction of 
$\omega_{\phi}^{\mathrm{inertial}} $
ensures consistency in the spin-aligned limit: it is $k$ (and not $k'$) 
that smoothly reduces to the standard periastron advance parameter $k_{\theta}$ of Eq.~\eqref{eq:ktheta}, 
to be seen later in Sec.~\ref{sec:spin_aligned}. 
 This is so because in the spin-aligned limit, the two planes in which $\phi_L$ and $\phi$ are measured, start to coincide, thereby turning $\phi_L + \phi$ into a geometrically meaningful angle.
 The reason we should think in the language of limits is because
in the  spin-aligned configuration, the NIF becomes undefined, 
as the basis unit vector $\vb{i}$ of the NIF must be parallel to the zero vector ($\vb{J} \times \vb{L}$).

Second, to  confirm that $\omega_{\phi}^{\text{inertial}}$ constitutes a valid frequency 
in the action-angle framework, one should realize that the action-angle variables, and hence the frequencies of a system are not unique. A valid transformation 
from one set of angle variables to another takes the form of a linear transformation where the transformation 
matrix $\mathcal{M}$ has integer entries and $\det(\mathcal{M}) = \pm 1$ (see Proposition~11.3 of 
Ref.~\cite{Fasano2006}). Naturally, this linear transformation applies also to the frequencies, which are 
the time derivatives of the angles. One can easily verify that the matrix

\begin{align}
\mathcal{M} = \begin{pmatrix}
1 & 1 & 0 & 0 \\
0 & 1 & 0 & 0 \\
0 & 0 & 1 & 0 \\
0 & 0 & 0 & 1
\end{pmatrix},
\end{align}
satisfies these two criteria. Applying $\mathcal{M}$ to our initial frequency vector 
$(\omega_{\phi}^{\text{orb}}, \omega_{\text{prec}}, \omega_{\text{nut}}, \omega_r)$ yields the 
transformed vector $(\omega_{\phi}^{\mathrm{inertial}}, \omega_{\text{prec}}, \omega_{\text{nut}}, \omega_r)$, 
confirming that $\omega_{\phi}^{\mathrm{inertial}}$ is a perfectly valid  frequency. All this makes the introduction of $\omega_{\phi}^{\mathrm{inertial}}$ and $k$ of Eq.~\eqref{eq:inertial_k_param} a very satisfying development.

%--------------------------------
\section{Limiting cases: circular and spin-aligned}
\label{sec:limits}
%--------------------------------

In this section, we demonstrate that our  2PN spinning quasi-Keplerian parametrization 
recovers the standard results for 
certain limiting cases (circular, spin-aligned) found in  Refs.~\cite{Kidder1995, Henry:2023tka}). We also discuss 
the possible agreement with
Ref.~\cite{Klein2010}.

\subsection{Circular limit}

 From 
Eqs.~\eqref{eq:QK_parametrization} and~\eqref{eq:QK_parameters},
we see that demanding 
$dr/dt = 0$ requires $h_r = 0$, in addition to $e_r = 0$; the latter condition alone is sufficient to guarantee $dr/dt = 0$ for the nonspinning case.
This means that $r = \ty{constant}$ orbit is not possible for general spin orientations at 2PN.
This conclusion is fully 2PN accurate as it is derived from our fully 2PN $r$ solution of Eq.~\eqref{eq:QK_parametrization}, which does not rely on the hybrid approximation. This is in line with the conclusions reached about the circular limit in Ref.~\cite{Kidder1995}.

\subsection{Spin-aligned limit} \label{sec:spin_aligned}

In the spin-aligned configuration, 
$\vb{L}, \vb{S}_1$ and $\vb{S}_2$ become parallel to each other and perpendicular to 
$\vb{R}, \vb{P}$, and the orbital plane, thus leading to vanishing of the angles $(\kappa_1, \kappa_2, \gamma)$ and $s_{0p}$.
The radial EOM Eq.~\eqref{eq:drdt_final} is 
regular in the spin-aligned limit. 
When we set $s_{0p}= 0$
in Eqs.~\eqref{eq:QKPt}
and~\eqref{eq:QKPr}, along with the 
quasi-Keplerian parameters 
($n, a_r, h_r, e_r, e_t, f_t, g_t$) in Eqs.~\eqref{eq:QK_parameters}, and $\bar{l^2}$ of Eq.~\eqref{eq:lbar_squared},
the resulting QKP solution for $r$
reduces directly to that derived 
in Ref.~\cite{Tessmer2010}.

The azimuthal sector requires more careful treatment, however, since the noninertial phase
$\phi(t)$ of Eq.~\eqref{eq:QKPphi} (defined in a noninertial frame) cannot be directly
compared with the inertial phase parameters ($k, e_\phi, f_\phi, g_\phi$) established in the
aligned-spin literature \cite{Tessmer2010}. We must therefore first project our orbital phase
$\phi(t)$ onto the inertial frame.

Our parametrization in Eq.~\eqref{eq:rvecNIF} tracks the binary's position in the NIF, where
the orbit is instantaneously confined to the $ij$-plane,
\begin{equation}
	\mathbf{r} = r\begin{bmatrix}\cos \phi \\ \sin \phi \\ 0\end{bmatrix}_{\text{NIF}} \,.
\end{equation}
To obtain the position in the fixed inertial frame (IF) used in the aligned-spin literature,
we apply the inverse Euler rotation $\Lambda_{\text{NIF}\to\text{IF}}
\equiv \ty{Transpose}(\Lambda_{\text{IF}\to\text{NIF}})$ to $\mathbf{r}$ in the NIF.
To avoid ill-defined quantities\footnote{For example, $\phi_L$ becomes undefined in the
aligned-spin configuration.}, rather than setting $\theta_L = 0$ directly in the spin-aligned
limit, we instead apply the $\theta_L \to 0$ limit to $\Lambda_{\text{NIF}\to\text{IF}}$. This
yields $\mathbf{r}$ in the IF as
\begin{equation}    \label{eq:long_peri_spin_alinged}
	\mathbf{r} = r\begin{bmatrix} \cos\phi_L \cos\phi - \sin\phi_L \left(1 - \mathcal{O}(\theta_L^2)\right) \sin\phi \\ \sin\phi_L \cos\phi + \cos\phi_L \left(1 - \mathcal{O}(\theta_L^2)\right) \sin\phi \\ \mathcal{O}(\theta_L) \end{bmatrix}_{\text{IF}} 
	= r\begin{bmatrix} \cos(\phi + \phi_L) \\ \sin(\phi + \phi_L) \\ 0 \end{bmatrix}_{\text{IF}} + \mathcal{O}(\theta_L) \,,
\end{equation}
which shows that the azimuthal phase of $\mathbf{r}$ in the IF, in the spin-aligned limit, is
$\Phi_{\text{spin-aligned}} \equiv \phi + \phi_L$, the combination identified in
Sec.~\ref{sec:four_frequencies} as the generalization of the sum of the longitude of periapsis
and the true anomaly.

To recover the aligned-spin azimuthal quasi-Keplerian parameters, we apply this limit directly
to the solutions for $\phi_L$ and $\phi$ given in Eq.~\eqref{eq:phiL_solution} (adapted to the
hybrid description) and Eq.~\eqref{eq:QKPphi}, respectively. We first note that, when the
inertial phase $\Phi_{\text{spin-aligned}} = \phi + \phi_L$ is assembled, the terms
proportional to $\beta_{2L}$ cancel identically. Moreover, since the aligned-spin limit also
implies $\kappa_1, s_{0p} \to 0$, the incomplete elliptic integrals multiplying $\beta_{1L}$
reduce to $\Upsilon^{(2,\mathrm{H})}$. Using the spin-aligned approximation
$(2\beta_{1L})/(\alpha_{1L}+x_-) \sim (3/2)(1-\lambda)(l+s_0)$, the phase then simplifies to
\begin{align}
	\Phi_{\text{spin-aligned}}(v_\phi,v_\theta) &= (1+k'') v_\phi + \frac{1}{c^4} \left[f_\phi \sin 2 v_\phi + g_\phi \sin 3 v_\phi\right] + \frac{3(1-\lambda)(l+s_0) + \nu(l+s_1+s_2)}{\sqrt{A(x_3-x_-)}}\Upsilon^{(2,\mathrm{H})}(v_\theta)\,,
\end{align}
where $\Upsilon^{(2,\mathrm{H})}(v_\theta)$ is defined in Eq.~\eqref{eq:kappa1_solution_hyb},
and $k''$ is $k'$ with its elliptic terms removed,
\begin{align}
    k'' & \equiv k' - \frac{1}{n c^2 d^3} \sum_{i=1}^{2} \frac{\beta_{iL}}{\alpha_{iL} + x_-} \frac{\Pi(n_i, \beta)}{K(\beta)} \,.
\end{align}

More generally the inertial phase $\Phi_{\text{spin-aligned}}(v_\phi,v_\theta)$ can be written as
\begin{align}   \label{eq:big_Phi_a}
\Phi_{\text{spin-aligned}}(v_\phi,v_\theta)  = A_0 v_\phi + B_0 v_\theta + B_1\sin v_\theta + A_2\sin(2v_\phi) + A_3\sin(3v_\phi) \,,
\end{align}
where the coefficients $A_i$ and $B_i$ depend on the parameters of the system.

Our next step is to express $\Phi_{\text{spin-aligned}}$ in terms of a single true anomaly,
$v_\phi$, thereby eliminating the dependence on $v_\theta$. To this end, we define the
eccentricity ($e_\theta, e_\phi$, or otherwise) associated with a given true anomaly ($v_\theta,
v_\phi$, etc.) to be functionally related to that true anomaly in the same way $v_\theta$ is
related to $e_\theta$ through Eqs.~\eqref{v_r} and~\eqref{beta-r}. With this convention, for any
two true anomalies $v_{\text{old}}$ and $v_{\text{new}}$ whose corresponding eccentricities
agree at leading (Newtonian) order and are related by
\begin{align}
e_{\text{new}} = e_{\text{old}}\left(1+\frac{N_2}{c^2}+\frac{N_3}{c^3}+\frac{N_4}{c^4}\right) \, ,
\end{align}
one can use Eqs.~\eqref{v_r} and~\eqref{beta-r} to show that the perturbative relation between
the two true anomalies, up to 2PN order, is (see also Ref.~\cite{Tessmer2010})
\begin{align}\label{eq:anomaly_shift}
v_{\text{old}} &= v_{\text{new}} -\frac{e_{\text{old}}}{1-e_{\text{old}}^2}\left(\frac{N_2}{c^2}+\frac{N_3}{c^3} \right)\sin v_{\text{new}}  -\frac{1}{c^4}\left[\left(N_4+\frac{N_2^2 e_{\text{old}}^2}{1-e_{\text{old}}^2}\right)\frac{e_{\text{old}} \sin v_{\text{new}}}{1-e_{\text{old}}^2} - \frac{N_2^2 e_{\text{old}}^2}{4(1-e_{\text{old}}^2)^2} \sin(2 v_{\text{new}}) \right] \,.
\end{align}
Using the above expression, we can
write
$\Phi_{\text{spin-aligned}}$ 
of Eq.~\eqref{eq:big_Phi_a} in terms of only 
one true anomaly $v_{\phi}$, and not
$v_{\theta}$, although we do not present it explicitly
here.

The standard quasi-Keplerian style of presentation
has been to write the orbital phase in terms of the true anomaly and its harmonics, with the first harmonic set to zero. To conform to this,
 we introduce a final true anomaly
$v_\parallel$, with a corresponding eccentricity $e_\parallel$, obtained by applying the
true-anomaly shift transformation
of Eq.~\eqref{eq:anomaly_shift}
once more, from $v_\phi$ to $v_\parallel$. The eccentricity
$e_\parallel$ is fixed by requiring that the coefficient of $\sin v_\parallel$ in
$\Phi_{\text{spin-aligned}}(v_\parallel)$ vanishes. Once this first harmonic is
suppressed and all remaining contributions are collected, the phase reduces to the standard
nonprecessing form
\begin{align}
\Phi_{\text{spin-aligned}}(v_\parallel)= (1+k_\parallel)v_\parallel + f_\parallel \sin(2v_\parallel) + g_\parallel \sin(3v_\parallel) \,  ,    \label{eq:phi_aligned}
\end{align}
with coefficients
\begin{align}    
k_\parallel &= \frac{1}{l^2 c^2} \left[ 3 - \frac{2\lseff}{l^2} + \frac{3 s_0^2}{2l^2} \right] + \frac{1}{4 l^4 c^4} \left[ 105-30\nu+2(15-6\nu)h l^2 \right] \,\label{eq:ktheta}, \\[6pt]
f_\parallel &= \frac{1}{c^4} \frac{\nu(1-3\nu)}{8l^4} \left(1+2 h l^2\right) \,, \\[6pt]
g_\parallel &= -\frac{1}{c^4} \frac{3\nu^2}{32l^4} \left(1+2 h l^2\right)^{3/2} \,, \\[6pt]
e_\parallel^2 &= 1 + 2hl^2 - \frac{h}{c^2} \left[ 12 + (15-\nu)hl^2 - \frac{8(1+hl^2)\lseff}{l^2} + \frac{2(3+4hl^2)s_0^2}{l^2} \right] \nn\\ &\quad - \frac{h}{8l^2c^4} \Big[ 408-232\nu-15\nu^2 +4(16-88\nu-9\nu^2)hl^2 - 4(160-30\nu+3\nu^2)h^2l^4 \Big] \,.
\end{align}   \label{eq:QKP_limit_spin}
These expressions reproduce, up to 2PN order for a BBH system, the aligned-spin parameters derived in
Ref.~\cite{Tessmer2010} (Eqs.~(68)--(81)), confirming that the processes of integrating the spin-precessing EOMs and taking the spin-aligned limit commute.

We also correctly recover the two degrees of freedom expected in the nonspinning limit:
although the nutation frequency of Eq.~\eqref{eq:freq_nut} remains nonzero for spin-aligned
configurations, the corresponding nutation amplitude of Eq.~\eqref{app:nutamp} vanishes since $\sin\kappa_a = 0$ whenever $\mathbf{s}_a$ is parallel to $\mathbf{l}$. Any
slight misalignment perturbs this configuration and immediately excites the otherwise latent nutation
frequency. Furthermore, as shown above, in the spin-aligned limit the inertial azimuthal phase of 
$\vb{r}$
degenerates to $\phi+\phi_L$, giving the azimuthal frequency
$\omega_\phi^{\mathrm{inertial}} \equiv \omega_\phi^{\mathrm{orb}} + \omega_{\text{prec}}$.
Since the radial frequency is unaffected in this limit, the system ultimately exhibits exactly
two active frequencies, $\omega_r$ and $\omega_\phi^{\mathrm{inertial}}$, in the
spin-aligned limit.

\subsection{Overlap with other works}

If we start out with a Hamiltonian whose PN orbit terms have been set to zero ($h_{1pn}=h_{2pn}=0$), and during the course of our calculation, if we set
 $\dot{\phi}_L = 0$ in Eq.~\eqref{eq:dphidtApp}, our solution almost reduces to the quasi-Keplerian parametrization 
of Ref.~\cite{Klein2010}\footnote{We are unsure about the validity and meaningfulness of assumptions like
$\dot{\phi}_L = 0$ to recover the results of Ref.~\cite{Klein2010}.}. 
While the frequency parameters
$n$ and $k$ agree, other
orbital elements exhibit differences, possibly due to 
differences in  coordinate choices (ADM versus harmonic) 
and the SSCs.

%--------------------------------
%--------------------------------	
\section{Concluding remarks}\label{sec:conclusion}
%--------------------------------
%--------------------------------

\subsection{Scope and novelty of the present work}

The primary objective of this work was to provide a 2PN analytical solution of the conservative 
dynamics of eccentric and spin-precessing  BBHs. 
Working within the ADM Hamiltonian framework with the NWP 
SSC, 
we have extended the 1.5PN solution of Ref.~\cite{Cho2019} to nearly 2PN order. 
Our main contributions can be summarized as follows

\begin{itemize}
    \item Hybrid 2PN spin solution and orbital solution. We have developed a hybrid analytical solution 
    for the three angular momenta evolution that incorporates nearly all conservative effects up to 2PN order, 
    including nonspinning corrections, 1.5PN spin-orbit coupling, and 2PN 
    long timescale spin-spin 
    interactions. By introducing the hybrid precession velocity $\boldsymbol{\Omega}^{(2,\mathrm{H})}$ 
    in Eq.~\eqref{eq:angular2H}, we capture both the 2PN accurate slow evolution and the 
    fast leading-order 1.5PN oscillations. The only missing ingredient are the subdominant 2PN 
    spin-spin induced fast orbital timescale oscillations. The spin dynamics are encapsulated in the final solution of 
    $\cos\kappa_1^{(2,\mathrm{H})}$ in Eq.~\eqref{eq:kappa1_solution_hyb}.

	 We also derived a quasi-Keplerian solution 
	for the relative separation vector $\mathbf{R}$, providing explicit solutions for the 
	radial motion $R(t)$ and the phase $\phi(t)$, given in Eq.~\eqref{eq:QK_parametrization}, 
	where the radial solution is fully 2PN accurate while the phase omits only the 
	2PN spin-spin oscillations arising from spin evolution. To ensure an observable solution, 
	we consistently tracked the rotation of the noninertial frame relative to the fixed inertial 
	frame via time-dependent Euler angles $\theta_L(t)$ and $\phi_L(t)$, keeping the total 
	azimuthal phase well-defined at all times. Interestingly at 2PN order, the nonspinning and spinning 
	sectors are coupled through products of 1PN terms, appearing in the azimuthal QKP 
	parameters in Eq.~\eqref{eq:QK_parameters}, where the mass term $\Delta$ interacts 
	with 1PN orbital contributions to produce novel 2PN cross-terms absent at lower orders. 
    We also sketch the straightforward way to construct the solution for $\vb{P}$, although we do not construct it explicitly.

\item Plots.
In Figs.~\ref{plot:S1x}
and~\ref{fig:r_vector_plot}, we provide plots
of our solution along with various other analytical solution and the full 2PN numerical one. This brings out in a visual manner the strengths and weaknesses of various analytical solutions against the 2PN numerical one. It is evident that our proposed 2PN solution performs the best among all other candidates.

\item Frequencies.  
In Sec.~\ref{sec:four_frequencies},
we give all four frequencies of the 2PN spinning BBH system. The frequencies are the by-products of our solution. These include 
two  frequencies capturing the angular momentum dynamics ($\omega_{\text{nut}}$ and $\omega_{\text{prec}}$),
the basic orbital frequency $n$ (mean motion), and a fourth frequency
$\omega_{\phi}^{\text{inertial}}$
that captures the generalization of periastron advance to spin-precessing systems.
In fact, this last frequency can be seen as a definitional way to quantify ``periastron advance'' for spinning BBHs.

\item  Limiting cases. In Sec.~\ref{sec:limits}, we verify that our solution agrees with the older results in the literature both in the circular and spin-aligned limits.

\item  Software implementation. For plotting purposes, the analytical solutions and the symbolic framework developed in this work are provided in a  public GitHub repository \cite{Tom_Colin_2026_BBH_Notebook}. To ensure broad accessibility, the results are hosted within the \texttt{2PN\_BBH\_solution} directory in multiple formats: the primary \textit{Mathematica} notebook (\texttt{.nb}), a Wolfram Language package (\texttt{.m}), a plain-text version (\texttt{.txt}) for algorithmic inspection, and a static PDF for read-only reference. These resources allow for the reproduction of many of our plots.

Interestingly, plotting the analytical and numerical solutions using this 
\textit{Mathematica} notebook makes it abundantly clear that the numerical solutions (obtained by the numerical integration of PN EOMs)
are slower than the analytical solutions by a vast order of magnitude, thus serving as additional motivation for our work.

\end{itemize}

We will close the subsection with a few remarks. 
Our solution 
sheds light on
how to solve a system which may not be exactly integrable, but rather only perturbatively so. One key requirement is that we need the perturbative COMs, just as we need exact COMs to solve an exactly integrable system exactly.
Another by-product of our solution has been the mathematical framework needed to compute the scaling of how the solutions under two different models of evolution (each with its own separate set of EOMs) diverge. This mathematical framework has been introduced and exploited in Appendix~\ref{App:errorbounds} to see how two 
spin solutions diverge.
There it is shown that our hybrid spin solutions diverge the least from the full 2PN model, among other models considered.

\subsection{Future perspectives}

An obvious further extension of our current work could be to make it completely 2PN accurate, although we think that such
 modifications will not be of much consequence as they are subdominant with respect to the other 2PN contributions that we have considered and addressed in this article.

Our solution is  conservative, describing the bound motion of the binary system 
under the 2PN Hamiltonian dynamics. 
Another natural and necessary extension of this framework is the inclusion of dissipative radiation-reaction effects, which first enter at 2.5PN order. These effects are responsible for the loss of energy and angular momentum from the binary system, driving the secular shrinking and circularization of the orbit. On the other hand, we work to improve the conservative solution to 2.5PN in order to incorporate the next-to-leading-order spin-orbit coupling. We expect that the inclusion of the next-to-leading-order spin-spin coupling will be considerably more challenging due to its increased complexity.

Future work can therefore focus on deriving the secular time evolution of the quasi-Keplerian parameters due to radiation reaction. 
Once the time dependence of the orbital elements is established, one can try to compute the  GW signal. 
The resulting waveforms will be crucial for accurate parameter estimation, as they will consistently capture both the amplitude 
modulations induced by spin precession and the frequency chirping driven by radiation reaction.

Taking a step back, we note that in this work we have tried to extend the 1.5PN conservative spinning BBH solution to 2PN.
Since the 1.5PN system is an integrable one with its action-angle variables and frequencies already computed~\cite{Tanay2020, Tanay2023}, there is hope to reconstruct the 2PN solution (starting from the 1.5PN baseline system) using spectral methods, 
taking inspiration from Refs.~\cite{Drummond:2022xej, Drummond:2022efc}, where bound orbits were computed for spinning EMRIs (extreme mass ratio inspirals).

In this artificial-intelligence-dominated age, PN work, which has traditionally been theoretical, can also benefit from embracing machine learning.
A prospect could be to apply a physics-informed machine learning approach to construct a ``nearly'' 2PN-accurate solution. The ``physics-informed'' part comes from the 
understanding of the action-angle variables and frequencies of the baseline 1.5PN integrable system. From the dynamical systems literature, we know that the most important aspect of the 2PN perturbation is the shift in the four frequencies. One can then try to train a physics-informed deep neural network to learn the four frequency shifts as a function of parameters of the baseline 1.5PN system and the 2PN perturbation.
The two nontraditional approaches to constructing the 2PN spinning BBH solution from the baseline 1.5PN solution advocated for in this and the above  paragraph could be useful when we try to go to 3PN order or higher where the traditional theoretical approaches might become too unwieldy to make any progress.

%--------------------------------
%--------------------------------
%--------------------------------
%--------------------------------
\begin{acknowledgments}
	We thank Guillaume Faye, 
	David Trestini, 
	Soham Bhattacharyya,
	Thibault Damour,	
	Quentin Henry, 
	and Leo Stein
	for useful discussions.
	L. B. acknowledges financial support from the ANR PRoGRAM project, Grant No. ANR-21-CE31-0003-001. L. B., T. C., and S. T. acknowledge financial support from the EU Horizon 2020 Research
	and Innovation Programme under the Marie Sklodowska-Curie Grant Agreement No. 101007855.
	S. T. was supported by the PSL postdoctoral fellowship during this work.
\end{acknowledgments}

\section*{Data Availability}

The Mathematica notebook and associated scripts that support the findings
of this article are openly available in the GitHub repository of
Ref.~\cite{Tom_Colin_2026_BBH_Notebook}. The notebook
(\texttt{2PN\_BBH\_solution.nb}) and its plain-text Mathematica package
version (\texttt{2PN\_BBH\_solution.m}) are also provided as ancillary
files with the arXiv version of this article.

%--------------------------------
%--------------------------------

%--------------------------------
%--------------------------------	
\appendix
%--------------------------------
%--------------------------------

%--------------------------------

%--------------------------------

\section{Roots of the cubic polynomial}
\label{app:rootsx}
%--------------------------------
We compute the roots of the cubic polynomial appearing in Eq.~\eqref{eq:dx_separated} following the approach of Appendix A.1 of Ref.~\cite{Tanay2023}. The polynomial reads
\begin{align}\label{eq:Px}
	P(x) = a_3 \,x^3 + a_2\, x^2 + a_1\, x + a_0 \, , 
\end{align}
with coefficients
\begin{subequations}
\begin{align}
	a_3&=2 \,m_2(m_1-m_2)\, l \,s_1\, ,\\
	a_2&=-\Big((m_1-m_2)^2\,l^2 + m_2^2\,s_1^2 + m_1^2\,s_2^2 + 2 m_1 m_2 \, s_1 \, s_2 \, \Sigma_1 + 2m_1(m_1-m_2) \, l \,s_2\, \Sigma_2 \Big) \, ,\\
	a_1&=2\,m_1 \,s_2\,\Big((m_1-m_2)\,l\,\Sigma_1+\big(m_2 \, s_1 +m_1\,s_2\,\Sigma_1 \big)\,\Sigma_2 \Big)\, ,\\
	a_0&=m_1^2 \,s_2^2\,\big(1-\Sigma_1^2 -\Sigma_2^2 \big) \, .
\end{align}
\end{subequations}
Note that $a_3=A$ of Eq.~\eqref{eq:A_constant}, and is positive since $m_1 > m_2$ by convention. The polynomial factorizes as
\begin{align}
	P(x)=a_3(x-x_-)(x-x_+)(x-x_3) \,.
\end{align} 

To obtain the roots analytically, we perform the standard depressed cubic transformation via the change of variables
\begin{align}
	y\equiv x +\frac{a_2}{3a_3}\, ,
\end{align}
which eliminates the quadratic term and yields the reduced polynomial
\begin{align}
	P(x)\equiv Q(y)=a_3\big(y^3+p y +q \big)\,,
\end{align}
where the reduced coefficients are
\begin{subequations}
\begin{align}
	p&=\frac{3 a_1 a_3-a_2^2}{3 a_3^2}\, , \\
	q&=\frac{2 a_2^3-9 a_1 a_2 a_3 + 27 a_0 a_3^2}{27 a_3^3} \, .
\end{align}
\end{subequations}
The discriminant of the depressed cubic is $\Delta = -4p^3 - 27q^2$. When $\Delta > 0$, all three roots are real and distinct, which corresponds to $p < 0$. In this regime, the roots can be expressed using the trigonometric expression
\begin{align}
x_k = -\frac{a_2}{3 a_3} + 2\sqrt{\frac{-p}{3}}\cos \left[\frac{1}{3} \arccos{\left(\frac{3q}{2p}\sqrt{\frac{-3}{p}} \right)} + \delta_k \right],
\end{align}
for $k \in \{ -, +,3\}$, with the corresponding phase shifts $\delta_3 = 0$, $\delta_- = 2\pi/3$, and $\delta_+ = 4\pi/3$. This expression is valid when the argument of the arccosine lies in $[-1,1]$, which is guaranteed by the condition $\Delta > 0$. Modifying the convention of Ref.~\cite{Cho2019}, these solutions naturally yield the ordering $x_- \leq x_+ < x_3$.

In the PN regime, where $s_a/l \sim \mathcal{O}(c^{-1})$, the three roots exhibit distinct behaviors. The largest root scales as
\begin{align}
    x_3 \sim \frac{m_1 - m_2}{6\, m_2}\frac{l}{s_1} \sim \mathcal{O}(c),
\end{align}
placing it outside the physical interval $[-1,1]$. The other two roots remain finite and nearly degenerate. They scale as
\begin{align}
x_\pm = \cos \kappa_1 + \frac{m_1}{m_1-m_2}\frac{s_2}{l}(\cos\gamma -\cos\kappa_1\cos \kappa_2 \pm |\sin \kappa_1 \sin \kappa_2 |) + \mathcal{O}(c^{-2}),
\end{align}
and both are physically relevant. From
Eq.~\eqref{eq:kappa1_solution},
the spin nutation amplitude then becomes
\begin{align}\label{app:nutamp}
x_+ - x_- &\sim \frac{2\,m_1}{m_1-m_2}\frac{s_2}{l} |\sin \kappa_1 \sin \kappa_2 | \sim \mathcal{O}(c^{-1}),
\end{align}
and $\cos \kappa_1$ oscillates between these two values over each nutation. We note that if either spin is aligned or antialigned with the orbital angular momentum, the nutation amplitude vanishes, while the nutation frequency defined in Eq.~\eqref{eq:freq_nut} remains nonzero. This implies that for configurations slightly deviating from the spin-aligned limit, the system will nutate at that frequency.
%--------------------------------
\section{Azimuthal equation of motion of the orbital angular momentum}
\label{app:AzeomL}
%--------------------------------

We project the definition of the total angular momentum $\mathbf{j}=\mathbf{l}+\mathbf{s}_1+\mathbf{s}_2$ onto the three orthogonal directions of the NIF. Using the vector decomposition given in Eq.~\eqref{eq:vecNIF}, we obtain
\begin{subequations}\label{eq:proj_tot}
\begin{align}
	0 &= s_1 \sin \kappa_1 \cos \phi_{s_1} + s_2  \sin \kappa_2 \cos \phi_{s_2} \, ,  \label{eq:proj_tot1}\\
	j \sin \theta_L &= s_1 \sin \kappa_1 \sin \phi_{s_1} + s_2 \sin \kappa_2 \sin \phi_{s_2}\, ,  \label{eq:proj_tot2}\\
	j \cos \theta_L &= l + s_1 \cos \kappa_1 + s_2 \cos \kappa_2\, . \label{eq:proj_tot3}
\end{align}
\end{subequations}
Substituting Eq.~\eqref{eq:Sigmas} into Eq.~\eqref{eq:proj_tot3}, we can express $\cos \theta_L$ as a function of $\cos \kappa_1$
\begin{align}
	\cos \theta_L = \frac{l+s_2 \Sigma_2}{j} + \frac{m_1-m_2}{m_1}\frac{s_1}{j}\cos \kappa_1 \, .
	\label{eq:costheta_L-b}
\end{align}
Similarly, from Eqs.~\eqref{eq:proj_tot1} and~\eqref{eq:proj_tot2}, we derive the azimuthal angle $\phi_{s_1}$ in terms of the other geometric variables
\begin{subequations}
\label{eq:phi_s1}
\begin{align}
	\sin \phi_{s_1} &= \frac{s_1 \sin \kappa_1 + s_2 \sin \kappa_2 \cos \Delta \phi_s}{j \sin \theta_L} \, , \label{eq:sin_phi_s1}\\
	\cos \phi_{s_1} &= \frac{s_2 \sin \kappa_2 \sin \Delta \phi_s}{j \sin \theta_L} \, 
	\label{eq:cos_phi_s1},
\end{align}
\end{subequations}
where $\Delta \phi_s \equiv \phi_{s_1}-\phi_{s_2}$ denotes the difference between the azimuthal angles of the two spin vectors projected onto the orbital plane, with $\phi_{s_a}$ defined in Eq.~\eqref{eq:vecNIF}. We obtained $\Delta \phi_s$ through the spherical law of cosines applied to the angular momentum triangle
\begin{align}
	\cos \Delta \phi_s = \frac{\cos \gamma - \cos \kappa_1 \cos \kappa_2}{\sin \kappa_1 \sin \kappa_2}\,.
	\label{eq:cos_Deltaphi}
\end{align} 
These are all the relationships we need to obtain Eq.~\eqref{eq:dthetaL_dt}.

\section{Error bounds between various models of angular momenta evolution}
\label{App:errorbounds}

In this paper, we have discussed various dynamical models of angular momenta evolution (1.5PN (1.5), 2PN (2), 2PN-averaged (2, A), 2PN-hybrid (2, H)). To compactly represent the state of the system, we define the collective phase-space vector $\mathbf{V}^X \equiv (\mathbf{s}_1^X, \mathbf{s}_2^X, \mathbf{l}^X, \mathbf{r}^X)$ and $\mathbf{U}^X \equiv (\mathbf{s}_1^X, \mathbf{s}_2^X, \mathbf{l}^X)$, with
the superscript $X$
denoting the model under
consideration:
$X \in $  $\{$(1.5),  (2), (2, A),  (2, H)$\}$.

Schematically, the EOMs for the angular momenta under a certain model $X$ can be represented as ($A \in \{ 1,2\}$ for the two BHs)
\begin{align} \label{eq:psi_spin_eom}
\frac{d \mathbf{s}_A^X}{dt} = \boldsymbol{\Psi}_{\mathbf{s}_A}^X(\mathbf{V}^X) , \quad \frac{d \mathbf{l}^X}{dt} = \boldsymbol{\Psi}_{\mathbf{l}}^X(\mathbf{V}^X) .
\end{align}
This notation is to be carefully interpreted. $\boldsymbol{\Psi}_{\mathbf{s}_A}^X$ and $\boldsymbol{\Psi}_{\mathbf{l}}^X$ are vector functions of the phase space variables. Performing the label change $X \rightarrow Y$ on the $\boldsymbol{\Psi}$ function changes the function (e.g. switching from Eqs.~\eqref{eq:spin_precession} to Eqs.~\eqref{eq:OA-1PN-b}). Changing the label $X \rightarrow Y$ on the state vector $\mathbf{V}$ selects the phase space variable solution under a different evolution scheme. It is understood that the initial values under all models are identical; i.e., $\mathbf{V}^X(t= 0) = \mathbf{V}^Y(t = 0) \equiv \mathbf{V}_0$.

For $\mathbf{s}_A$, under two different evolution models $X$ and $Y$, we define $\delta \boldsymbol{\Psi}_{\mathbf{s}_A} \equiv \boldsymbol{\Psi}_{\mathbf{s}_A}^Y - \boldsymbol{\Psi}_{\mathbf{s}_A}^X$, yielding the relations
\begin{subequations}  \label{eq:evolution_breakdown}
\begin{align}       
\frac{d \mathbf{s}_A^X}{dt} &= \boldsymbol{\Psi}_{\mathbf{s}_A}^X (\mathbf{V}^X) , \\
\frac{d \mathbf{s}_A^Y}{dt} &= \boldsymbol{\Psi}_{\mathbf{s}_A}^Y (\mathbf{V}^Y) \equiv \boldsymbol{\Psi}_{\mathbf{s}_A}^X (\mathbf{V}^Y) + \delta \boldsymbol{\Psi}_{\mathbf{s}_A} (\mathbf{V}^Y) . \label{eq:break_Psi}
\end{align}
\end{subequations}

From the EOMs, the evolution of the difference $\delta \mathbf{s}_A \equiv \mathbf{s}_A^Y - \mathbf{s}_A^X$ is governed by the differential equation
\begin{align} \label{eq:exact_diff_eq}
\frac{d\,\delta \mathbf{s}_A}{dt} &= \boldsymbol{\Psi}_{\mathbf{s}_A}^Y(\mathbf{V}^Y) - \boldsymbol{\Psi}_{\mathbf{s}_A}^X(\mathbf{V}^X) \nonumber \\
&= \left[ \boldsymbol{\Psi}_{\mathbf{s}_A}^Y(\mathbf{V}^Y) - \boldsymbol{\Psi}_{\mathbf{s}_A}^X(\mathbf{V}^Y) \right] + \left[ \boldsymbol{\Psi}_{\mathbf{s}_A}^X(\mathbf{V}^Y) - \boldsymbol{\Psi}_{\mathbf{s}_A}^X(\mathbf{V}^X) \right] \nonumber \\
&= \delta \boldsymbol{\Psi}_{\mathbf{s}_A}(\mathbf{V}^Y) + \left[ \boldsymbol{\Psi}_{\mathbf{s}_A}^X(\mathbf{V}^Y) - \boldsymbol{\Psi}_{\mathbf{s}_A}^X(\mathbf{V}^X) \right] .
\end{align}

Because $\mathbf{V}^X(t= 0) = \mathbf{V}^Y(t = 0) \equiv \mathbf{V}_0$,
and under the assumption that the evolution of $\mathbf{s}_A$ is continuous in time for all models, there exists a $t_{\delta} > 0$ (to be determined later) such that for all $t < t_{\delta}$, the phase space divergence remains small enough to satisfy
\begin{subequations} \label{eq:small_time_system}
\begin{align}
    \lVert \boldsymbol{\Psi}_{\mathbf{s}_A}^X (\mathbf{V}^Y(t)) - \boldsymbol{\Psi}_{\mathbf{s}_A}^X (\mathbf{V}^X(t)) \rVert &\lesssim \lVert \delta \boldsymbol{\Psi}_{\mathbf{s}_A} (\mathbf{V}^Y(t)) \rVert , \label{eq:small_time_a}  \\
    \left\lVert \frac{\partial \boldsymbol{\Psi}_{\mathbf{s}_A}^X(\mathbf{V}^X)}{\partial \mathbf{s}_B^X} \delta \mathbf{s}_B + \frac{\partial \boldsymbol{\Psi}_{\mathbf{s}_A}^X(\mathbf{V}^X)}{\partial \mathbf{l}^X} \delta \mathbf{l} + \frac{\partial \boldsymbol{\Psi}_{\mathbf{s}_A}^X(\mathbf{V}^X)}{\partial \mathbf{r}^X} \delta \mathbf{r} \right\rVert &\lesssim \lVert \delta \boldsymbol{\Psi}_{\mathbf{s}_A} (\mathbf{V}^Y) \rVert  , \label{eq:small_time_b}
\end{align}
\end{subequations}
where $(\delta \mathbf{s}_B, \delta \mathbf{l} , \delta \mathbf{r} ) \equiv (\mathbf{s}_B^Y - \mathbf{s}_B^X, \mathbf{l}^Y - \mathbf{l}^X , \mathbf{r}^Y - \mathbf{r}^X )$, and a Taylor series expansion has been employed around $\mathbf{V}^X$. Here, we adopt the notation where $({\partial \boldsymbol{\Psi}}/{\partial \mathbf{l}}) \delta \mathbf{l}$ denotes a vector with $i$th component $({\partial \Psi^i}/{\partial l^j}) \delta l^j$\footnote{We adopt the standard Einstein summation convention for repeated indices. We also implicitly extend this summation convention to the spin label index $B \in \{1, 2\}$, eliminating the explicit sum symbol.}.
For $t < t_{\delta}$,
due to Eq.~\eqref{eq:small_time_a}, Eq.~\eqref{eq:exact_diff_eq} reduces to
\begin{align}     \label{eq:div_eqn}
\frac{d\,\delta \mathbf{s}_A}{dt}  \sim \delta \boldsymbol{\Psi}_{\mathbf{s}_A}(\mathbf{V}^Y) .
\end{align}

Below we study the evolution of $\delta \mathbf{s}_A$ for $t \lesssim t_{\delta}$ and determine this bound $t_{\delta}$. 

\begin{enumerate}

\item 1.5PN vs 2PN. With $(X, Y) = ((1.5), (2))$, we define the 1.5PN and  full 2PN $\boldsymbol{\Psi}_{\mathbf{s}_A}$ from Eqs.~\eqref{eq:spin_precession} 
\begin{align}
\boldsymbol{\Psi}_{\mathbf{s}_A}^X &= \frac{1}{c^2 r^3} \delta_A \mathbf{l} \times \mathbf{s}_A , \label{eq:psi_X_15} \\
\boldsymbol{\Psi}_{\mathbf{s}_A}^Y &= \frac{1}{c^2 r^3} \left\{ \delta_A \mathbf{l} + \nu_A \left[ 3(\mathbf{n} \cdot \mathbf{s}_0)\mathbf{n} - \mathbf{s}_0 \right] \right\} \times \mathbf{s}_A , \label{eq:psi_Y_2PN}
\end{align}
which yields the difference
\begin{align} \label{eq:delta_psi_15_2PN}
\delta \boldsymbol{\Psi}_{\mathbf{s}_A} &= \frac{1}{c^2 r^3} \nu_A \left[ 3(\mathbf{n} \cdot \mathbf{s}_0)\mathbf{n} - \mathbf{s}_0 \right] \times \mathbf{s}_A .
\end{align}
This, along with Eq.~\eqref{eq:div_eqn}, 
the assumption that $\delta \boldsymbol{\Psi}_{\mathbf{s}_A}$ has a nonzero time average, and recalling that the spin vectors themselves scale as $s \sim c^{-1}$, gives for $t \lesssim t_{\delta}$\footnote{For a 3D-Cartesian vector $\mathbf{A}$, $\mathbf{A} \sim \epsilon$ means that each component $A^i \sim \epsilon$, where $i = 1, 2, 3$.}
\begin{align} \label{eq:div_scaling_1a}
\delta \mathbf{s}_A \sim \int \delta \boldsymbol{\Psi}_{\mathbf{s}_A} dt \sim \frac{t}{c^4} .
\end{align}
The results for the other deviations are obtained in a similar manner, yielding
\begin{align} \label{eq:div_scaling_1b}
\delta \mathbf{l} \sim \frac{t}{c^4} , \quad \delta \mathbf{r} \sim \frac{t}{c^4} .
\end{align}

To determine the scaling of $t_{\delta}$, we substitute the scaling of the deviations into Eq.~\eqref{eq:small_time_b} evaluated at $t = t_{\delta}$. From Eq.~\eqref{eq:psi_X_15}, the Jacobian components scale as
\begin{align} \label{eq:jacobian_scaling_1}
\frac{\partial \boldsymbol{\Psi}_{\mathbf{s}_A}^X (\mathbf{V}^X)}{\partial \mathbf{s}_A^X} \sim c^{-2}, \quad \frac{\partial \boldsymbol{\Psi}_{\mathbf{s}_A}^X (\mathbf{V}^X)}{\partial \mathbf{l}^X} \sim c^{-3}, \quad \frac{\partial \boldsymbol{\Psi}_{\mathbf{s}_A}^X (\mathbf{V}^X)}{\partial \mathbf{r}^X} \sim c^{-3} .
\end{align}
By plugging Eqs.~\eqref{eq:div_scaling_1a}, \eqref{eq:div_scaling_1b}, \eqref{eq:jacobian_scaling_1}, 
and
$\delta \boldsymbol{\Psi}_{\mathbf{s}_A} \sim 1/c^4$
(easy to check)
into  Eq.~\eqref{eq:small_time_b}, we obtain the balance condition
\begin{align}
\frac{1}{c^2} \frac{t_{\delta}}{c^4} + \frac{1}{c^3} \frac{t_{\delta}}{c^4} + \frac{1}{c^3} \frac{t_{\delta}}{c^4} \sim \frac{1}{c^4} \implies t_{\delta} \sim c^2 ,
\end{align}
meaning $\mathbf{s}_A$ solutions from 1.5PN and 2PN models diverge as $t/c^4$ within $t \lesssim c^2$.

\item 2PN vs 2PN averaged. Here $(X, Y) = ((2,\mathrm{A}), (2))$, and $\boldsymbol{\Psi}_{\mathbf{s}_A}^Y$ is the 2PN function given in Eq.~\eqref{eq:psi_Y_2PN}. The (2, A) model does the orbit-averaging (see Eqs.~\eqref{eq:OA-1PN-b})
\begin{align} \label{eq:Psi_2PN_2PN_X}
\boldsymbol{\Psi}_{\mathbf{s}_A}^X(\mathbf{U}) &= \frac{1}{c^2 d^3} \left\{ \delta_A \mathbf{l} + \frac{\nu_A}{2} \left[ \mathbf{s}_0 - 3 \frac{(\mathbf{l} \cdot \mathbf{s}_0)}{l^2} \mathbf{l} \right] \right\} \times \mathbf{s}_A ,
\end{align}
and the difference is defined as $\delta \boldsymbol{\Psi}_{\mathbf{s}_A} \equiv \boldsymbol{\Psi}_{\mathbf{s}_A}^Y - \boldsymbol{\Psi}_{\mathbf{s}_A}^X $. 
Expanding Eq.~\eqref{eq:div_eqn} via Taylor series about the initial time yields
\begin{align} \label{eq:evol_eq_div_2PN_2PN}
\frac{d\,\delta \mathbf{s}_A}{dt} &\sim \delta \boldsymbol{\Psi}_{\mathbf{s}_A}(\mathbf{V}^Y) = \boldsymbol{\Psi}_{\mathbf{s}_A}^Y(\mathbf{U}^Y, \mathbf{r}^Y) - \boldsymbol{\Psi}_{\mathbf{s}_A}^X(\mathbf{U}^Y) \nonumber \\
&\approx \left[ \boldsymbol{\Psi}_{\mathbf{s}_A}^Y(\mathbf{U}_0, \mathbf{r}^Y(t)) - \boldsymbol{\Psi}_{\mathbf{s}_A}^X(\mathbf{U}_0) \right] + \left( \left. \frac{\partial \boldsymbol{\Psi}_{\mathbf{s}_A}^Y}{\partial \mathbf{l}} \right|_{\mathbf{U}_0, \mathbf{r}^Y} - \left. \frac{\partial \boldsymbol{\Psi}_{\mathbf{s}_A}^X}{\partial \mathbf{l}} \right|_{\mathbf{U}_0} \right) \Delta \mathbf{l}^Y + \left( \left. \frac{\partial \boldsymbol{\Psi}_{\mathbf{s}_A}^Y}{\partial \mathbf{s}_B} \right|_{\mathbf{U}_0, \mathbf{r}^Y} - \left. \frac{\partial \boldsymbol{\Psi}_{\mathbf{s}_A}^X}{\partial \mathbf{s}_B} \right|_{\mathbf{U}_0} \right) \Delta \mathbf{s}_B^Y,
\end{align}
where  the deviations are given by $\Delta \mathbf{s}_B^Y \equiv \mathbf{s}_B^Y(t) - \mathbf{s}_B^Y(t=0)$ and $\Delta \mathbf{l}^Y \equiv \mathbf{l}^Y(t) - \mathbf{l}^Y(t=0)$. 
From Eqs.~\eqref{eq:psi_Y_2PN} and \eqref{eq:Psi_2PN_2PN_X}, it is easy to see that the scaling of the partial derivatives (the Jacobian components) for both models is given by
\begin{subequations} \label{eq:2PN-2PN_scaling_a}
\begin{align}
\frac{\partial \boldsymbol{\Psi}_{\mathbf{s}_A}^Y}{\partial \mathbf{l}} \sim \frac{\partial \boldsymbol{\Psi}_{\mathbf{s}_A}^X}{\partial \mathbf{l}} &\sim c^{-3} , \\
\frac{\partial \boldsymbol{\Psi}_{\mathbf{s}_A}^Y}{\partial \mathbf{s}_B} \sim \frac{\partial \boldsymbol{\Psi}_{\mathbf{s}_A}^X}{\partial \mathbf{s}_B} &\sim c^{-2} ,
\end{align}
\end{subequations}
whereas from Eqs.~\eqref{eq:spin_precession}, the deviations in the angular momenta with respect to their initial values scale as
\begin{align} \label{eq:2PN-2PN_scaling_b}
 \Delta \mathbf{l}^{Y} \sim \Delta \mathbf{s}_A^{Y} \sim \frac{t}{c^3} .
\end{align}

Integrating Eq.~\eqref{eq:evol_eq_div_2PN_2PN} with respect to time from $0$ to $t = N \tau_{\text{orb}} + \delta t$, where $N = \lfloor t / \tau_{\text{orb}} \rfloor$ and $\delta t < \tau_{\text{orb}}$, we find the scaling of the leading-order difference (in the square brackets) and the linearization terms. Focusing first on the square brackets, integrating over the time variable $\tau$ yields
\begin{align} \label{eq:integral_decomposition}
\int_0^t \left[ \boldsymbol{\Psi}_{\mathbf{s}_A}^Y(\tau) - \boldsymbol{\Psi}_{\mathbf{s}_A}^X \right] d\tau &= \sum_{k=0}^{N-1} \int_{k \tau_{\text{orb}}}^{(k+1)\tau_{\text{orb}}} \left[ \boldsymbol{\Psi}_{\mathbf{s}_A}^Y(\tau) - \boldsymbol{\Psi}_{\mathbf{s}_A}^X \right] d\tau + \int_{N \tau_{\text{orb}}}^{t} \left[ \boldsymbol{\Psi}_{\mathbf{s}_A}^Y(\tau) - \boldsymbol{\Psi}_{\mathbf{s}_A}^X \right] d\tau \,.
\end{align}
Because $\boldsymbol{\Psi}_{\mathbf{s}_A}^X = \boldsymbol{\Psi}_{\mathbf{s}_A}^{(2\mathrm{A})}$ is the orbit average of $\boldsymbol{\Psi}_{\mathbf{s}_A}^Y = \boldsymbol{\Psi}_{\mathbf{s}_A}^{(2)}$ evaluated at fixed angular momenta, the integral over any complete orbital period vanishes, leaving the fractional interval $\delta t$ to contribute a scaling of $\mathcal{O}(\delta t / c^3)$. The scaling of the linearization terms in the parentheses of Eq.~\eqref{eq:evol_eq_div_2PN_2PN} via Eqs.~\eqref{eq:2PN-2PN_scaling_a} and \eqref{eq:2PN-2PN_scaling_b} yields $\mathcal{O}(t/c^5)$, thereby making its time integral to scale as $\mathcal{O}(t^2 / c^5)$. Combining these contributions gives the overall scaling for the spin deviation,
\begin{align} \label{eq:s_a_deviation_2PN_2PN}
\delta \mathbf{s}_A \sim \frac{\delta t}{c^3} + \frac{t^2}{c^5} \,.
\end{align}
An analogous derivation provides the same scaling for the orbital angular momentum,
\begin{align} \label{eq:l_deviation_2PN_2PN}
\delta \mathbf{l} \sim \frac{\delta t}{c^3} + \frac{t^2}{c^5} \,.
\end{align}

Just as before, to determine the scaling of $t_{\delta}$, we substitute the relations of Eqs.~\eqref{eq:2PN-2PN_scaling_a},
\eqref{eq:s_a_deviation_2PN_2PN} and, \eqref{eq:l_deviation_2PN_2PN} into Eq.~\eqref{eq:small_time_b} evaluated at $t = t_{\delta}$, while noting that the last term on the lhs of Eq.~\eqref{eq:small_time_b} vanishes since $\partial \boldsymbol{\Psi}_{\mathbf{s}_A}^X / \partial \mathbf{r} = 0$.
With all this, Eq.~\eqref{eq:small_time_b} when stripped down to scaling, becomes
\begin{align} \label{eq:scaling_balance}
\frac{1}{c^2} \left( \frac{\delta t}{c^3} + \frac{t_{\delta}^2}{c^5} \right) + \frac{1}{c^3} \left( \frac{\delta t}{c^3} + \frac{t_{\delta}^2}{c^5} \right) \sim \frac{1}{c^3} \implies t_{\delta} \sim c^2 ,
\end{align} 
which means that the $\mathbf{s}_A$ solutions for evolution under full 2PN dynamics vs 2PN averaged dynamics diverge as $(\delta t/c^3 + t^2/c^5)$ for $t < t_{\delta}$, where $t_{\delta} \sim c^2$.

\item 2PN vs 2PN hybrid.
Here $(X, Y) = ((2,\mathrm{H}), (2))$, and $\boldsymbol{\Psi}_{\mathbf{s}_A}^Y$ remains the 2PN function from Eq.~\eqref{eq:psi_Y_2PN}. We decompose $\boldsymbol{\Psi}_{\mathbf{s}_A}^Y$ into its spin-orbit ($\mathrm{SO}$) and spin-spin ($\mathrm{SS}$) components
\begin{subequations} \label{eq:decomp_2PN}
\begin{align}
\boldsymbol{\Psi}_{\mathbf{s}_A}^Y &= \boldsymbol{\Psi}_{\mathrm{SO}} + \boldsymbol{\Psi}_{\mathrm{SS}} , \\
\boldsymbol{\Psi}_{\mathrm{SO}} &= \frac{1}{c^2 r^3} \delta_A \mathbf{l} \times \mathbf{s}_A , \\
\boldsymbol{\Psi}_{\mathrm{SS}} &= \frac{1}{c^2 r^3} \nu_A \left[ 3(\mathbf{n} \cdot \mathbf{s}_0)\mathbf{n} - \mathbf{s}_0 \right] \times \mathbf{s}_A .
\end{align}
\end{subequations}
We further decompose the spin-spin interaction into its orbit average ($\bar{\boldsymbol{\Psi}}_{\mathrm{SS}}$) and an oscillatory remainder ($\widetilde{\boldsymbol{\Psi}}_{\mathrm{SS}}$);
see Eq.~\eqref{eq:OA-1PN-b}
\begin{subequations} \label{eq:decomp_SS}
\begin{align}
\boldsymbol{\Psi}_{\mathrm{SS}} &= \bar{\boldsymbol{\Psi}}_{\mathrm{SS}} + \widetilde{\boldsymbol{\Psi}}_{\mathrm{SS}} , \\
\bar{\boldsymbol{\Psi}}_{\mathrm{SS}} &= \frac{1}{c^2 d^3} \frac{\nu_A}{2} \left[ \mathbf{s}_0 - 3\lambda\mathbf{l} \right] \times \mathbf{s}_A .
\end{align}
\end{subequations}

The hybrid model EOMs' rhs are obtained by multiplying the OA EOMs' rhs by $d^3/r^3$; see Eqs.~\eqref{eq:hybrid-EOM}
\begin{align} \label{eq:psi_X_hybrid}
\boldsymbol{\Psi}_{\mathbf{s}_A}^X(\mathbf{V}) &= \boldsymbol{\Psi}_{\mathrm{SO}} + \frac{d^3}{r^3} \bar{\boldsymbol{\Psi}}_{\mathrm{SS}} = \frac{1}{c^2 r^3} \delta_A \mathbf{l} \times \mathbf{s}_A + \frac{1}{c^2 r^3} \frac{\nu_A}{2} \left[ \mathbf{s}_0 - 3\lambda \mathbf{l} \right] \times \mathbf{s}_A .
\end{align}
Eqs.~\eqref{eq:decomp_2PN},  \eqref{eq:decomp_SS}, and \eqref{eq:psi_X_hybrid}
give
\begin{align} \label{eq:delta_psi_hybrid}
\delta \boldsymbol{\Psi}_{\mathbf{s}_A} &= 
\boldsymbol{\Psi}^Y_{\mathbf{s}_A} - \boldsymbol{\Psi}^X_{\mathbf{s}_A}  = 
\widetilde{\boldsymbol{\Psi}}_{\mathrm{SS}} + \left(1 - \frac{d^3}{r^3}\right) \bar{\boldsymbol{\Psi}}_{\mathrm{SS}} .
\end{align}

Expanding Eq.~\eqref{eq:div_eqn} via Taylor series about the initial time yields
\begin{align} \label{eq:evol_eq_div_2PN_2H}
\frac{d\,\delta \mathbf{s}_A}{dt} & \sim \delta \boldsymbol{\Psi}_{\mathbf{s}_A}(\mathbf{V}^Y) = \boldsymbol{\Psi}_{\mathbf{s}_A}^Y(\mathbf{V}^Y) - \boldsymbol{\Psi}_{\mathbf{s}_A}^X(\mathbf{V}^Y) \nonumber \\
&\approx \left[ \boldsymbol{\Psi}_{\mathbf{s}_A}^Y(\mathbf{U}_0, \mathbf{r}^Y(t)) - \boldsymbol{\Psi}_{\mathbf{s}_A}^X(\mathbf{U}_0, \mathbf{r}^Y(t)) \right] \nonumber \\
&\quad + \left( \left. \frac{\partial \boldsymbol{\Psi}_{\mathbf{s}_A}^Y}{\partial \mathbf{l}} \right|_{\mathbf{U}_0, \vb{r}^Y } - \left. \frac{\partial \boldsymbol{\Psi}_{\mathbf{s}_A}^X}{\partial \mathbf{l}} \right|_{\mathbf{U}_0, \vb{r}^Y} \right) \Delta \mathbf{l}^Y + \left( \left. \frac{\partial \boldsymbol{\Psi}_{\mathbf{s}_A}^Y}{\partial \mathbf{s}_B} \right|_{\mathbf{U}_0, \vb{r}^Y} - \left. \frac{\partial \boldsymbol{\Psi}_{\mathbf{s}_A}^X}{\partial \mathbf{s}_B} \right|_{\mathbf{U}_0, \vb{r}^Y} \right) \Delta \mathbf{s}_B^Y . 
\end{align}

With the aid of Eq.~\eqref{eq:delta_psi_hybrid}, we 
integrate the leading-order difference
in the square brackets
of Eq.~\eqref{eq:evol_eq_div_2PN_2H}
over $t = N \tau_{\text{orb}} + \delta t$ to yield
\begin{align} \label{eq:integral_decomposition_hybrid}
&\int_0^t \left[ \boldsymbol{\Psi}_{\mathbf{s}_A}^Y( \vb{U}_0 ,  \tau) - \boldsymbol{\Psi}_{\mathbf{s}_A}^X(\vb{U}_0, \tau) \right] d\tau \nonumber \\
&= \sum_{k=0}^{N-1} \int_{k \tau_{\text{orb}}}^{(k+1)\tau_{\text{orb}}} \left[ \widetilde{\boldsymbol{\Psi}}_{\mathrm{SS}}(\tau) + \left(1 - \frac{d^3}{r^3(\tau)}\right) \bar{\boldsymbol{\Psi}}_{\mathrm{SS}} \right] d\tau + \int_{N \tau_{\text{orb}}}^{t} \left[ \widetilde{\boldsymbol{\Psi}}_{\mathrm{SS}}(\tau) + \left(1 - \frac{d^3}{r^3(\tau)}\right) \bar{\boldsymbol{\Psi}}_{\mathrm{SS}} \right] d\tau .
\end{align}
The integral over any complete orbital period vanishes: $\langle \widetilde{\boldsymbol{\Psi}}_{\mathrm{SS}} \rangle = 0$ by definition, and the orbital average of $(1 - d^3/r^3)$ is zero since $\langle r^{-3} \rangle = d^{-3}$.
Hence, the sum over $N-1$ periods drops out. Therefore, only the second square bracket of Eq.~\eqref{eq:integral_decomposition_hybrid} contributes. Due to $\boldsymbol{\Psi}_{\text{SS}} \sim c^{-4}$, the surviving integrand within the fractional period $\delta t$ scales as $c^{-4}$, leaving a residual of $\delta t / c^4$.

We still need to account for the linearization terms in Eq.~\eqref{eq:evol_eq_div_2PN_2H}. While the individual Jacobians for each model scale as
\begin{subequations} \label{eq:hybrid_jacobian_scaling}
\begin{align}
\frac{\partial \boldsymbol{\Psi}_{\mathbf{s}_A}^Y}{\partial \mathbf{l}} \sim \frac{\partial \boldsymbol{\Psi}_{\mathbf{s}_A}^X}{\partial \mathbf{l}} &\sim c^{-3} \,, \\
\frac{\partial \boldsymbol{\Psi}_{\mathbf{s}_A}^Y}{\partial \mathbf{s}_B} \sim \frac{\partial \boldsymbol{\Psi}_{\mathbf{s}_A}^X}{\partial \mathbf{s}_B} &\sim c^{-2} \,,
\end{align}
\end{subequations}
the leading-order spin-orbit term $\boldsymbol{\Psi}_{\mathrm{SO}}$ is identical in both the full 2PN 
$\vb{\Psi}^Y$
and hybrid
$\vb{\Psi}^X$
models. Consequently, it cancels  upon subtraction in the parentheses of 
Eq.~\eqref{eq:evol_eq_div_2PN_2H}. The surviving terms within the two parentheses of Eq.~\eqref{eq:evol_eq_div_2PN_2H} arise purely from the spin-spin interactions. Evaluating the derivatives of these spin-spin terms with respect to $\mathbf{l}$ and $\mathbf{s}_B$ causes the two parentheses to scale as $\mathcal{O}(c^{-4})$ and $\mathcal{O}(c^{-3})$. Multiplying these by the angular momentum deviations $\Delta \mathbf{l}^Y \sim \Delta \mathbf{s}_B^Y \sim \mathcal{O}(t/c^3)$ from Eq.~\eqref{eq:2PN-2PN_scaling_b}, we find that the 
dominant parenthesis of 
Eq.~\eqref{eq:evol_eq_div_2PN_2H}
 scales  as $\mathcal{O}(t/c^6)$.

Finally, integrating Eq.~\eqref{eq:evol_eq_div_2PN_2H} with respect to time yields
\begin{align} \label{eq:delta_scaling_hybrid_sa}
\delta \mathbf{s}_A \sim \frac{\delta t}{c^4} + \frac{t^2}{c^6} \,.
\end{align}
An analogous relation can be derived for $\mathbf{l}$ as well,
\begin{align} \label{eq:delta_scaling_hybrid_l}
\delta \mathbf{l} \sim \frac{\delta t}{c^4} + \frac{t^2}{c^6} \,.
\end{align}

Substituting Eqs.~\eqref{eq:delta_scaling_hybrid_sa} and \eqref{eq:delta_scaling_hybrid_l} along with the individual Jacobian scalings from Eq.~\eqref{eq:hybrid_jacobian_scaling} into the balance condition Eq.~\eqref{eq:small_time_b} gives
\begin{align} \label{eq:hybrid_balance_condition}
    \frac{1}{c^2} \left( \frac{\delta t}{c^4} + \frac{t_{\delta}^2}{c^6} \right) + \frac{1}{c^3} \left( \frac{\delta t}{c^4} + \frac{t_{\delta}^2}{c^6} \right) \sim \frac{1}{c^4} \implies t_{\delta} \sim c^2 \,.
\end{align}
\end{enumerate}

\section{Justification of the quasi-Keplerian ansatz}
\label{app:ansatz}

In this Appendix, we justify the functional form of the quasi-Keplerian
parametrization introduced in Sec.~\ref{sec:QKP} by analyzing the harmonic
content of the EOMs. We first identify which coefficients in
the radial EOM require an ansatz treatment, and then determine the minimal
set of oscillatory corrections that must be added to the known solution.

%--------------------------------
\subsection{From direct integration to the ansatz approach}
\label{app:strategy}
%--------------------------------

\textit{Nonspinning and spin-orbit cases.} In the absence of spin-spin interactions, 
all coefficients appearing in the
radial EOM~\eqref{eq:drdt_final} are constant: in the nonspinning case,
they depend only on $h$ and $l$, while the inclusion of spin-orbit 
effects also adds $\lseff$, which is a COM at that order. The radial EOM 
therefore reduces to a polynomial in $r^{-1}$ with constant coefficients, 
which can be integrated in closed form. This is the approach of  
Sch\"{a}fer and Wex~\cite{Schafer1993} in the nonspinning case, and of Cho and 
Lee~\cite{Cho2019} at 1.5PN order. The resulting parametrization reads\footnote{
This parametrization of $\phi$ in 
Eqs.~\eqref{app:known_solution}
corresponds to the
azimuthal differential evolution Eq.~\eqref{eq:dphidt} with the parentheses
involving $\cos \kappa_1$ ignored for now. The quantity in the parentheses  can be separately integrated using
elliptic functions, as in Ref.~\cite{Cho2019}
and then added to the above parametrization.}
\begin{subequations}\label{app:known_solution}
\begin{align}
n(t - t_0) &= u - e_t \sin u
  + \frac{1}{c^4}\left[ f_t \sin v_\phi + g_t (v_\phi - u) \right], \\
r &= a_r (1 - e_r \cos u), \\
v_\phi &= u + 2\arctan\!\left(\frac{\beta_\phi \sin u}{1 - \beta_\phi \cos u}\right),
  \quad \beta_\phi = \frac{e_\phi}{1 + \sqrt{1 - e_\phi^2}}, \\
\phi(t) - \phi(t_0) &= (1 + k') v_\phi
  + \frac{1}{c^4}\left[ f_\phi \sin 2v_\phi + g_\phi \sin 3v_\phi \right],
\end{align}
\end{subequations}
where all orbital elements ($n$, $a_r$, $e_r$, $e_t$, $e_\phi$, $k'$,
$f_t$, $g_t$, $f_\phi$, $g_\phi$) are constants, and $v_\phi$ is the
quasi-Keplerian true anomaly. 

\textit{Spin-spin interactions and $\phi$-dependent coefficients.}
The situation changes qualitatively once 2PN spin-spin interactions are
included. As shown in Sec.~\ref{sec:eomR}, two coefficients 
$C$ and $D_1$
in the radial
EOM~\eqref{eq:drdt_final} acquire $\phi$-dependent oscillatory corrections
\begin{enumerate}
\item $C$ acquires an oscillatory correction through the evolution of $l^2$.
  From Eq.~\eqref{eq:l_decomposition}, $l^2$ oscillates with three harmonics of
  $\phi$ (at $\phi - 2\phi_{sp}$, $2\phi - 2\phi_{sp}$, and $3\phi -
  2\phi_{sp}$).
\item $D_{1}$ acquires a $\phi$-dependent correction through
  $(\mathbf{n} \cdot \mathbf{s}_0)^2 = s_{0p}^2 \cos^2(\phi - \phi_{sp})$,
  introducing two oscillatory harmonic at $2\phi - 2\phi_{sp}$, as given in
  Eq.~\eqref{eq:radial_coeffs_final}.
\end{enumerate}
Since $C$ and $D_1$ are the coefficients of $r^{-2}$ and $r^{-3}$
respectively in Eq.~\eqref{eq:drdt_final}, these corrections take the
explicit form
\begin{subequations}\label{app:osc_CD}
\begin{align}
\frac{\delta C(\phi)}{r^2}
&\propto \frac{s_{0p}^2}{\,c^2\,\bar{l}\,^2\,r^2}\left[
  3e\cos(\phi - 2\phi_{sp})
  + 3\cos(2\phi - 2\phi_{sp})
  + e\cos(3\phi - 2\phi_{sp})
\right], \\[4pt]
\frac{\delta D_{1}(\phi)}{r^3}
&\propto \frac{s_{0p}^2}{c^2\,r^3}\big(1 - \cos(2\phi - 2\phi_{sp})\big).
\end{align}
\end{subequations}
The constant term in the numerator of $\delta D_1$ can be eliminated 
by absorbing it into a redefinition of the orbital elements. Meanwhile, 
the oscillatory components of Eqs.~\eqref{app:osc_CD} act as new 
source terms on the rhs of the radial EOM~Eq.~\eqref{eq:drdt_final},
necessitating the addition of corresponding oscillatory functions to the 
ansatz for $r$, whose derivative
the lhs of Eq.~\eqref{eq:drdt_final} is built out of.

To evaluate the oscillatory components of Eqs.~\eqref{app:osc_CD}, we first substitute the Newtonian radial relation 
connecting $r$ and $\phi$ 
\begin{equation}\label{app:r_newton}
  r = \frac{a(1-e^2)}{1 + e \cos v_{\phi}}\,, \qquad \phi = v_{\phi},  
\end{equation}
in Eq.~\eqref{eq:drdt_final},
thereby giving us
 new source terms on the 
 rhs 
\begin{equation}\label{app:Delta_r}
  \frac{1}{c^2}\sum_{k=0}^{5} \alpha_k\cos(k\,v_\phi - 2\phi_{sp})\,,
\end{equation}
where the coefficients $\alpha_k$ depend on $h$, $\bar{l}$, and $s_{0p}^2$,
which are all constant on the orbital timescale. The series terminates at $k = 5$
because
\begin{itemize}
\item the oscillatory part of $C$ contains harmonics up to
  $3\phi - 2\phi_{sp}$, and expanding $r^{-2}$ introduces up to two
  additional harmonics, giving a maximum of $3 + 2 = 5$;
\item the oscillatory part of $D_1$ contains a single harmonic at
  $2\phi - 2\phi_{sp}$, and expanding $r^{-3}$ introduces up to three
  additional harmonics, giving a maximum of $2 + 3 = 5$.
\end{itemize}

%--------------------------------
\subsection{Harmonic analysis of the radial equation}
\label{app:radial_harmonics}
%--------------------------------

We perturb the known parametrization~\eqref{app:known_solution} by adding
2PN spin-spin corrections, leading to the following proposed ansatz\footnote{The overall $c^{-2}$ factor combined
with the proportionality of $h_r^{(m)}$ and $h_t^{(n)}$ to $s_{0p}^2$
ensures the correct 2PN scaling.}
\begin{subequations}\label{app:spin_corrections}
\begin{align}
r & = a_r(1 - e_r\cos u)
  + \frac{1}{c^2}\sum_{m=1}^{M} h_{rm}\cos(m\,v_\phi - 2\phi_{sp}), \label{app:rAnz}\\
n(t - t_0) & =  u - e_t\sin u
  + \frac{1}{c^4}\left[f_t\sin v_\phi + g_t(v_\phi - u)\right]
  + \frac{1}{c^2}\sum_{n=1}^{N} h_{tn}\sin(n\,v_\phi - 2\phi_{sp})\, ,\label{app:tAnz} \\
  v_\phi &= u + 2 \arctan \left( \frac{\beta_\phi \sin u}{1 - \beta_\phi \cos u} \right), \quad \text{ with }\  \beta_\phi = \frac{e_\phi}{1 + \sqrt{1 - e_\phi^2}}    \label{app:v_phi} .
\end{align}
\end{subequations}
The upper limits $M$ and $N$ are fixed by the requirement that the
lhs of Eq.~\eqref{eq:drdt_final}, computed via the chain rule\footnote{To compute various derivatives in Eq.~\eqref{app:chain_rule}, 
we use Eq.~\eqref{app:spin_corrections}; all occurrences of $v_\phi$ are to be considered functions of $u$ as per Eq.~\eqref{app:v_phi}.}
\begin{equation}\label{app:chain_rule}
\left(\frac{dr}{dt}\right)^2
= \left(\frac{d r}{d u}\right)^2 
\!\left(\frac{dt}{du}\right)^{-2},
\end{equation}
generates no harmonics of order $k > 5$, 
as dictated by
Eq.~\eqref{app:Delta_r}.
Adding corrections of higher harmonic order
would introduce spurious harmonics on the lhs with no
counterpart in the source. This leads to $M = 2$ in Eq.~\eqref{app:rAnz} 
and $N = 1$ in Eq.~\eqref{app:tAnz}. The radial and temporal parametrizations are therefore
\begin{subequations}\label{app:rt_final}
\begin{align}
r &= a_r(1 - e_r\cos u)
  + \frac{1}{c^2}\Big[ h_{r1}\cos(v_\phi - 2\phi_{sp}) + h_{r2}\cos(2v_\phi - 2\phi_{sp}) \Big],
  \label{app:r_final} \\
n(t - t_0) &= u - e_t\sin u
  + \frac{1}{c^4}\left[f_t\sin v_\phi + g_t(v_\phi - u)\right]
  + \frac{1}{c^2}\,h_{t1}\sin(v_\phi - 2\phi_{sp}).
  \label{app:t_final}
\end{align}
\end{subequations}

%--------------------------------
\subsection{Harmonic analysis of the azimuthal equation}
\label{app:azimuthal_harmonics}
%--------------------------------

Our strategy to integrate the azimuthal evolution 
Eq.~\eqref{eq:dphidt} is to split this equation in two segments: first segment consists of the first four terms on the rhs, and the second segment is the parentheses. We integrate the first segment with the ansatz-based approach, and integrate the second segment directly using elliptic functions as in Ref.~\cite{Cho2019}. Finally, we add the two resulting integrations to construct the full $\phi$.

So, we proceed to
rewrite the azimuthal evolution Eq.~\eqref{eq:dphidt} as 
(with the parentheses term dropped for now)
\begin{equation}\label{app:dphidt_schematic}
r^2\frac{d\phi}{dt} = F(\phi) + \frac{I_{1o} + I_{1s}}{r} + \frac{I_2}{r^2} + \frac{I_3}{r^3}\,.
\end{equation}
Because $F(\phi)$ depends on $l$ at Newtonian order, it is the only coefficient that acquires 
an oscillatory correction at 2PN order
\begin{equation}\label{app:osc_F}
\delta F(\phi)
\propto\frac{s_{0p}^2}{c^2\,\bar{l}}\Big[
  3e\cos(\phi - 2\phi_{sp})
  + 3\cos(2\phi - 2\phi_{sp})
  + e\cos(3\phi - 2\phi_{sp})
\Big]\,.
\end{equation}

Because $\delta F$ introduces 3 harmonics and is not scaled by any inverse powers 
of $r$ in Eq.~\eqref{app:dphidt_schematic}, the rhs  source harmonics reach 
a maximum order of $k=3$. To match these terms, we add a spin-spin oscillatory sum to 
the known azimuthal parametrization
\begin{equation}\label{app:phi_correction}
\phi = (1 + k')v_\phi
  + \frac{1}{c^4}\left[f_\phi\sin 2v_\phi + g_\phi\sin 3v_\phi\right]
  + \frac{1}{c^2}\sum_{\ell=1}^{L} h_{\phi\ell}\sin(\ell\,v_\phi - 2\phi_{sp})\,.
\end{equation}
Similar to the radial analysis, the upper limit $L$ is fixed by requiring the constructed left-hand side, $r^2 d\phi/dt$, 
to generate no harmonics beyond the source maximum of $k=3$. This leads to $L=3$.

%--------------------------------
\subsection{Complete quasi-Keplerian ansatz}
\label{app:full_ansatz}
%--------------------------------

Synthesizing the harmonic analysis above, the full ansatz required to capture all 2PN spin-spin corrections is
\begin{subequations}\label{app:full_QKP}
\begin{align}
n(t - t_0) &= u - e_t\sin u
  + \frac{1}{c^4}\left[f_t\sin v_\phi + g_t(v_\phi - u)\right]
  + \frac{1}{c^2}\,h_{t1}\sin(v_\phi - 2\phi_{sp}),
  \label{app:t_ansatz} \\[4pt]
r &= a_r(1 - e_r\cos u)
  + \frac{1}{c^2}\Big[
    h_{r1}\cos(v_\phi - 2\phi_{sp})
    + h_{r2}\cos(2v_\phi - 2\phi_{sp})
  \Big],
  \label{app:r_ansatz} \\[4pt]
\phi(t) - \phi(t_0) &= (1 + k')v_\phi
  + \frac{1}{c^4}\left[f_\phi\sin 2v_\phi + g_\phi\sin 3v_\phi\right]
  \nonumber \\
&\quad
  + \frac{1}{c^2}\Big[
    h_{\phi1}\sin(v_\phi - 2\phi_{sp})
    + h_{\phi2}\sin(2v_\phi - 2\phi_{sp})
    + h_{\phi3}\sin(3v_\phi - 2\phi_{sp})
  \Big].
  \label{app:phi_ansatz}
\end{align}
\end{subequations}

This QKP ansatz contains the fewest harmonic corrections necessary to represent the 2PN spin dynamics, 
recovering the functional structure introduced by Klein and Jetzer (Eqs.~(A1)–(A4), Ref.~\cite{Klein2010}). 
To determine the final solution, we add to the standard quasi-Keplerian parameters ($n$, $a_r$, $e_r$, $e_t$, 
$e_\phi$, $\tilde{k}$, $f_t$, $g_t$, $f_\phi$, $g_\phi$) a free, undetermined 2PN spin-spin contributions alongside 
their known nonspinning and spin-orbit expressions. Crucially, these modified parameters, along with the new 
spin-dependent coefficients ($h_t$, $h_{r1}$, $h_{r2}$, $h_{\phi1}$, $h_{\phi2}$, $h_{\phi3}$), are treated as 
constants on the short orbital timescale, though they are free to undergo secular evolution on the much longer 
precession timescale. All undetermined coefficients are then fixed by substituting Eqs.~\eqref{app:full_QKP}
into the EOMs Eqs.~\eqref{eq:drdt_final}
and~\eqref{eq:dphidt}  and requiring them to be satisfied, yielding the solution presented in 
Eqs.~\eqref{eq:QK_parametrization} and~\eqref{eq:QK_parameters}.
As already mentioned, all this is done while ignoring the parentheses term in Eq.~\eqref{eq:dphidt}. 
This parentheses term is integrated separately, and the final
total result is displayed in
Eqs.~\eqref{eq:QK_parametrization}.

%--------------------------------
%--------------------------------	
%\section{Bibliography}
%--------------------------------
%--------------------------------

\bibliography{article_f}

\begin{thebibliography}{62}%
\makeatletter
\providecommand \@ifxundefined [1]{%
 \@ifx{#1\undefined}
}%
\providecommand \@ifnum [1]{%
 \ifnum #1\expandafter \@firstoftwo
 \else \expandafter \@secondoftwo
 \fi
}%
\providecommand \@ifx [1]{%
 \ifx #1\expandafter \@firstoftwo
 \else \expandafter \@secondoftwo
 \fi
}%
\providecommand \natexlab [1]{#1}%
\providecommand \enquote  [1]{``#1''}%
\providecommand \bibnamefont  [1]{#1}%
\providecommand \bibfnamefont [1]{#1}%
\providecommand \citenamefont [1]{#1}%
\providecommand \href@noop [0]{\@secondoftwo}%
\providecommand \href [0]{\begingroup \@sanitize@url \@href}%
\providecommand \@href[1]{\@@startlink{#1}\@@href}%
\providecommand \@@href[1]{\endgroup#1\@@endlink}%
\providecommand \@sanitize@url [0]{\catcode `\\12\catcode `\$12\catcode
  `\&12\catcode `\#12\catcode `\^12\catcode `\_12\catcode `\%12\relax}%
\providecommand \@@startlink[1]{}%
\providecommand \@@endlink[0]{}%
\providecommand \url  [0]{\begingroup\@sanitize@url \@url }%
\providecommand \@url [1]{\endgroup\@href {#1}{\urlprefix }}%
\providecommand \urlprefix  [0]{URL }%
\providecommand \Eprint [0]{\href }%
\providecommand \doibase [0]{https://doi.org/}%
\providecommand \selectlanguage [0]{\@gobble}%
\providecommand \bibinfo  [0]{\@secondoftwo}%
\providecommand \bibfield  [0]{\@secondoftwo}%
\providecommand \translation [1]{[#1]}%
\providecommand \BibitemOpen [0]{}%
\providecommand \bibitemStop [0]{}%
\providecommand \bibitemNoStop [0]{.\EOS\space}%
\providecommand \EOS [0]{\spacefactor3000\relax}%
\providecommand \BibitemShut  [1]{\csname bibitem#1\endcsname}%
\let\auto@bib@innerbib\@empty
%</preamble>
\bibitem [{\citenamefont {Hannam}\ \emph {et~al.}(2022)\citenamefont {Hannam},
  \citenamefont {Hoy}, \citenamefont {Thompson}, \citenamefont {Fairhurst},
  \citenamefont {Raymond} \emph {et~al.}}]{Hannam:2022pit}%
  \BibitemOpen
  \bibfield  {author} {\bibinfo {author} {\bibfnamefont {M.}~\bibnamefont
  {Hannam}}, \bibinfo {author} {\bibfnamefont {C.}~\bibnamefont {Hoy}},
  \bibinfo {author} {\bibfnamefont {J.~E.}\ \bibnamefont {Thompson}}, \bibinfo
  {author} {\bibfnamefont {S.}~\bibnamefont {Fairhurst}}, \bibinfo {author}
  {\bibfnamefont {V.}~\bibnamefont {Raymond}}, \emph {et~al.},\ }\bibfield
  {title} {\bibinfo {title} {{General-relativistic precession in a black-hole
  binary}},\ }\bibfield  {journal} {\bibinfo  {journal} {Nature}\ }\href
  {https://doi.org/10.1038/s41586-022-05212-z} {10.1038/s41586-022-05212-z}
  (\bibinfo {year} {2022}),\ \Eprint {https://arxiv.org/abs/2112.11300}
  {arXiv:2112.11300 [gr-qc]} \BibitemShut {NoStop}%
\bibitem [{\citenamefont {Abbott}\ \emph {et~al.}(2020)\citenamefont {Abbott}
  \emph {et~al.}}]{LIGOScientific:2020stg}%
  \BibitemOpen
  \bibfield  {author} {\bibinfo {author} {\bibfnamefont {R.}~\bibnamefont
  {Abbott}} \emph {et~al.} (\bibinfo {collaboration} {LIGO Scientific,
  Virgo}),\ }\bibfield  {title} {\bibinfo {title} {{GW190412: Observation of a
  Binary-Black-Hole Coalescence with Asymmetric Masses}},\ }\href
  {https://doi.org/10.1103/PhysRevD.102.043015} {\bibfield  {journal} {\bibinfo
   {journal} {Phys. Rev. D}\ }\textbf {\bibinfo {volume} {102}},\ \bibinfo
  {pages} {043015} (\bibinfo {year} {2020})},\ \Eprint
  {https://arxiv.org/abs/2004.08342} {arXiv:2004.08342 [astro-ph.HE]}
  \BibitemShut {NoStop}%
\bibitem [{\citenamefont {Romero-Shaw}\ \emph {et~al.}(2020)\citenamefont
  {Romero-Shaw}, \citenamefont {Lasky}, \citenamefont {Thrane},\ and\
  \citenamefont {Bustillo}}]{Romero-Shaw:2020thy}%
  \BibitemOpen
  \bibfield  {author} {\bibinfo {author} {\bibfnamefont {I.~M.}\ \bibnamefont
  {Romero-Shaw}}, \bibinfo {author} {\bibfnamefont {P.~D.}\ \bibnamefont
  {Lasky}}, \bibinfo {author} {\bibfnamefont {E.}~\bibnamefont {Thrane}},\ and\
  \bibinfo {author} {\bibfnamefont {J.~C.}\ \bibnamefont {Bustillo}},\
  }\bibfield  {title} {\bibinfo {title} {{GW190521: orbital eccentricity and
  signatures of dynamical formation in a binary black hole merger signal}},\
  }\href {https://doi.org/10.3847/2041-8213/abbe26} {\bibfield  {journal}
  {\bibinfo  {journal} {Astrophys. J. Lett.}\ }\textbf {\bibinfo {volume}
  {903}},\ \bibinfo {pages} {L5} (\bibinfo {year} {2020})},\ \Eprint
  {https://arxiv.org/abs/2009.04771} {arXiv:2009.04771 [astro-ph.HE]}
  \BibitemShut {NoStop}%
\bibitem [{\citenamefont {Gerosa}\ and\ \citenamefont
  {Fishbach}(2021)}]{Gerosa:2021mno}%
  \BibitemOpen
  \bibfield  {author} {\bibinfo {author} {\bibfnamefont {D.}~\bibnamefont
  {Gerosa}}\ and\ \bibinfo {author} {\bibfnamefont {M.}~\bibnamefont
  {Fishbach}},\ }\bibfield  {title} {\bibinfo {title} {{Hierarchical mergers of
  stellar-mass black holes and their gravitational-wave signatures}},\ }\href
  {https://doi.org/10.1038/s41550-021-01398-w} {\bibfield  {journal} {\bibinfo
  {journal} {Nature Astron.}\ }\textbf {\bibinfo {volume} {5}},\ \bibinfo
  {pages} {749} (\bibinfo {year} {2021})},\ \Eprint
  {https://arxiv.org/abs/2105.03439} {arXiv:2105.03439 [astro-ph.HE]}
  \BibitemShut {NoStop}%
\bibitem [{\citenamefont {Mandel}\ and\ \citenamefont
  {Farmer}(2022)}]{Mandel:2018hfr}%
  \BibitemOpen
  \bibfield  {author} {\bibinfo {author} {\bibfnamefont {I.}~\bibnamefont
  {Mandel}}\ and\ \bibinfo {author} {\bibfnamefont {A.}~\bibnamefont
  {Farmer}},\ }\bibfield  {title} {\bibinfo {title} {{Merging stellar-mass
  binary black holes}},\ }\href {https://doi.org/10.1016/j.physrep.2022.01.003}
  {\bibfield  {journal} {\bibinfo  {journal} {Phys. Rept.}\ }\textbf {\bibinfo
  {volume} {955}},\ \bibinfo {pages} {1} (\bibinfo {year} {2022})},\ \Eprint
  {https://arxiv.org/abs/1806.05820} {arXiv:1806.05820 [astro-ph.HE]}
  \BibitemShut {NoStop}%
\bibitem [{\citenamefont {Romero-Shaw}\ \emph {et~al.}(2025)\citenamefont
  {Romero-Shaw}, \citenamefont {Stegmann}, \citenamefont {Tagawa},
  \citenamefont {Gerosa}, \citenamefont {Samsing}, \citenamefont {Gupte},\ and\
  \citenamefont {Green}}]{Romero-Shaw:2025vbc}%
  \BibitemOpen
  \bibfield  {author} {\bibinfo {author} {\bibfnamefont {I.}~\bibnamefont
  {Romero-Shaw}}, \bibinfo {author} {\bibfnamefont {J.}~\bibnamefont
  {Stegmann}}, \bibinfo {author} {\bibfnamefont {H.}~\bibnamefont {Tagawa}},
  \bibinfo {author} {\bibfnamefont {D.}~\bibnamefont {Gerosa}}, \bibinfo
  {author} {\bibfnamefont {J.}~\bibnamefont {Samsing}}, \bibinfo {author}
  {\bibfnamefont {N.}~\bibnamefont {Gupte}},\ and\ \bibinfo {author}
  {\bibfnamefont {S.~R.}\ \bibnamefont {Green}},\ }\bibfield  {title} {\bibinfo
  {title} {{GW200208{\_}222617 as an eccentric black-hole binary merger:
  Properties and astrophysical implications}},\ }\href
  {https://doi.org/10.1103/jj7m-x66y} {\bibfield  {journal} {\bibinfo
  {journal} {Phys. Rev. D}\ }\textbf {\bibinfo {volume} {112}},\ \bibinfo
  {pages} {063052} (\bibinfo {year} {2025})},\ \Eprint
  {https://arxiv.org/abs/2506.17105} {arXiv:2506.17105 [astro-ph.HE]}
  \BibitemShut {NoStop}%
\bibitem [{\citenamefont {Romero-Shaw}\ \emph {et~al.}(2023)\citenamefont
  {Romero-Shaw}, \citenamefont {Gerosa},\ and\ \citenamefont
  {Loutrel}}]{10.1093/mnras/stad031}%
  \BibitemOpen
  \bibfield  {author} {\bibinfo {author} {\bibfnamefont {I.~M.}\ \bibnamefont
  {Romero-Shaw}}, \bibinfo {author} {\bibfnamefont {D.}~\bibnamefont
  {Gerosa}},\ and\ \bibinfo {author} {\bibfnamefont {N.}~\bibnamefont
  {Loutrel}},\ }\bibfield  {title} {\bibinfo {title} {Eccentricity or spin
  precession? distinguishing subdominant effects in gravitational-wave data},\
  }\href {https://doi.org/10.1093/mnras/stad031} {\bibfield  {journal}
  {\bibinfo  {journal} {Monthly Notices of the Royal Astronomical Society}\
  }\textbf {\bibinfo {volume} {519}},\ \bibinfo {pages} {5352} (\bibinfo {year}
  {2023})},\ \Eprint
  {https://arxiv.org/abs/https://academic.oup.com/mnras/article-pdf/519/4/5352/50423218/stad031.pdf}
  {https://academic.oup.com/mnras/article-pdf/519/4/5352/50423218/stad031.pdf}
  \BibitemShut {NoStop}%
\bibitem [{\citenamefont {Tibrewal}\ \emph {et~al.}(2026)\citenamefont
  {Tibrewal}, \citenamefont {Zimmerman}, \citenamefont {Lange},\ and\
  \citenamefont {Shoemaker}}]{Tibrewal:2026jci}%
  \BibitemOpen
  \bibfield  {author} {\bibinfo {author} {\bibfnamefont {S.}~\bibnamefont
  {Tibrewal}}, \bibinfo {author} {\bibfnamefont {A.}~\bibnamefont {Zimmerman}},
  \bibinfo {author} {\bibfnamefont {J.}~\bibnamefont {Lange}},\ and\ \bibinfo
  {author} {\bibfnamefont {D.}~\bibnamefont {Shoemaker}},\ }\href@noop {}
  {\bibinfo {title} {{Misinterpreting Spin Precession as Orbital Eccentricity
  in Gravitational-Wave Signals}}} (\bibinfo {year} {2026}),\ \Eprint
  {https://arxiv.org/abs/2601.02260} {arXiv:2601.02260 [gr-qc]} \BibitemShut
  {NoStop}%
\bibitem [{\citenamefont {Stegmann}\ \emph {et~al.}(2025)\citenamefont
  {Stegmann} \emph {et~al.}}]{Stegmann_2025}%
  \BibitemOpen
  \bibfield  {author} {\bibinfo {author} {\bibfnamefont {J.}~\bibnamefont
  {Stegmann}} \emph {et~al.},\ }\href
  {https://doi.org/10.3847/2041-8213/ae1d66} {\bibfield  {journal} {\bibinfo
  {journal} {Astrophys. J. Lett.}\ }\textbf {\bibinfo {volume} {994}},\
  \bibinfo {pages} {L47} (\bibinfo {year} {2025})}\BibitemShut {NoStop}%
\bibitem [{\citenamefont {Baibhav}(2025)}]{Baibhav:2025mzw}%
  \BibitemOpen
  \bibfield  {author} {\bibinfo {author} {\bibfnamefont {V.}~\bibnamefont
  {Baibhav}},\ }\href@noop {} {\bibinfo {title} {{Inferring Eccentricity of
  Binary Black Holes from Spin-Orbit Misalignment}}} (\bibinfo {year} {2025}),\
  \Eprint {https://arxiv.org/abs/2512.22044} {arXiv:2512.22044 [astro-ph.HE]}
  \BibitemShut {NoStop}%
\bibitem [{\citenamefont {Lehner}(2001)}]{Lehner2001}%
  \BibitemOpen
  \bibfield  {author} {\bibinfo {author} {\bibfnamefont {L.}~\bibnamefont
  {Lehner}},\ }\bibfield  {title} {\bibinfo {title} {Numerical relativity: a
  review},\ }\href@noop {} {\bibfield  {journal} {\bibinfo  {journal}
  {Classical and Quantum Gravity}\ }\textbf {\bibinfo {volume} {18}},\ \bibinfo
  {pages} {R25} (\bibinfo {year} {2001})}\BibitemShut {NoStop}%
\bibitem [{\citenamefont {Sperhake}(2015)}]{Sperhake2015}%
  \BibitemOpen
  \bibfield  {author} {\bibinfo {author} {\bibfnamefont {U.}~\bibnamefont
  {Sperhake}},\ }\bibfield  {title} {\bibinfo {title} {The numerical relativity
  breakthrough for binary black holes},\ }\href@noop {} {\bibfield  {journal}
  {\bibinfo  {journal} {Classical and Quantum Gravity}\ }\textbf {\bibinfo
  {volume} {32}},\ \bibinfo {pages} {124011} (\bibinfo {year}
  {2015})}\BibitemShut {NoStop}%
\bibitem [{\citenamefont {Kokkotas}\ and\ \citenamefont
  {Schmidt}(1999)}]{Kokkotas1999}%
  \BibitemOpen
  \bibfield  {author} {\bibinfo {author} {\bibfnamefont {K.~D.}\ \bibnamefont
  {Kokkotas}}\ and\ \bibinfo {author} {\bibfnamefont {B.~G.}\ \bibnamefont
  {Schmidt}},\ }\bibfield  {title} {\bibinfo {title} {Quasi-normal modes of
  stars and black holes},\ }\href@noop {} {\bibfield  {journal} {\bibinfo
  {journal} {Living Reviews in Relativity}\ }\textbf {\bibinfo {volume} {2}},\
  \bibinfo {pages} {2} (\bibinfo {year} {1999})}\BibitemShut {NoStop}%
\bibitem [{\citenamefont {Berti}\ \emph {et~al.}(2009)\citenamefont {Berti},
  \citenamefont {Cardoso},\ and\ \citenamefont {Starinets}}]{Berti2009}%
  \BibitemOpen
  \bibfield  {author} {\bibinfo {author} {\bibfnamefont {E.}~\bibnamefont
  {Berti}}, \bibinfo {author} {\bibfnamefont {V.}~\bibnamefont {Cardoso}},\
  and\ \bibinfo {author} {\bibfnamefont {A.~O.}\ \bibnamefont {Starinets}},\
  }\bibfield  {title} {\bibinfo {title} {Quasinormal modes of black holes and
  black branes},\ }\href@noop {} {\bibfield  {journal} {\bibinfo  {journal}
  {Classical and Quantum Gravity}\ }\textbf {\bibinfo {volume} {26}},\ \bibinfo
  {pages} {163001} (\bibinfo {year} {2009})}\BibitemShut {NoStop}%
\bibitem [{\citenamefont {Blanchet}(2014)}]{Blanchet2014}%
  \BibitemOpen
  \bibfield  {author} {\bibinfo {author} {\bibfnamefont {L.}~\bibnamefont
  {Blanchet}},\ }\bibfield  {title} {\bibinfo {title} {Gravitational radiation
  from post-newtonian sources and inspiralling compact binaries},\ }\href@noop
  {} {\bibfield  {journal} {\bibinfo  {journal} {Living reviews in relativity}\
  }\textbf {\bibinfo {volume} {17}},\ \bibinfo {pages} {2} (\bibinfo {year}
  {2014})}\BibitemShut {NoStop}%
\bibitem [{\citenamefont {Buonanno}\ and\ \citenamefont
  {Damour}(1999)}]{Buonanno1999}%
  \BibitemOpen
  \bibfield  {author} {\bibinfo {author} {\bibfnamefont {A.}~\bibnamefont
  {Buonanno}}\ and\ \bibinfo {author} {\bibfnamefont {T.}~\bibnamefont
  {Damour}},\ }\bibfield  {title} {\bibinfo {title} {Effective one-body
  approach to general relativistic two-body dynamics},\ }\href@noop {}
  {\bibfield  {journal} {\bibinfo  {journal} {Physical Review D}\ }\textbf
  {\bibinfo {volume} {59}},\ \bibinfo {pages} {084006} (\bibinfo {year}
  {1999})}\BibitemShut {NoStop}%
\bibitem [{\citenamefont {Khan}\ \emph {et~al.}(2016)\citenamefont {Khan},
  \citenamefont {Husa}, \citenamefont {Hannam}, \citenamefont {Ohme},
  \citenamefont {P{\"u}rrer}, \citenamefont {Forteza},\ and\ \citenamefont
  {Boh{\'e}}}]{Khan2016}%
  \BibitemOpen
  \bibfield  {author} {\bibinfo {author} {\bibfnamefont {S.}~\bibnamefont
  {Khan}}, \bibinfo {author} {\bibfnamefont {S.}~\bibnamefont {Husa}}, \bibinfo
  {author} {\bibfnamefont {M.}~\bibnamefont {Hannam}}, \bibinfo {author}
  {\bibfnamefont {F.}~\bibnamefont {Ohme}}, \bibinfo {author} {\bibfnamefont
  {M.}~\bibnamefont {P{\"u}rrer}}, \bibinfo {author} {\bibfnamefont {X.~J.}\
  \bibnamefont {Forteza}},\ and\ \bibinfo {author} {\bibfnamefont
  {A.}~\bibnamefont {Boh{\'e}}},\ }\bibfield  {title} {\bibinfo {title}
  {Frequency-domain gravitational waves from nonprecessing black-hole binaries.
  ii. a phenomenological model for the advanced detector era},\ }\href@noop {}
  {\bibfield  {journal} {\bibinfo  {journal} {Physical Review D}\ }\textbf
  {\bibinfo {volume} {93}},\ \bibinfo {pages} {044007} (\bibinfo {year}
  {2016})}\BibitemShut {NoStop}%
\bibitem [{\citenamefont {Schmidt}\ \emph {et~al.}(2011)\citenamefont
  {Schmidt}, \citenamefont {Hannam}, \citenamefont {Husa},\ and\ \citenamefont
  {Ajith}}]{Schmidt:2010it}%
  \BibitemOpen
  \bibfield  {author} {\bibinfo {author} {\bibfnamefont {P.}~\bibnamefont
  {Schmidt}}, \bibinfo {author} {\bibfnamefont {M.}~\bibnamefont {Hannam}},
  \bibinfo {author} {\bibfnamefont {S.}~\bibnamefont {Husa}},\ and\ \bibinfo
  {author} {\bibfnamefont {P.}~\bibnamefont {Ajith}},\ }\bibfield  {title}
  {\bibinfo {title} {{Tracking the precession of compact binaries from their
  gravitational-wave signal}},\ }\href
  {https://doi.org/10.1103/PhysRevD.84.024046} {\bibfield  {journal} {\bibinfo
  {journal} {Phys. Rev. D}\ }\textbf {\bibinfo {volume} {84}},\ \bibinfo
  {pages} {024046} (\bibinfo {year} {2011})},\ \Eprint
  {https://arxiv.org/abs/1012.2879} {arXiv:1012.2879 [gr-qc]} \BibitemShut
  {NoStop}%
\bibitem [{\citenamefont {Schmidt}\ \emph {et~al.}(2012)\citenamefont
  {Schmidt}, \citenamefont {Hannam},\ and\ \citenamefont
  {Husa}}]{Schmidt:2012rh}%
  \BibitemOpen
  \bibfield  {author} {\bibinfo {author} {\bibfnamefont {P.}~\bibnamefont
  {Schmidt}}, \bibinfo {author} {\bibfnamefont {M.}~\bibnamefont {Hannam}},\
  and\ \bibinfo {author} {\bibfnamefont {S.}~\bibnamefont {Husa}},\ }\bibfield
  {title} {\bibinfo {title} {{Towards models of gravitational waveforms from
  generic binaries: A simple approximate mapping between precessing and
  non-precessing inspiral signals}},\ }\href
  {https://doi.org/10.1103/PhysRevD.86.104063} {\bibfield  {journal} {\bibinfo
  {journal} {Phys. Rev. D}\ }\textbf {\bibinfo {volume} {86}},\ \bibinfo
  {pages} {104063} (\bibinfo {year} {2012})},\ \Eprint
  {https://arxiv.org/abs/1207.3088} {arXiv:1207.3088 [gr-qc]} \BibitemShut
  {NoStop}%
\bibitem [{\citenamefont {Boyle}\ \emph {et~al.}(2011)\citenamefont {Boyle},
  \citenamefont {Owen},\ and\ \citenamefont {Pfeiffer}}]{Boyle:2011gg}%
  \BibitemOpen
  \bibfield  {author} {\bibinfo {author} {\bibfnamefont {M.}~\bibnamefont
  {Boyle}}, \bibinfo {author} {\bibfnamefont {R.}~\bibnamefont {Owen}},\ and\
  \bibinfo {author} {\bibfnamefont {H.~P.}\ \bibnamefont {Pfeiffer}},\
  }\bibfield  {title} {\bibinfo {title} {{A geometric approach to the
  precession of compact binaries}},\ }\href
  {https://doi.org/10.1103/PhysRevD.84.124011} {\bibfield  {journal} {\bibinfo
  {journal} {Phys. Rev. D}\ }\textbf {\bibinfo {volume} {84}},\ \bibinfo
  {pages} {124011} (\bibinfo {year} {2011})},\ \Eprint
  {https://arxiv.org/abs/1110.2965} {arXiv:1110.2965 [gr-qc]} \BibitemShut
  {NoStop}%
\bibitem [{\citenamefont {Chatziioannou}\ \emph {et~al.}(2017)\citenamefont
  {Chatziioannou}, \citenamefont {Klein}, \citenamefont {Yunes},\ and\
  \citenamefont {Cornish}}]{Klein2017b}%
  \BibitemOpen
  \bibfield  {author} {\bibinfo {author} {\bibfnamefont {K.}~\bibnamefont
  {Chatziioannou}}, \bibinfo {author} {\bibfnamefont {A.}~\bibnamefont
  {Klein}}, \bibinfo {author} {\bibfnamefont {N.}~\bibnamefont {Yunes}},\ and\
  \bibinfo {author} {\bibfnamefont {N.}~\bibnamefont {Cornish}},\ }\bibfield
  {title} {\bibinfo {title} {Constructing gravitational waves from generic
  spin-precessing compact binary inspirals},\ }\href@noop {} {\bibfield
  {journal} {\bibinfo  {journal} {Physical Review D}\ }\textbf {\bibinfo
  {volume} {95}},\ \bibinfo {pages} {104004} (\bibinfo {year}
  {2017})}\BibitemShut {NoStop}%
\bibitem [{\citenamefont {Arredondo}\ \emph {et~al.}(2024)\citenamefont
  {Arredondo}, \citenamefont {Klein},\ and\ \citenamefont
  {Yunes}}]{Arredondo:2024nsl}%
  \BibitemOpen
  \bibfield  {author} {\bibinfo {author} {\bibfnamefont {J.~N.}\ \bibnamefont
  {Arredondo}}, \bibinfo {author} {\bibfnamefont {A.}~\bibnamefont {Klein}},\
  and\ \bibinfo {author} {\bibfnamefont {N.}~\bibnamefont {Yunes}},\ }\bibfield
   {title} {\bibinfo {title} {{Efficient gravitational-wave model for
  fully-precessing and moderately eccentric, compact binary inspirals}},\
  }\href {https://doi.org/10.1103/PhysRevD.110.044044} {\bibfield  {journal}
  {\bibinfo  {journal} {Phys. Rev. D}\ }\textbf {\bibinfo {volume} {110}},\
  \bibinfo {pages} {044044} (\bibinfo {year} {2024})},\ \Eprint
  {https://arxiv.org/abs/2402.06804} {arXiv:2402.06804 [gr-qc]} \BibitemShut
  {NoStop}%
\bibitem [{\citenamefont {Morras}\ \emph {et~al.}(2025)\citenamefont {Morras},
  \citenamefont {Pratten},\ and\ \citenamefont {Schmidt}}]{Morras:2025nlp}%
  \BibitemOpen
  \bibfield  {author} {\bibinfo {author} {\bibfnamefont {G.}~\bibnamefont
  {Morras}}, \bibinfo {author} {\bibfnamefont {G.}~\bibnamefont {Pratten}},\
  and\ \bibinfo {author} {\bibfnamefont {P.}~\bibnamefont {Schmidt}},\
  }\bibfield  {title} {\bibinfo {title} {{Improved post-Newtonian waveform
  model for inspiralling precessing-eccentric compact binaries}},\ }\href
  {https://doi.org/10.1103/PhysRevD.111.084052} {\bibfield  {journal} {\bibinfo
   {journal} {Phys. Rev. D}\ }\textbf {\bibinfo {volume} {111}},\ \bibinfo
  {pages} {084052} (\bibinfo {year} {2025})},\ \Eprint
  {https://arxiv.org/abs/2502.03929} {arXiv:2502.03929 [gr-qc]} \BibitemShut
  {NoStop}%
\bibitem [{\citenamefont {Damour}\ and\ \citenamefont
  {Deruelle}(1985)}]{Damour1985}%
  \BibitemOpen
  \bibfield  {author} {\bibinfo {author} {\bibfnamefont {T.}~\bibnamefont
  {Damour}}\ and\ \bibinfo {author} {\bibfnamefont {N.}~\bibnamefont
  {Deruelle}},\ }\bibfield  {title} {\bibinfo {title} {General relativistic
  celestial mechanics of binary systems. i. the post-newtonian motion},\ }\href
  {https://www.numdam.org/item/AIHPA_1985__43_1_107_0/} {\bibfield  {journal}
  {\bibinfo  {journal} {Annales de l'I.H.P. Physique th\'eorique}\ }\textbf
  {\bibinfo {volume} {43}},\ \bibinfo {pages} {107} (\bibinfo {year}
  {1985})}\BibitemShut {NoStop}%
\bibitem [{\citenamefont {Damour}\ \emph {et~al.}(2014)\citenamefont {Damour},
  \citenamefont {Jaranowski},\ and\ \citenamefont {Sch{\"a}fer}}]{Damour2014}%
  \BibitemOpen
  \bibfield  {author} {\bibinfo {author} {\bibfnamefont {T.}~\bibnamefont
  {Damour}}, \bibinfo {author} {\bibfnamefont {P.}~\bibnamefont {Jaranowski}},\
  and\ \bibinfo {author} {\bibfnamefont {G.}~\bibnamefont {Sch{\"a}fer}},\
  }\bibfield  {title} {\bibinfo {title} {Nonlocal-in-time action for the fourth
  post-newtonian conservative dynamics of two-body systems},\ }\href@noop {}
  {\bibfield  {journal} {\bibinfo  {journal} {Physical Review D}\ }\textbf
  {\bibinfo {volume} {89}},\ \bibinfo {pages} {064058} (\bibinfo {year}
  {2014})}\BibitemShut {NoStop}%
\bibitem [{\citenamefont {Cho}\ \emph {et~al.}(2022)\citenamefont {Cho},
  \citenamefont {Tanay}, \citenamefont {Gopakumar},\ and\ \citenamefont
  {Lee}}]{Cho2022}%
  \BibitemOpen
  \bibfield  {author} {\bibinfo {author} {\bibfnamefont {G.}~\bibnamefont
  {Cho}}, \bibinfo {author} {\bibfnamefont {S.}~\bibnamefont {Tanay}}, \bibinfo
  {author} {\bibfnamefont {A.}~\bibnamefont {Gopakumar}},\ and\ \bibinfo
  {author} {\bibfnamefont {H.~M.}\ \bibnamefont {Lee}},\ }\bibfield  {title}
  {\bibinfo {title} {Generalized quasi-keplerian solution for eccentric,
  nonspinning compact binaries at 4pn order and the associated
  inspiral-merger-ringdown waveform},\ }\href@noop {} {\bibfield  {journal}
  {\bibinfo  {journal} {Physical Review D}\ }\textbf {\bibinfo {volume}
  {105}},\ \bibinfo {pages} {064010} (\bibinfo {year} {2022})}\BibitemShut
  {NoStop}%
\bibitem [{\citenamefont {Tessmer}\ \emph {et~al.}(2010)\citenamefont
  {Tessmer}, \citenamefont {Hartung},\ and\ \citenamefont
  {Schafer}}]{Tessmer2010}%
  \BibitemOpen
  \bibfield  {author} {\bibinfo {author} {\bibfnamefont {M.}~\bibnamefont
  {Tessmer}}, \bibinfo {author} {\bibfnamefont {J.}~\bibnamefont {Hartung}},\
  and\ \bibinfo {author} {\bibfnamefont {G.}~\bibnamefont {Schafer}},\
  }\bibfield  {title} {\bibinfo {title} {{Motion and gravitational wave forms
  of eccentric compact binaries with orbital-angular-momentum-aligned spins
  under next-to-leading order in spin-orbit and leading order in
  spin(1)-spin(2) and spin-squared couplings}},\ }\href
  {https://doi.org/10.1088/0264-9381/27/16/165005} {\bibfield  {journal}
  {\bibinfo  {journal} {Class. Quant. Grav.}\ }\textbf {\bibinfo {volume}
  {27}},\ \bibinfo {pages} {165005} (\bibinfo {year} {2010})},\ \Eprint
  {https://arxiv.org/abs/1003.2735} {arXiv:1003.2735 [gr-qc]} \BibitemShut
  {NoStop}%
\bibitem [{\citenamefont {Tessmer}\ \emph {et~al.}(2013)\citenamefont
  {Tessmer}, \citenamefont {Hartung},\ and\ \citenamefont
  {Sch{\"a}fer}}]{Tessmer2013}%
  \BibitemOpen
  \bibfield  {author} {\bibinfo {author} {\bibfnamefont {M.}~\bibnamefont
  {Tessmer}}, \bibinfo {author} {\bibfnamefont {J.}~\bibnamefont {Hartung}},\
  and\ \bibinfo {author} {\bibfnamefont {G.}~\bibnamefont {Sch{\"a}fer}},\
  }\bibfield  {title} {\bibinfo {title} {Aligned spins: orbital elements,
  decaying orbits, and last stable circular orbit to high post-newtonian
  orders},\ }\href@noop {} {\bibfield  {journal} {\bibinfo  {journal}
  {Classical and Quantum Gravity}\ }\textbf {\bibinfo {volume} {30}},\ \bibinfo
  {pages} {015007} (\bibinfo {year} {2013})}\BibitemShut {NoStop}%
\bibitem [{\citenamefont {Cho}\ and\ \citenamefont {Lee}(2019)}]{Cho2019}%
  \BibitemOpen
  \bibfield  {author} {\bibinfo {author} {\bibfnamefont {G.}~\bibnamefont
  {Cho}}\ and\ \bibinfo {author} {\bibfnamefont {H.~M.}\ \bibnamefont {Lee}},\
  }\bibfield  {title} {\bibinfo {title} {Analytic keplerian-type
  parametrization for general spinning compact binaries with leading order
  spin-orbit interactions},\ }\href
  {https://doi.org/10.1103/PhysRevD.100.044046} {\bibfield  {journal} {\bibinfo
   {journal} {Phys. Rev. D}\ }\textbf {\bibinfo {volume} {100}},\ \bibinfo
  {pages} {044046} (\bibinfo {year} {2019})},\ \Eprint
  {https://arxiv.org/abs/1908.02927} {arXiv:1908.02927 [gr-qc]} \BibitemShut
  {NoStop}%
\bibitem [{\citenamefont {{Wagoner}}\ and\ \citenamefont
  {{Will}}(1976)}]{1976ApJ...210..764W}%
  \BibitemOpen
  \bibfield  {author} {\bibinfo {author} {\bibfnamefont {R.~V.}\ \bibnamefont
  {{Wagoner}}}\ and\ \bibinfo {author} {\bibfnamefont {C.~M.}\ \bibnamefont
  {{Will}}},\ }\bibfield  {title} {\bibinfo {title} {{Post-Newtonian
  gravitational radiation from orbiting point masses.}},\ }\href
  {https://doi.org/10.1086/154886} {\bibfield  {journal} {\bibinfo  {journal}
  {\apj}\ }\textbf {\bibinfo {volume} {210}},\ \bibinfo {pages} {764} (\bibinfo
  {year} {1976})}\BibitemShut {NoStop}%
\bibitem [{\citenamefont {Sch{\"a}fer}\ and\ \citenamefont
  {Wex}(1993)}]{Schafer1993}%
  \BibitemOpen
  \bibfield  {author} {\bibinfo {author} {\bibfnamefont {G.}~\bibnamefont
  {Sch{\"a}fer}}\ and\ \bibinfo {author} {\bibfnamefont {N.}~\bibnamefont
  {Wex}},\ }\bibfield  {title} {\bibinfo {title} {Second post-newtonian motion
  of compact binaries},\ }\href@noop {} {\bibfield  {journal} {\bibinfo
  {journal} {Physics Letters A}\ }\textbf {\bibinfo {volume} {174}},\ \bibinfo
  {pages} {196} (\bibinfo {year} {1993})}\BibitemShut {NoStop}%
\bibitem [{\citenamefont {Memmesheimer}\ \emph {et~al.}(2004)\citenamefont
  {Memmesheimer}, \citenamefont {Gopakumar},\ and\ \citenamefont
  {Schaefer}}]{Memmesheimer:2004cv}%
  \BibitemOpen
  \bibfield  {author} {\bibinfo {author} {\bibfnamefont {R.-M.}\ \bibnamefont
  {Memmesheimer}}, \bibinfo {author} {\bibfnamefont {A.}~\bibnamefont
  {Gopakumar}},\ and\ \bibinfo {author} {\bibfnamefont {G.}~\bibnamefont
  {Schaefer}},\ }\bibfield  {title} {\bibinfo {title} {{Third post-Newtonian
  accurate generalized quasi-Keplerian parametrization for compact binaries in
  eccentric orbits}},\ }\href {https://doi.org/10.1103/PhysRevD.70.104011}
  {\bibfield  {journal} {\bibinfo  {journal} {Phys. Rev. D}\ }\textbf {\bibinfo
  {volume} {70}},\ \bibinfo {pages} {104011} (\bibinfo {year} {2004})},\
  \Eprint {https://arxiv.org/abs/gr-qc/0407049} {arXiv:gr-qc/0407049}
  \BibitemShut {NoStop}%
\bibitem [{\citenamefont {Arnowitt}\ \emph {et~al.}(2008)\citenamefont
  {Arnowitt}, \citenamefont {Deser},\ and\ \citenamefont
  {Misner}}]{Arnowitt2008}%
  \BibitemOpen
  \bibfield  {author} {\bibinfo {author} {\bibfnamefont {R.}~\bibnamefont
  {Arnowitt}}, \bibinfo {author} {\bibfnamefont {S.}~\bibnamefont {Deser}},\
  and\ \bibinfo {author} {\bibfnamefont {C.~W.}\ \bibnamefont {Misner}},\
  }\bibfield  {title} {\bibinfo {title} {Republication of: The dynamics of
  general relativity},\ }\href {https://doi.org/10.1007/s10714-008-0661-1}
  {\bibfield  {journal} {\bibinfo  {journal} {General Relativity and
  Gravitation}\ }\textbf {\bibinfo {volume} {40}},\ \bibinfo {pages}
  {1997–2027} (\bibinfo {year} {2008})}\BibitemShut {NoStop}%
\bibitem [{\citenamefont {Samanta}\ \emph {et~al.}(2023)\citenamefont
  {Samanta}, \citenamefont {Tanay},\ and\ \citenamefont {Stein}}]{Samanta2022}%
  \BibitemOpen
  \bibfield  {author} {\bibinfo {author} {\bibfnamefont {R.}~\bibnamefont
  {Samanta}}, \bibinfo {author} {\bibfnamefont {S.}~\bibnamefont {Tanay}},\
  and\ \bibinfo {author} {\bibfnamefont {L.~C.}\ \bibnamefont {Stein}},\
  }\bibfield  {title} {\bibinfo {title} {Closed-form solutions of spinning,
  eccentric binary black holes at 1.5 post-newtonian order},\ }\href
  {https://doi.org/10.1103/PhysRevD.108.124039} {\bibfield  {journal} {\bibinfo
   {journal} {Phys. Rev. D}\ }\textbf {\bibinfo {volume} {108}},\ \bibinfo
  {pages} {124039} (\bibinfo {year} {2023})},\ \Eprint
  {https://arxiv.org/abs/2210.01605} {arXiv:2210.01605 [gr-qc]} \BibitemShut
  {NoStop}%
\bibitem [{\citenamefont {Schnittman}(2004)}]{Schnittman2004}%
  \BibitemOpen
  \bibfield  {author} {\bibinfo {author} {\bibfnamefont {J.~D.}\ \bibnamefont
  {Schnittman}},\ }\bibfield  {title} {\bibinfo {title} {Spin-orbit resonance
  and the evolution of compact binary systems},\ }\href
  {https://doi.org/10.1103/PhysRevD.70.124020} {\bibfield  {journal} {\bibinfo
  {journal} {Phys. Rev. D}\ }\textbf {\bibinfo {volume} {70}},\ \bibinfo
  {pages} {124020} (\bibinfo {year} {2004})},\ \Eprint
  {https://arxiv.org/abs/astro-ph/0409174} {arXiv:astro-ph/0409174}
  \BibitemShut {NoStop}%
\bibitem [{\citenamefont {Gerosa}\ \emph
  {et~al.}(2015{\natexlab{a}})\citenamefont {Gerosa}, \citenamefont {Kesden},
  \citenamefont {O'Shaughnessy}, \citenamefont {Klein}, \citenamefont {Berti},
  \citenamefont {Sperhake},\ and\ \citenamefont
  {Trifir{\`o}}}]{Gerosa:2015hba}%
  \BibitemOpen
  \bibfield  {author} {\bibinfo {author} {\bibfnamefont {D.}~\bibnamefont
  {Gerosa}}, \bibinfo {author} {\bibfnamefont {M.}~\bibnamefont {Kesden}},
  \bibinfo {author} {\bibfnamefont {R.}~\bibnamefont {O'Shaughnessy}}, \bibinfo
  {author} {\bibfnamefont {A.}~\bibnamefont {Klein}}, \bibinfo {author}
  {\bibfnamefont {E.}~\bibnamefont {Berti}}, \bibinfo {author} {\bibfnamefont
  {U.}~\bibnamefont {Sperhake}},\ and\ \bibinfo {author} {\bibfnamefont
  {D.}~\bibnamefont {Trifir{\`o}}},\ }\bibfield  {title} {\bibinfo {title}
  {{Precessional instability in binary black holes with aligned spins}},\
  }\href {https://doi.org/10.1103/PhysRevLett.115.141102} {\bibfield  {journal}
  {\bibinfo  {journal} {Phys. Rev. Lett.}\ }\textbf {\bibinfo {volume} {115}},\
  \bibinfo {pages} {141102} (\bibinfo {year} {2015}{\natexlab{a}})},\ \Eprint
  {https://arxiv.org/abs/1506.09116} {arXiv:1506.09116 [gr-qc]} \BibitemShut
  {NoStop}%
\bibitem [{\citenamefont {Gerosa}\ \emph
  {et~al.}(2015{\natexlab{b}})\citenamefont {Gerosa}, \citenamefont {Kesden},
  \citenamefont {Sperhake}, \citenamefont {Berti},\ and\ \citenamefont
  {O'Shaughnessy}}]{Gerosa:2015tea}%
  \BibitemOpen
  \bibfield  {author} {\bibinfo {author} {\bibfnamefont {D.}~\bibnamefont
  {Gerosa}}, \bibinfo {author} {\bibfnamefont {M.}~\bibnamefont {Kesden}},
  \bibinfo {author} {\bibfnamefont {U.}~\bibnamefont {Sperhake}}, \bibinfo
  {author} {\bibfnamefont {E.}~\bibnamefont {Berti}},\ and\ \bibinfo {author}
  {\bibfnamefont {R.}~\bibnamefont {O'Shaughnessy}},\ }\bibfield  {title}
  {\bibinfo {title} {{Multi-timescale analysis of phase transitions in
  precessing black-hole binaries}},\ }\href
  {https://doi.org/10.1103/PhysRevD.92.064016} {\bibfield  {journal} {\bibinfo
  {journal} {Phys. Rev. D}\ }\textbf {\bibinfo {volume} {92}},\ \bibinfo
  {pages} {064016} (\bibinfo {year} {2015}{\natexlab{b}})},\ \Eprint
  {https://arxiv.org/abs/1506.03492} {arXiv:1506.03492 [gr-qc]} \BibitemShut
  {NoStop}%
\bibitem [{\citenamefont {Tanay}\ \emph {et~al.}(2021)\citenamefont {Tanay},
  \citenamefont {Stein},\ and\ \citenamefont {G\'alvez~Ghersi}}]{Tanay2020}%
  \BibitemOpen
  \bibfield  {author} {\bibinfo {author} {\bibfnamefont {S.}~\bibnamefont
  {Tanay}}, \bibinfo {author} {\bibfnamefont {L.~C.}\ \bibnamefont {Stein}},\
  and\ \bibinfo {author} {\bibfnamefont {J.~T.}\ \bibnamefont
  {G\'alvez~Ghersi}},\ }\bibfield  {title} {\bibinfo {title} {{Integrability of
  eccentric, spinning black hole binaries up to second post-Newtonian order}},\
  }\href {https://doi.org/10.1103/PhysRevD.103.064066} {\bibfield  {journal}
  {\bibinfo  {journal} {Phys. Rev. D}\ }\textbf {\bibinfo {volume} {103}},\
  \bibinfo {pages} {064066} (\bibinfo {year} {2021})},\ \Eprint
  {https://arxiv.org/abs/2012.06586} {arXiv:2012.06586 [gr-qc]} \BibitemShut
  {NoStop}%
\bibitem [{\citenamefont {Racine}(2008)}]{Racine2008}%
  \BibitemOpen
  \bibfield  {author} {\bibinfo {author} {\bibfnamefont {E.}~\bibnamefont
  {Racine}},\ }\bibfield  {title} {\bibinfo {title} {Analysis of spin
  precession in binary black hole systems including quadrupole-monopole
  interaction},\ }\href {https://doi.org/10.1103/PhysRevD.78.044021} {\bibfield
   {journal} {\bibinfo  {journal} {Phys. Rev. D}\ }\textbf {\bibinfo {volume}
  {78}},\ \bibinfo {pages} {044021} (\bibinfo {year} {2008})},\ \Eprint
  {https://arxiv.org/abs/0803.1820} {arXiv:0803.1820 [gr-qc]} \BibitemShut
  {NoStop}%
\bibitem [{\citenamefont {Gergely}(2000)}]{Gergely:1999pd}%
  \BibitemOpen
  \bibfield  {author} {\bibinfo {author} {\bibfnamefont {L.~A.}\ \bibnamefont
  {Gergely}},\ }\bibfield  {title} {\bibinfo {title} {{Spin spin effects in
  radiating compact binaries}},\ }\href
  {https://doi.org/10.1103/PhysRevD.61.024035} {\bibfield  {journal} {\bibinfo
  {journal} {Phys. Rev. D}\ }\textbf {\bibinfo {volume} {61}},\ \bibinfo
  {pages} {024035} (\bibinfo {year} {2000})},\ \Eprint
  {https://arxiv.org/abs/gr-qc/9911082} {arXiv:gr-qc/9911082} \BibitemShut
  {NoStop}%
\bibitem [{\citenamefont {Klein}\ and\ \citenamefont
  {Jetzer}(2010)}]{Klein2010}%
  \BibitemOpen
  \bibfield  {author} {\bibinfo {author} {\bibfnamefont {A.}~\bibnamefont
  {Klein}}\ and\ \bibinfo {author} {\bibfnamefont {P.}~\bibnamefont {Jetzer}},\
  }\bibfield  {title} {\bibinfo {title} {Spin effects in the phasing of
  gravitational waves from binaries on eccentric orbits},\ }\href@noop {}
  {\bibfield  {journal} {\bibinfo  {journal} {Physical Review D—Particles,
  Fields, Gravitation, and Cosmology}\ }\textbf {\bibinfo {volume} {81}},\
  \bibinfo {pages} {124001} (\bibinfo {year} {2010})}\BibitemShut {NoStop}%
\bibitem [{\citenamefont {Kidder}(1995)}]{Kidder1995}%
  \BibitemOpen
  \bibfield  {author} {\bibinfo {author} {\bibfnamefont {L.~E.}\ \bibnamefont
  {Kidder}},\ }\bibfield  {title} {\bibinfo {title} {Coalescing binary systems
  of compact objects to (post) 5/2-newtonian order. v. spin effects},\
  }\href@noop {} {\bibfield  {journal} {\bibinfo  {journal} {Physical Review
  D}\ }\textbf {\bibinfo {volume} {52}},\ \bibinfo {pages} {821} (\bibinfo
  {year} {1995})}\BibitemShut {NoStop}%
\bibitem [{\citenamefont {Barker}\ and\ \citenamefont
  {O'Connell}(1974)}]{Barker1974}%
  \BibitemOpen
  \bibfield  {author} {\bibinfo {author} {\bibfnamefont {B.~M.}\ \bibnamefont
  {Barker}}\ and\ \bibinfo {author} {\bibfnamefont {R.~F.}\ \bibnamefont
  {O'Connell}},\ }\bibfield  {title} {\bibinfo {title} {Nongeodesic motion in
  general relativity},\ }\href {https://doi.org/10.1007/BF02451397} {\bibfield
  {journal} {\bibinfo  {journal} {General Relativity and Gravitation}\ }\textbf
  {\bibinfo {volume} {5}},\ \bibinfo {pages} {539} (\bibinfo {year}
  {1974})}\BibitemShut {NoStop}%
\bibitem [{\citenamefont {Hartl}\ and\ \citenamefont
  {Buonanno}(2005)}]{Hartl:2004xr}%
  \BibitemOpen
  \bibfield  {author} {\bibinfo {author} {\bibfnamefont {M.~D.}\ \bibnamefont
  {Hartl}}\ and\ \bibinfo {author} {\bibfnamefont {A.}~\bibnamefont
  {Buonanno}},\ }\bibfield  {title} {\bibinfo {title} {{The Dynamics of
  precessing binary black holes using the post-Newtonian approximation}},\
  }\href {https://doi.org/10.1103/PhysRevD.71.024027} {\bibfield  {journal}
  {\bibinfo  {journal} {Phys. Rev. D}\ }\textbf {\bibinfo {volume} {71}},\
  \bibinfo {pages} {024027} (\bibinfo {year} {2005})},\ \Eprint
  {https://arxiv.org/abs/gr-qc/0407091} {arXiv:gr-qc/0407091} \BibitemShut
  {NoStop}%
\bibitem [{\citenamefont {Gopakumar}\ and\ \citenamefont
  {Konigsdorffer}(2005)}]{konigs:2005zz}%
  \BibitemOpen
  \bibfield  {author} {\bibinfo {author} {\bibfnamefont {A.}~\bibnamefont
  {Gopakumar}}\ and\ \bibinfo {author} {\bibfnamefont {C.}~\bibnamefont
  {Konigsdorffer}},\ }\bibfield  {title} {\bibinfo {title} {{The Deterministic
  nature of conservative post-Newtonian accurate dynamics of compact binaries
  with leading order spin-orbit interaction}},\ }\href
  {https://doi.org/10.1103/PhysRevD.72.121501} {\bibfield  {journal} {\bibinfo
  {journal} {Phys. Rev. D}\ }\textbf {\bibinfo {volume} {72}},\ \bibinfo
  {pages} {121501} (\bibinfo {year} {2005})},\ \Eprint
  {https://arxiv.org/abs/gr-qc/0511009} {arXiv:gr-qc/0511009} \BibitemShut
  {NoStop}%
\bibitem [{\citenamefont {Colin}\ and\ \citenamefont
  {Tanay}(2026)}]{Tom_Colin_2026_BBH_Notebook}%
  \BibitemOpen
  \bibfield  {author} {\bibinfo {author} {\bibfnamefont {T.}~\bibnamefont
  {Colin}}\ and\ \bibinfo {author} {\bibfnamefont {S.}~\bibnamefont {Tanay}},\
  }\href@noop {} {\bibinfo {title} {{Mathematica notebook for the analytical
  2PN spinning, eccentric binary black hole solution}}},\ \bibinfo
  {howpublished} {\url{https://github.com/sashwattanay/BBH-PN-Toolkit}}
  (\bibinfo {year} {2026}),\ \bibinfo {note} {directory:
  \texttt{2PN\_BBH\_solution}. Main file: \texttt{2PN\_BBH\_solution.nb} (also
  provided in .m, .pdf, and .txt formats).}\BibitemShut {Stop}%
\bibitem [{\citenamefont {Pryce}(1948)}]{Pryce1948}%
  \BibitemOpen
  \bibfield  {author} {\bibinfo {author} {\bibfnamefont {M.~H.~L.}\
  \bibnamefont {Pryce}},\ }\bibfield  {title} {\bibinfo {title} {The
  mass-centre in the restricted theory of relativity and its connexion with the
  quantum theory of elementary particles},\ }\href
  {https://doi.org/10.1098/rspa.1948.0103} {\bibfield  {journal} {\bibinfo
  {journal} {Proceedings of the Royal Society of London. Series A. Mathematical
  and Physical Sciences}\ }\textbf {\bibinfo {volume} {195}},\ \bibinfo {pages}
  {62} (\bibinfo {year} {1948})}\BibitemShut {NoStop}%
\bibitem [{\citenamefont {Newton}\ and\ \citenamefont
  {Wigner}(1949)}]{Newton1949}%
  \BibitemOpen
  \bibfield  {author} {\bibinfo {author} {\bibfnamefont {T.~D.}\ \bibnamefont
  {Newton}}\ and\ \bibinfo {author} {\bibfnamefont {E.~P.}\ \bibnamefont
  {Wigner}},\ }\bibfield  {title} {\bibinfo {title} {Localized states for
  elementary systems},\ }\href {https://doi.org/10.1103/RevModPhys.21.400}
  {\bibfield  {journal} {\bibinfo  {journal} {Rev. Mod. Phys.}\ }\textbf
  {\bibinfo {volume} {21}},\ \bibinfo {pages} {400} (\bibinfo {year}
  {1949})}\BibitemShut {NoStop}%
\bibitem [{\citenamefont {Damour}(2001)}]{Damour2001}%
  \BibitemOpen
  \bibfield  {author} {\bibinfo {author} {\bibfnamefont {T.}~\bibnamefont
  {Damour}},\ }\bibfield  {title} {\bibinfo {title} {Coalescence of two
  spinning black holes: An effective one-body approach},\ }\href
  {https://doi.org/10.1103/PhysRevD.64.124013} {\bibfield  {journal} {\bibinfo
  {journal} {Phys. Rev. D}\ }\textbf {\bibinfo {volume} {64}},\ \bibinfo
  {pages} {124013} (\bibinfo {year} {2001})}\BibitemShut {NoStop}%
\bibitem [{\citenamefont {Barker}\ \emph {et~al.}(1966)\citenamefont {Barker},
  \citenamefont {Gupta},\ and\ \citenamefont {Haracz}}]{Barker:1966zz}%
  \BibitemOpen
  \bibfield  {author} {\bibinfo {author} {\bibfnamefont {B.~M.}\ \bibnamefont
  {Barker}}, \bibinfo {author} {\bibfnamefont {S.~N.}\ \bibnamefont {Gupta}},\
  and\ \bibinfo {author} {\bibfnamefont {R.~D.}\ \bibnamefont {Haracz}},\
  }\bibfield  {title} {\bibinfo {title} {{One-Graviton Exchange Interaction of
  Elementary Particles}},\ }\href {https://doi.org/10.1103/PhysRev.149.1027}
  {\bibfield  {journal} {\bibinfo  {journal} {Phys. Rev.}\ }\textbf {\bibinfo
  {volume} {149}},\ \bibinfo {pages} {1027} (\bibinfo {year}
  {1966})}\BibitemShut {NoStop}%
\bibitem [{\citenamefont {Barker}\ and\ \citenamefont
  {O'Connell}(1975)}]{PhysRevD.12.329}%
  \BibitemOpen
  \bibfield  {author} {\bibinfo {author} {\bibfnamefont {B.~M.}\ \bibnamefont
  {Barker}}\ and\ \bibinfo {author} {\bibfnamefont {R.~F.}\ \bibnamefont
  {O'Connell}},\ }\bibfield  {title} {\bibinfo {title} {Gravitational two-body
  problem with arbitrary masses, spins, and quadrupole moments},\ }\href
  {https://doi.org/10.1103/PhysRevD.12.329} {\bibfield  {journal} {\bibinfo
  {journal} {Phys. Rev. D}\ }\textbf {\bibinfo {volume} {12}},\ \bibinfo
  {pages} {329} (\bibinfo {year} {1975})}\BibitemShut {NoStop}%
\bibitem [{\citenamefont {Steinhoff}(2011)}]{Steinhoff:2011sya}%
  \BibitemOpen
  \bibfield  {author} {\bibinfo {author} {\bibfnamefont {J.}~\bibnamefont
  {Steinhoff}},\ }\bibfield  {title} {\bibinfo {title} {{Canonical formulation
  of spin in general relativity}},\ }\href
  {https://doi.org/10.1002/andp.201000178} {\bibfield  {journal} {\bibinfo
  {journal} {Annalen Phys.}\ }\textbf {\bibinfo {volume} {523}},\ \bibinfo
  {pages} {296} (\bibinfo {year} {2011})},\ \Eprint
  {https://arxiv.org/abs/1106.4203} {arXiv:1106.4203 [gr-qc]} \BibitemShut
  {NoStop}%
\bibitem [{\citenamefont {Poisson}\ and\ \citenamefont
  {Will}(2014)}]{Poisson_Will_2014}%
  \BibitemOpen
  \bibfield  {author} {\bibinfo {author} {\bibfnamefont {E.}~\bibnamefont
  {Poisson}}\ and\ \bibinfo {author} {\bibfnamefont {C.~M.}\ \bibnamefont
  {Will}},\ }\href@noop {} {\emph {\bibinfo {title} {Gravity: Newtonian,
  Post-Newtonian, Relativistic}}}\ (\bibinfo  {publisher} {Cambridge University
  Press},\ \bibinfo {year} {2014})\BibitemShut {NoStop}%
\bibitem [{\citenamefont {Spitale}\ and\ \citenamefont
  {Greenberg}(2002)}]{SPITALE2002211}%
  \BibitemOpen
  \bibfield  {author} {\bibinfo {author} {\bibfnamefont {J.}~\bibnamefont
  {Spitale}}\ and\ \bibinfo {author} {\bibfnamefont {R.}~\bibnamefont
  {Greenberg}},\ }\bibfield  {title} {\bibinfo {title} {Numerical evaluation of
  the general yarkovsky effect: Effects on eccentricity and longitude of
  periapse},\ }\href {https://doi.org/https://doi.org/10.1006/icar.2001.6774}
  {\bibfield  {journal} {\bibinfo  {journal} {Icarus}\ }\textbf {\bibinfo
  {volume} {156}},\ \bibinfo {pages} {211} (\bibinfo {year}
  {2002})}\BibitemShut {NoStop}%
\bibitem [{\citenamefont {{Scheeres}}\ and\ \citenamefont
  {{Hu}}(2001)}]{Scheeres}%
  \BibitemOpen
  \bibfield  {author} {\bibinfo {author} {\bibfnamefont {D.~J.}\ \bibnamefont
  {{Scheeres}}}\ and\ \bibinfo {author} {\bibfnamefont {W.}~\bibnamefont
  {{Hu}}},\ }\bibfield  {title} {\bibinfo {title} {{Secular Motion in a 2nd
  Degree and Order-Gravity Field with no Rotation}},\ }\href
  {https://doi.org/10.1023/A:1017555005699} {\bibfield  {journal} {\bibinfo
  {journal} {Celestial Mechanics and Dynamical Astronomy}\ }\textbf {\bibinfo
  {volume} {79}},\ \bibinfo {pages} {183} (\bibinfo {year} {2001})}\BibitemShut
  {NoStop}%
\bibitem [{\citenamefont {Tanay}\ \emph {et~al.}(2023)\citenamefont {Tanay},
  \citenamefont {Stein},\ and\ \citenamefont {Cho}}]{Tanay2023}%
  \BibitemOpen
  \bibfield  {author} {\bibinfo {author} {\bibfnamefont {S.}~\bibnamefont
  {Tanay}}, \bibinfo {author} {\bibfnamefont {L.~C.}\ \bibnamefont {Stein}},\
  and\ \bibinfo {author} {\bibfnamefont {G.}~\bibnamefont {Cho}},\ }\bibfield
  {title} {\bibinfo {title} {Action-angle variables of a binary black hole with
  arbitrary eccentricity, spins, and masses at 1.5 post-newtonian order},\
  }\href@noop {} {\bibfield  {journal} {\bibinfo  {journal} {Physical Review
  D}\ }\textbf {\bibinfo {volume} {107}},\ \bibinfo {pages} {103040} (\bibinfo
  {year} {2023})}\BibitemShut {NoStop}%
\bibitem [{\citenamefont {Jos{\'e}}\ and\ \citenamefont
  {Saletan}(1998)}]{Jose1998}%
  \BibitemOpen
  \bibfield  {author} {\bibinfo {author} {\bibfnamefont {J.}~\bibnamefont
  {Jos{\'e}}}\ and\ \bibinfo {author} {\bibfnamefont {E.}~\bibnamefont
  {Saletan}},\ }\href {https://books.google.com/books?id=Eql9dRQDgvQC} {\emph
  {\bibinfo {title} {Classical Dynamics: A Contemporary Approach}}}\ (\bibinfo
  {publisher} {Cambridge University Press},\ \bibinfo {year}
  {1998})\BibitemShut {NoStop}%
\bibitem [{\citenamefont {da~Silva}(2001)}]{cannas}%
  \BibitemOpen
  \bibfield  {author} {\bibinfo {author} {\bibfnamefont {A.}~\bibnamefont
  {da~Silva}},\ }\href {https://books.google.com.pe/books?id=r9i6pXc7GEQC}
  {\emph {\bibinfo {title} {Lectures on Symplectic Geometry}}},\ \bibinfo
  {series} {Lecture Notes in Mathematics}\ No.\ \bibinfo {number} {no. 1764}\
  (\bibinfo  {publisher} {Springer},\ \bibinfo {year} {2001})\BibitemShut
  {NoStop}%
\bibitem [{\citenamefont {Fasano}\ and\ \citenamefont
  {Marmi}(2006)}]{Fasano2006}%
  \BibitemOpen
  \bibfield  {author} {\bibinfo {author} {\bibfnamefont {A.}~\bibnamefont
  {Fasano}}\ and\ \bibinfo {author} {\bibfnamefont {S.}~\bibnamefont {Marmi}},\
  }\href {https://books.google.ci/books?id=zDslMYMEGvoC} {\emph {\bibinfo
  {title} {Analytical Mechanics: An Introduction}}},\ Oxford Graduate Texts\
  (\bibinfo  {publisher} {OUP Oxford},\ \bibinfo {year} {2006})\BibitemShut
  {NoStop}%
\bibitem [{\citenamefont {Henry}\ and\ \citenamefont
  {Khalil}(2023)}]{Henry:2023tka}%
  \BibitemOpen
  \bibfield  {author} {\bibinfo {author} {\bibfnamefont {Q.}~\bibnamefont
  {Henry}}\ and\ \bibinfo {author} {\bibfnamefont {M.}~\bibnamefont {Khalil}},\
  }\bibfield  {title} {\bibinfo {title} {{Spin effects in gravitational
  waveforms and fluxes for binaries on eccentric orbits to the third
  post-Newtonian order}},\ }\href {https://doi.org/10.1103/PhysRevD.108.104016}
  {\bibfield  {journal} {\bibinfo  {journal} {Phys. Rev. D}\ }\textbf {\bibinfo
  {volume} {108}},\ \bibinfo {pages} {104016} (\bibinfo {year} {2023})},\
  \Eprint {https://arxiv.org/abs/2308.13606} {arXiv:2308.13606 [gr-qc]}
  \BibitemShut {NoStop}%
\bibitem [{\citenamefont {Drummond}\ and\ \citenamefont
  {Hughes}(2022{\natexlab{a}})}]{Drummond:2022xej}%
  \BibitemOpen
  \bibfield  {author} {\bibinfo {author} {\bibfnamefont {L.~V.}\ \bibnamefont
  {Drummond}}\ and\ \bibinfo {author} {\bibfnamefont {S.~A.}\ \bibnamefont
  {Hughes}},\ }\bibfield  {title} {\bibinfo {title} {{Precisely computing bound
  orbits of spinning bodies around black holes. I. General framework and
  results for nearly equatorial orbits}},\ }\href
  {https://doi.org/10.1103/PhysRevD.105.124040} {\bibfield  {journal} {\bibinfo
   {journal} {Phys. Rev. D}\ }\textbf {\bibinfo {volume} {105}},\ \bibinfo
  {pages} {124040} (\bibinfo {year} {2022}{\natexlab{a}})},\ \Eprint
  {https://arxiv.org/abs/2201.13334} {arXiv:2201.13334 [gr-qc]} \BibitemShut
  {NoStop}%
\bibitem [{\citenamefont {Drummond}\ and\ \citenamefont
  {Hughes}(2022{\natexlab{b}})}]{Drummond:2022efc}%
  \BibitemOpen
  \bibfield  {author} {\bibinfo {author} {\bibfnamefont {L.~V.}\ \bibnamefont
  {Drummond}}\ and\ \bibinfo {author} {\bibfnamefont {S.~A.}\ \bibnamefont
  {Hughes}},\ }\bibfield  {title} {\bibinfo {title} {{Precisely computing bound
  orbits of spinning bodies around black holes. II. Generic orbits}},\ }\href
  {https://doi.org/10.1103/PhysRevD.105.124041} {\bibfield  {journal} {\bibinfo
   {journal} {Phys. Rev. D}\ }\textbf {\bibinfo {volume} {105}},\ \bibinfo
  {pages} {124041} (\bibinfo {year} {2022}{\natexlab{b}})},\ \Eprint
  {https://arxiv.org/abs/2201.13335} {arXiv:2201.13335 [gr-qc]} \BibitemShut
  {NoStop}%
\end{thebibliography}%

\end{document}